\documentclass[aps,pra,pdf,superscriptaddress,twocolumn,showpacs,nofootinbib]{revtex4-2}
\usepackage{amsmath,amssymb,amsfonts}
\usepackage{eqnarray}
\usepackage{graphics,float}
\usepackage{bm}
\usepackage{braket}
\usepackage{amsmath}
\usepackage{physics}
\usepackage{mathtools}
\usepackage{gensymb}
\usepackage{color,xcolor,colortbl}
\usepackage{tikz}
\usepackage{soul}
\usetikzlibrary{shapes}
\usepackage[colorlinks=true,
linkcolor=blue,
filecolor=magenta,      
urlcolor=blue,
citecolor=blue]{hyperref}

\usepackage{lipsum}

\definecolor{Gray}{gray}{0.85}
\definecolor{LightCyan}{rgb}{0.25,0.9,0.95}
\definecolor{LightMagenta}{rgb}{0.91,0.21,0.96}

\graphicspath{{Plots/}}

\begin{document}
	\title{Roadmap to vortex nucleation below critical rotation frequency in a dipolar Bose-Einstein condensate}
	\author{Soumyadeep Halder}
	\email{soumya.hhs@gmail.com}
	\affiliation{Department of Physics, Indian Institute of Technology Kharagpur, Kharagpur 721302, India}
	
	\author{Hari Sadhan Ghosh}
	\affiliation{Department of Physics, Indian Institute of Technology Kharagpur, Kharagpur 721302, India}

	\author{Arpana Saboo}
	\affiliation{Department of Physics, Indian Institute of Technology Kharagpur, Kharagpur 721302, India}

	\author{Andy M. Martin}
	\affiliation{School of Physics, University of Melbourne, Parkville, Victoria 3010, Australia}
	
    \author{Sonjoy Majumder}
	\affiliation{Department of Physics, Indian Institute of Technology Kharagpur, Kharagpur 721302, India}
	
	\date{\today}
\begin{abstract}
The formation of quantized vortices in a superfluid above a certain critical trap rotation frequency serves as a hallmark signature of superfluidity. Based on the beyond mean field framework, crucial for the formation of exotic supersolid and droplet states, we investigate dynamic protocols for vortex nucleation in the superfluid and supersolid states of a dipolar Bose-Einstein condensate (BEC), at a significantly lower trap rotation frequency. We find that the critical rotation frequency of the trap varies with the dipole-dipole interaction strength and the polarization direction of the external magnetic field. Leveraging these characteristics of dipolar BECs, we demonstrate three dynamic protocols for vortex nucleation even when rotating below the critical rotation frequency \emph{viz}.: (i) varying the $s$-wave scattering length, (ii) changing the polarizing angle, and (iii) successive modulation of both the scattering length and polarizing angle. These dynamic vortex seeding protocols could serve as important benchmarks for future experimental studies.
\end{abstract} 
	
\maketitle

\section{Introduction}
Superfluidity is an intriguing phenomenon characterized by the ability to flow without viscosity, first observed in liquid helium and later in ultracold atomic Bose-Einstein condensates (BECs) \cite{london_1938_boseeinstein, london_1938_lphenomenon, landau_1941_theory, anderson_1995_observation}. The uniform phase across the condensate and the single-valued macroscopic wavefunction results in an irrotational velocity field. This leads to the quantization of circulation and the formation of quantized vortices \cite{yarmchuk_1979_observation, abo-shaeer_2001_observation, madison_2000_vortex}. These vortices are topologically protected singularities, characterized by a $2\pi$ phase winding and serve as a hallmark property of superfluidity \cite{onsager_1949_statistical, feynman_1955_chapter}. Since the first experimental realization of gaseous BECs \cite{anderson_1995_observation}, the study of the formation of quantized vortices and their dynamical evolution has become a major research focus. Quantized vortices in BECs can be nucleated using several distinct methods \emph{viz}., by rotating a slightly deformed trap \cite{abo-shaeer_2001_observation, haljan_2001_driving}, stirring the condensate with laser or a localized obstacle above certain critical velocity \cite{madison_2000_vortex, raman_2001_vortex, neely_2010_observation, kwon_2021_sound}, following a rapid temperature quench across the transition temperature for the onset of condensation (Kibble-Zurek mechanism) \cite{kibble_1976_topology, zurek_1985_cosmological, weiler_2008_spontaneous, donadello_2014_observation}, or directly imprinting the optical phase \cite{matthews_1999_vortices, leanhardt_2002_imprinting, kumar_2018_producing, andersen_2006_quantized}, etc.\par 
While earlier experimental works and theoretical investigations considered nondipolar BECs, dipolar bosonic gases composed of highly magnetic atoms such as Dy and Er, have attracted significant attention in recent years \cite{aikawa_2012_boseeinstein, tang_2015_bose, tang_2015_wave, lucioni_2018_dysprosium, chomaz_2022_dipolar}. The coexistence of long-range anisotropic dipole-dipole interactions (DDI) and short-range isotropic contact interactions provides a rich platform for studying distinctive vortex properties in dipolar BECs. Apart from all the standard methods, vortices in a dipolar BEC can also be nucleated by rotating the polarization direction, known as magnetostirring \cite{prasad_2019_vortex, prasad_2021_arbitraryangle, klaus_2022_observationa}. A vortex in a dipolar BEC produces a mesoscopic dipolar potential that alters the interaction among vortices, and inhibits vortex-antivortex annihilation \cite{mulkerin_2013_anisotropic, gautam_2014_dynamics}. Furthermore, a tilted magnetic field introduces an in-plane anisotropy, resulting in the formation of elliptical vortex cores and vortex stripes in a dipolar BEC \cite{yi_2006_vortex, prasad_2019_vortex, klaus_2022_observationa, cai_2018_vortex}. \par 
Dipolar BECs also have emerged as an intriguing system for studying various fascinating quantum phenomena such as anisotropic superfluidity \cite{ticknor_2011_anisotropic, bismut_2012_anisotropic, wenzel_2018_anisotropic}, observation of the quantum analog of Rosenzweig instability \cite{kadau_2016_observing}, appearance of roton excitations \cite{bisset_2013_roton, schmidt_2021_roton, blakie_2012_roton, petter_2019_probing, santos_2003_rotonmaxon, natale_2019_excitation, kirkby_2023_spin, chomaz_2018_observation, lee_2022_stability}, formation of self-bound quantum droplets \cite{baillie_2016_selfbound, ferrier-barbut_2016_observation, baillie_2017_collective, schmitt_2016_selfbound, chomaz_2016_quantumfluctuationdriven, wachtler_2016_groundstate, wachtler_2016_quantum, mishra_2020_selfbound, bisset_2021_quantum}, and supersolid states \cite{hertkorn_2021_supersolidity, tanzi_2019_observation, smith_2023_supersolidity, blakie_2020_supersolidity, poli_2021_maintaining, roccuzzo_2019_supersolid, norcia_2021_twodimensional, bland_2022_twodimensional,halder_2022_control}. The supersolid state is a unique state of matter that combines a crystalline structure with superfluidity and has been observed in a series of theoretical and experimental studies over the past few years, including molecular dipolar BECs \cite{schmidt_2022_selfbound, langen_2024_quantum, bigagli_2024_observation}, bubble-trapped dipolar BECs \cite{ghosh_2024_induced, sanchez-baena_2024_ring, ciardi_2024_supersolid, prestipino_2019_ground}, binary dipolar mixtures \cite{halder_2023_twodimensional, halder_2024_induced, scheiermann_2023_catalyzation, bland_2022_alternatingdomain, li_2022_longlifetime, arazo_2023_selfbound}, Rydberg atoms \cite{cinti_2010_supersolid, henkel_2010_threedimensional, henkel_2012_supersolid} and spin-orbit coupled BECs \cite{wang_2010_spinorbit, li_2013_superstripes, li_2016_spinorbit, li_2017_stripe, bersano_2019_experimental, putra_2020_spatial, sachdeva_2020_selfbound, geier_2021_exciting}. The rotational response of a supersolid state provides a crucial test to identify superfluid properties in a supersolid \cite{roccuzzo_2022_moment}. Although, the formation of vortices was predicted long ago in a model of the supersolid state \cite{pomeau_1994_dynamics}, their realization in a supersolid state has been the object of recent theoretical and experimental investigations. Theoretical investigations have predicted unique characteristics of these vortices in a supersolid, including their robust character \cite{sindik_2022_creation}, a deformed vortex core \cite{gallemi_2020_quantized, roccuzzo_2020_rotating}, vortex pinning-unpinning, and snaking of vortices between the periodic crystalline structure \cite{ancilotto_2021_vortex, poli_2023_glitches}. Despite the experimental challenges, vortices have recently been observed in a dipolar supersolid \cite{casotti_2024_observation}.\par  
A vortex in a superfluid state becomes energetically favorable above a certain critical threshold of external rotation frequency \cite{dalfovo_1996_bosons, sinha_1997_semiclassical, lundh_1997_zerotemperature}. However, dynamical vortex nucleation via sudden introduction of trap rotation requires a higher rotation frequency due to the presence of an energy barrier for the vortex to seed into the condensate \cite{madison_2000_vortex}. For a BEC confined in a slightly deformed harmonic trap and rotating with a rotation frequency $\Omega\approx 0.7\omega$, where $\omega$ is the transverse trapping frequency, the condensate spontaneously undergoes dynamical instability of quadrupole mode \cite{sinha_2001_dynamic, kasamatsu_2003_nonlinear, madison_2001_stationary, parker_2006_instabilities}. When the atoms of a BEC interact only through short-range contact interaction, this critical rotation frequency is influenced solely by the trap's ellipticity, and a specific critical ellipticity is required for dynamical vortex nucleation \cite{madison_2001_stationary, parker_2006_instabilities}. In contrast, for a dipolar BEC, the critical rotation frequency is affected by both the trap's ellipticity and the strength of the DDI \cite{odell_2007_vortex, martin_2017_vortices}. As the strength of the DDI increases, both the thermodynamic critical rotational frequency and the quadrupolar frequency decreases  \cite{kasamatsu_2003_nonlinear, vanbijnen_2007_dynamical, vanbijnen_2009_exact}. Moreover, in a supersolid state, the intermediate low-density region between the droplets helps in reducing the energy barrier for vortex pinning \cite{gallemi_2020_quantized, roccuzzo_2020_rotating}. The inherent anisotropy and long-range character of DDI enhance our understanding of vortex formation and stability in both superfluid and supersolid phases. Extensive research has been conducted to find efficient methods for generating vortices in dipolar BECs, highlighting their potential for uncovering new quantum phenomena.\par
 In this article, we first evaluate the critical rotation frequency threshold for the nucleation of vortices in the ground states of superfluid and supersolid phases as a function of the angle between the rotation axis and the polarization direction. Interestingly, this critical rotation frequency differs between superfluid and supersolid states and varies in a unique way with the polarization direction. Moreover, we find that vortices in both superfluid and supersolid states are topologically robust against changes in the $s$-wave scattering length and polarizing angle.  Utilizing these properties, we demonstrate dynamic protocols for the nucleation of vortices even when rotating an axially symmetric trap below its critical rotation frequency in both superfluid and supersolid phases.\par
The subsequent material in this paper is arranged as follows. In Sec. \ref{secii}, we introduce the theoretical model in the form of an extended Gross-Pitaevskii equation (eGPE) governing the dynamics of a dipolar BEC. We discuss the critical rotation frequencies for the superfluid and supersolid phases of a dipolar BEC in Sec. \ref{seciii}. We propose different dynamical routes for vortex nucleation in a dipolar BEC in Sec. \ref{seciv}. During the dynamical evolution of the system, we achieve the vortex nucleation in the following ways: (a) by tuning the scattering length as discussed in Sec. \ref{sseciva}, (b) by tuning the polarizing angle as discussed in Sec. \ref{ssecivb} and (c) by varying both the polarization direction and the scattering length as discussed in Sec. \ref{ssecivc}. We summarize this work in Sec. \ref{secv}. Appendix \ref{secvi} describes the formation of vortices by decreasing the scattering length. 

\begin{figure}[tb!]
	\centering
	\includegraphics[width=0.45\textwidth]{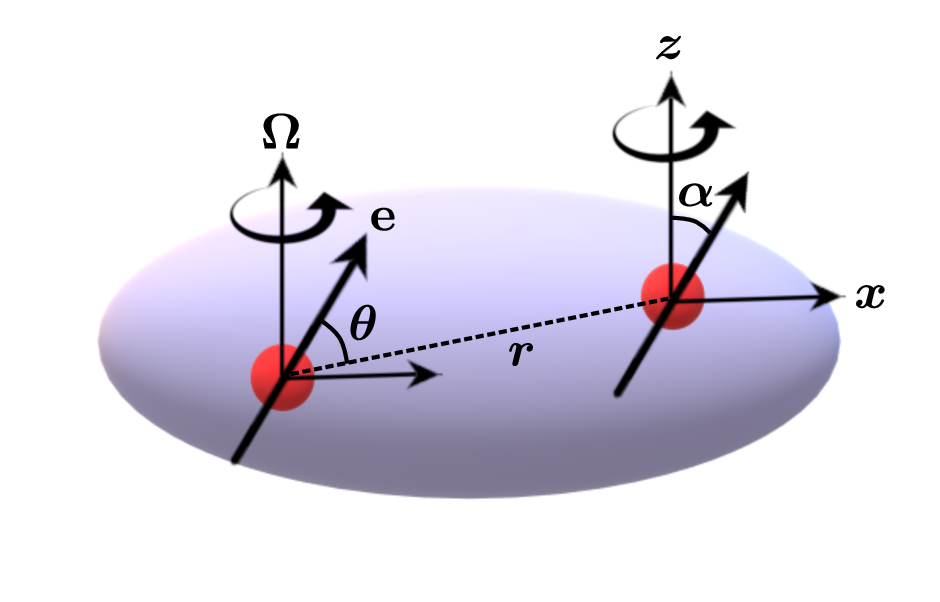}
	\caption{Schematic illustration of a dipolar BEC confined in a pancake-shaped harmonic trap rotating with an angular frequency $\Omega$ about the $z$-axis. The dipoles are polarized by an external magnetic field in the $x$-$z$ plane, making an angle $\alpha$ with the trap rotation axis ($z$-axis) and $\theta$ is the angle between the polarization direction $\vb{e}$ and the separation vector between the dipoles $\vb{r}$.}\label{fig:1}
\end{figure}
\section{Model}\label{secii}
We consider a dipolar BEC whose atoms possess a large magnetic dipole moment $\mu_m$ confined in a pancake-shaped harmonic trap geometry. The atoms are polarized by a uniform external magnetic field, oriented in the $x$-$z$ plane, making a tilt angle $\alpha$ with the $z$-axis (see Fig. \ref{fig:1}). The anisotropic long-range DDI between the atoms of the condensate takes the form 
\begin{equation}
    V_{\rm dd}(\textbf{r})=\frac{3g_{\rm dd}}{4\pi}\left(\frac{1-3\cos^2\theta}{r^3}\right),\label{dip_pot}
\end{equation}
 where $g_{\rm dd}=\mu_0\mu_m^2/3$ is the DDI strength and $\theta$ is the angle between the relative position vector $\vb{r}$ of the dipoles and the polarization direction $\vb{e}=\sin{\alpha}\vb{e}_x+\cos{\alpha} \vb{e}_z$ with $\vb{e}_x$ and $\vb{e}_z$ being the unit vector along the $x$ and $z$-direction, respectively. Additionally, we assume that the condensate is rotating with an angular frequency $\Omega$ about the $z$-axis. In the ultracold regime, the dipolar condensate is characterized by the macroscopic wave function $\psi$, whose temporal evolution in a rotating frame is described by the beyond mean field eGPE:
\begin{align}
	&i\hbar\frac{\partial\psi(\textbf{r},t)}{\partial t}=\Big[-\frac{\hbar^2}{2m}\nabla^2+V_t(\textbf{r})-\Omega L_z+g\abs{\psi(\bf{r},t)}^2\nonumber\\&+\int d\textbf{r}' V_{\rm dd}(\textbf{r}-\textbf{r}')\abs{\psi(\bf{r}^{\prime},t)}^2+\gamma(\epsilon_{\rm dd})\abs{\psi(\textbf{r},t)}^3\Big]\psi(\textbf{r},t).\label{egpe}
\end{align}
\begin{figure}[tb!]
	\centering
	\includegraphics[width=0.45\textwidth]{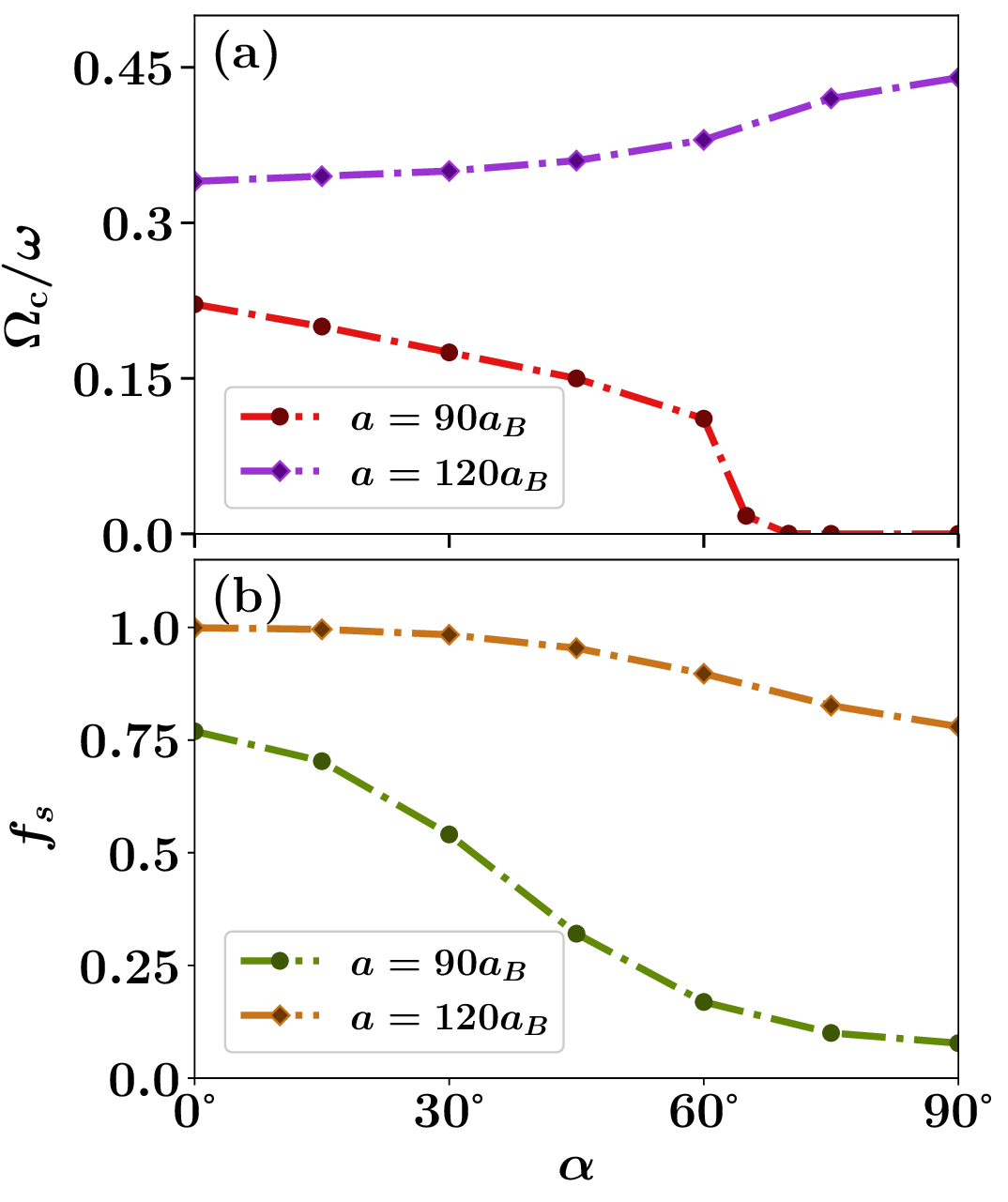}
	\caption{(a) Variation of critical rotation frequency $\Omega_c$ and (b) superfluid fraction $f_s$ as a function of polarizing angle $\alpha$ for a $^{164}$Dy condensate with $a=90a_B$ and $a=120a_B$ corresponds to a supersolid and superfluid state, respectively. Results are for the case of $N=6\times10^4$ number of $^{164}$Dy atoms confined in an axially symmetric pancake-shaped harmonic trap with $(\omega,\omega_z)=2\pi \times(45,133)$Hz.}\label{fig:2}
\end{figure}
Here, $V_t(\textbf{r})=\frac{1}{2}m\omega^2(x^2+y^2+\lambda^2z^2)$ is the axially symmetric harmonic trapping potential with angular frequencies $\omega_x=\omega_y=\omega,\omega_z$; with $\lambda=\omega_z/\omega$ being the trap aspect ratio; $m$ is the atomic mass. $L_z=-i\hbar(x\partial_y-y\partial_x)$ corresponds to the $z$-component of the angular momentum operator arising due to the rotation of the condensate about the $z$-axis. The atoms interact via a short-range contact interaction, characterized by the coupling constant $g=4\pi\hbar^2a/m$, with $a$ being the $s$-wave scattering length of the atoms. The DDI potential is evaluated in the momentum space due to the singularity at $r=0$, and we obtain the real space contribution by applying the convolution theorem. The resulting DDI in the Fourier space is 
\begin{equation}
    \tilde{V}_{\rm dd}(\textbf{k})=g_{\rm dd}\left[\frac{3(k_x\sin\alpha+k_z\cos\alpha)^2}{k^2}-1\right],
\end{equation}
where $k_i$, $i=x,y,z$ are the cartesian components of the momentum. The last term appearing in the Eq. ({\ref{egpe}}) represents the effect of quantum fluctuations in the form of dipolar Lee-Huang-Yang (LHY) correction with the coefficient $\gamma (\epsilon_{\rm dd})=\frac{32}{3}g\sqrt{\frac{a^3}{\pi}} \left(1+\frac{3}{2}\epsilon_{\rm dd}^2\right)$ \cite{schutzhold_2006_meanfield, lima_2011_quantum, lima_2012_meanfield, petrov_2015_quantum, bisset_2016_groundstate}, where the dimensionless parameter $\epsilon_{\rm dd}=a_{\rm dd}/a$ with $a_{\rm dd}=\mu_0\mu_m^2 m/12\pi\hbar^2$ being the dipolar length, quantifies the strength of DDI relative to the contact interaction. The order parameter of the condensate is normalized to the total number of atoms in the condensate, $N=\int d\textbf{r}\abs{\psi(\textbf{r})}^2$.

\section{Critical rotation frequency}\label{seciii}
We consider a dipolar BEC of $N=6\times10^4$ $^{164}$Dy atoms, confined in an axially symmetric pancake-shaped harmonic trap with trapping frequencies, $(\omega,\omega_z)=2\pi \times(45,133)$Hz \cite{bisset_2016_groundstate, wachtler_2016_quantum}. The ground state phases of the condensate are obtained numerically by the imaginary time evolution of the eGPE [Eq. (\ref{egpe})]. The condensate can exhibit a superfluid, supersolid, or droplet phase depending on the $s$-wave scattering length \cite{bisset_2016_groundstate, poli_2021_maintaining, halder_2022_control}. The superfluid to supersolid phase transition occurs below a critical value of the $s$-wave scattering length $a_c$. In our considered case with a static uniform magnetic field polarizing along the $z$-axis ($\alpha=0\degree$), this transition occurs at $a_c\approx 95a_B$. The ground state of a superfluid phase remains unaffected by the trap rotation below a critical rotation frequency $\Omega_c^{\rm SF}$ of the trap. However, once the rotation frequency $\Omega$ exceeds this critical threshold ($\Omega>\Omega_c^{\rm SF}$), vortices begin to form in the superfluid phase. On the contrary, both the supersolid and isolated droplet crystal phases respond to any external rotation as a result of the solid body response exhibited by the droplets and consequently, the angular momentum of the system grows linearly with the rotation frequency $\Omega$. Nevertheless, due to the superfluid background in the supersolid phase, the vortical solution also becomes energetically favorable beyond a critical rotation frequency $\Omega_c^{\rm SS}$.\par 
\begin{figure}[tb!]
	\centering
	\includegraphics[width=0.45\textwidth]{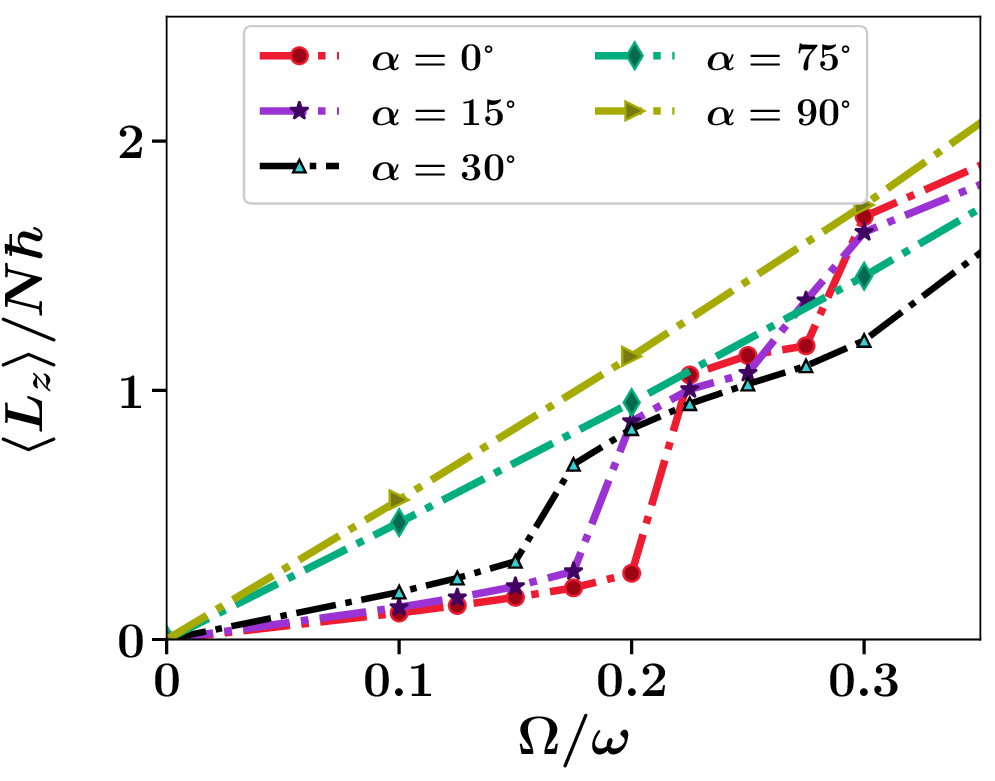}
	\caption{Variation of angular momentum per particle $\expval{L_z}/N\hbar$ of a supersolid state corresponding to $a=90a_B$ with trap rotation frequency $\Omega$ for different polarizing angle $\alpha=0\degree, ~15\degree, ~30\degree,~75\degree$ and $90\degree$, respectively. The other parameters are $N=6\times10^4$ number of $^{164}$Dy atoms confined in an axially symmetric pancake-shaped harmonic trap with $(\omega,\omega_z)=2\pi \times(45,133)$Hz.}\label{fig:3}
\end{figure}
 For a superfluid phase characterized by $a=120a_B$ and $\alpha=0$, we observe that the critical rotation frequency is $\Omega_c^{\rm SF}=0.34\omega$. As we increase the tilt angle $\alpha$ from $0\degree$ to $90\degree$, the growing attractive DDI along the $x$-direction induces an anisotropy in the condensate. In order to minimize the DDI energy, the atoms orient themselves along the polarization direction in a head-to-tail arrangement, resulting in the condensate being elongated along the $x$-direction. With the increased anisotropy and peak density in the $x$-$y$ plane, a larger rotation frequency $\Omega$ is required to initiate vortex nucleation. Therefore, for a superfluid dipolar BEC, the critical rotation frequency increases with the increase in the tilt angle $\alpha$, i.e., $\eval{\Omega_c^{\rm SF}}_{\alpha=0\degree}<\eval{\Omega_c^{\rm SF}}_{\alpha=90\degree}$ [see Fig. \ref{fig:2}(a)]. Furthermore, for the tilt angle $\alpha\neq 0\degree$, the vortex core becomes elliptical in shape with the major axis of the ellipse along the polarization direction and for larger rotation frequency, we observe a transition from a triangular vortex lattice pattern to a stripe vortex lattice pattern similar like \cite{klaus_2022_observationa, prasad_2019_vortex}.\par

In the supersolid state, periodic density peaks (droplets) are connected by regions of lower density. Owing to the lower intermediate density regions between the droplets and a reduced superfluid fraction \cite{firstnote} in the supersolid phase, a relatively lower critical rotation frequency $\Omega_c^{\rm SS}$ is required to pin vortices in the interstitial regions between the droplets $(\Omega_c^{\rm SS}<\Omega_c^{\rm SF}$). Specifically, in our considered case, the supersolid phase with $a=90a_B$ and $\alpha=0\degree$ corresponds to a superfluid fraction $f_s=0.76$, requires a critical rotation frequency $\Omega_c^{\rm SS}=0.225\omega$ to initiate vortex nucleation. As we increase the tilt angle $\alpha$, the dominant DDI causes a reduction in both the superfluid fraction of the condensate [see Fig. \ref{fig:2} (b)] and the density of the interstitial region between the droplets. This leads to a decrease in the value of $\Omega_c^{\rm SS}$, as illustrated in Fig. \ref{fig:2}(a). Beyond a tilt angle $\alpha=60\degree$, the superfluid fraction of the condensate drops below $f_s<0.15$ [see Fig. \ref{fig:2}(b)] and the condensate exhibits a macro-droplet phase elongated along the polarization direction. Consequently, for $\alpha>60\degree$ the ground state of the condensate can not exhibit a vortical solution due to the increased density and decreased superfluid fraction. Nonetheless, in this phase ($\alpha>60\degree$ and $a=90a_B$), the condensate showcases a solid body response to the external rotation of the trap and the angular momentum of the condensate varies linearly with $\Omega$ as shown in Fig. \ref{fig:3}. While for $\alpha<60\degree$, when the condensate exhibits a vortical solution, the angular momentum of the condensate experiences a sudden jump in value at $\Omega=\eval{\Omega_c^{\rm SS}}_{\alpha<60\degree}$ (see Fig. \ref{fig:3}).

\section{Dynamical routes to vortex nucleation}\label{seciv}
In general, dynamical vortex formation via the mechanical rotation of the trap requires a substantially higher rotation frequency than the critical rotation frequency. The dynamical vortex formation marks the onset of quadrupolar instability. This quadrupolar oscillation, driven by dissipative dynamics, leads to the emergence of vortices within the system \cite{kasamatsu_2003_nonlinear}. In the preceding section, we have demonstrated that a supersolid state necessitates a lower critical rotation frequency $\Omega_c^{\rm SS}$ to generate vortices in the ground state of the system ($\Omega_c^{\rm SS}<\Omega_c^{\rm SF}$), and $\Omega_c^{\rm SS}$ decreases as the tilt angle $\alpha$ increases. Moreover, quantized vortices are topological defects; therefore, they can not be eliminated by perturbations from a superfluid system \cite{sindik_2022_creation}. By using these two properties, in this section, we will
{\unskip\parfillskip 0pt\par}

\begin{figure}[H]
	\centering
	\includegraphics[width=0.46\textwidth]{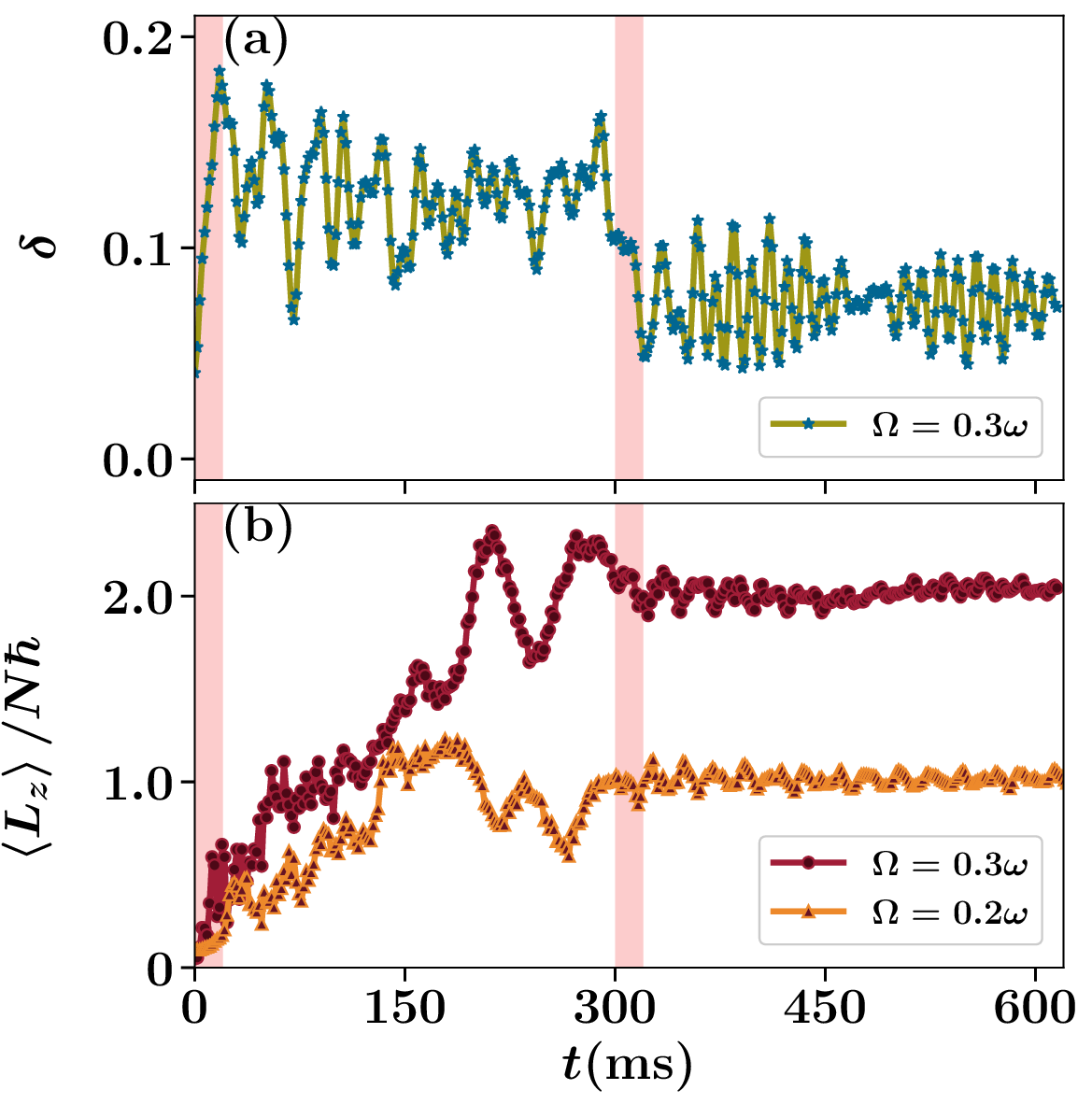}
	\caption{Time evolution of (a) the deformation parameter $\delta$ and (b) the angular momentum per particle $\expval{L_z}/N\hbar$ following the change in the $s$-wave scattering length from $a=120a_B$ to $a=90a_B$ and then back to the initial $a=120a_B$. The red-shaded region indicates the period during which the $s$-wave scattering length changes. Results are presented for the case of $N=6\times10^4$ $^{164}$Dy atoms with a fixed tilt angle $\alpha=45\degree$ trapped in an axially symmetric pancake-shaped trap, with trapping frequencies $(\omega,\omega_z)=2\pi\times(45,133)$Hz, rotating with rotation frequency $\Omega$ (see the legends).}\label{fig:4}
\end{figure}

\noindent demonstrate various dynamic routes that result in vortex nucleation in a dipolar condensate at rotation frequencies, below the critical threshold $\Omega_c$. In all these dynamic processes, we add a small noise to the initial wavefunction of the condensate to simulate the effects of quantum fluctuations \cite{baillie_2020_rotational}.

\subsection{Tuning scattering length}\label{sseciva}
\begin{figure}[H]
	\centering
	\includegraphics[width=0.96\textwidth]{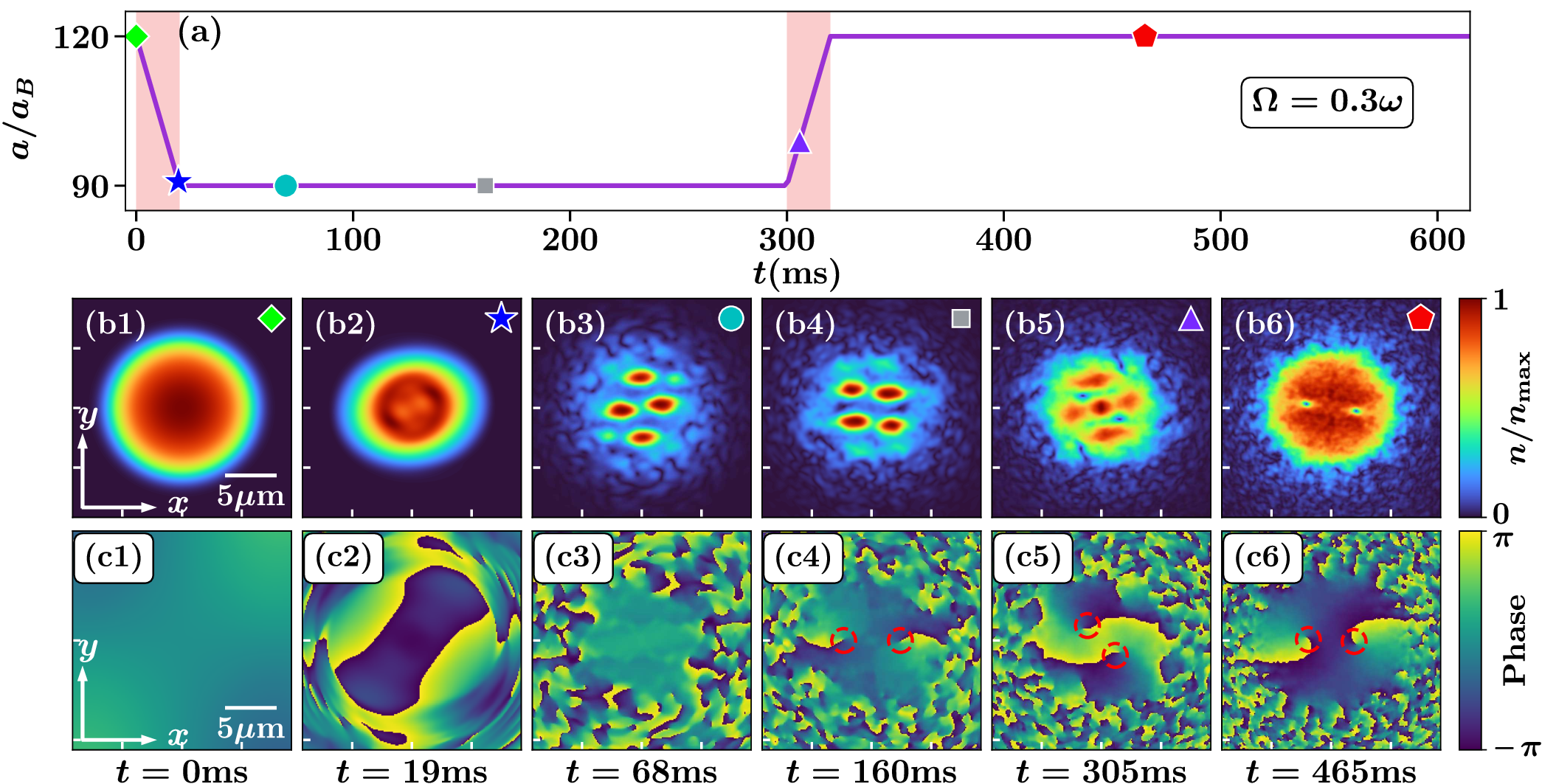}
	\caption{Nucleation of vortices in a superfluid state by varying the $s$-wave scattering length from $a=120a_B$ to $a=90a_B$ and then back to the initial value of $a=120a_B$, with a fixed polarization direction of $\alpha=45\degree$ and a trap rotation frequency $\Omega=0.3\omega<\eval{\Omega_c^{\rm SF}}_{\alpha=45\degree}$. Panel (a) illustrates the temporal variation of the scattering length $a$. The red-shaded region indicates the period during which the s-wave scattering length changes. Panels ($\rm b1-b6$) and ($\rm c1-c6$) present snapshots of the density and phase profiles in the $x$-$y$ plane, respectively, marked by the corresponding marker as marked in panel (a). The red circular patches in the phase profile indicate the positions of the vortices. Results are presented for the case of $N=6\times10^4$ $^{164}$Dy atoms trapped in an axially symmetric harmonic trap with $(\omega,\omega_z)=2\pi\times(45,133)$Hz.}\label{fig:5}
\end{figure}

We first consider a dipolar BEC of $N=6\times10^4$ number of particles, in the superfluid phase with $a=120a_B$ confined in an axially symmetric harmonic trap rotating at a frequency of $\Omega=0.3\omega$. The external polarizing magnetic field orients about the rotation axis of the trap ($z$-axis) with a tilt angle $\alpha=45\degree$. Due to the nonzero tilt angle, the initial ground state of the dipolar BEC experiences a slight deformation, elongating along the $x$-direction [see Fig. \ref{fig:4}(a)]. However, since the rotation frequency $\Omega=0.3\omega <\eval{\Omega_c^{\rm SF}}_{\alpha=45\degree}=0.36\omega$, the superfluid state does not respond to the external rotation. Consequently, the condensate does not possess vortices under this initial condition [see Figs. \ref{fig:5}(b1) and \ref{fig:5}(c1)] and thus has an angular momentum $\expval{L_z}=0$ [see Fig. \ref{fig:4}(b)]. Nevertheless, this rotation frequency exceeds the critical threshold required for initiating 
{\unskip\parfillskip 0pt\par}

\onecolumngrid

\twocolumngrid
\noindent vortex formation in the supersolid phase ($\Omega>\eval{\Omega_c^{\rm SS}}_{\alpha=45\degree}$). Therefore, we decrease the $s$-wave scattering length from $a=120a_B$ to $a=90a_B$ using a linear ramp over a duration of $20~\rm ms$. After that, the scattering length $a$ is kept constant and we let the system evolve up to $300~\rm ms$ as shown in Fig. \ref{fig:5}(a). As the scattering length decreases, the deformation of the condensate $\delta=\left(\expval{x^2}-\expval{y^2}\right)/\left(\expval{x^2}+\expval{y^2}\right)$ increases to its maximum extent $\delta\approx 0.2$ [see Fig. \ref{fig:4}(a)], leading to a periodic oscillation in shape accompanied by quadrupolar phase field as shown in Figs. \ref{fig:5}(b2) and \ref{fig:5}(c2). Subsequently, the condensate transitions into a supersolid state featuring four droplets with an unstable boundary arising due to the first-order phase transition [see Figs. \ref{fig:5}(b3) and \ref{fig:5}(c3)] \cite{biagioni_2022_dimensional}. At the boundary of the supersolid, surface ripples emerge in the form of density fluctuations and phase singularities. Since these surface ripples occur in regions of very low density, they do not contribute significantly to the angular momentum of the system. Over time, vortices penetrate into the condensate from the unstable boundary and nest within the interstitial regions between the droplets [see Figs. \ref{fig:5}(b4) and \ref{fig:5}(c4)], resulting in an increase in the angular momentum of the system [see Fig. \ref{fig:4}(b)]. Due to the snaking of vortex cores from one interstitial site to another, the angular momentum of the condensate fluctuates around $\expval{L_z}/N\hbar=2$. Moreover, the vortex cores undergo deformation due to the presence of the droplets and elongate along the $x$-direction owing to the tilted external polarizing magnetic field. As vortex cores form within the supersolid, the deformation of the condensate diminishes [Fig. \ref{fig:4}(a)]. \par 
After forming the vortices, at $t=300~\rm ms$, we linearly increase the scattering length from $a=90a_B$ back to its initial value $a=120a_B$ over a duration of $20~\rm ms$ [see Fig. \ref{fig:5}(a)]. As we increase the scattering length, the deformation is further reduced (oscillates with a smaller amplitude around $\delta=0.07$ [see Fig. \ref{fig:4} (a)]) and the system transitions back to a superfluid state. Of utmost significance, the vortices situated close to the rotation axis survive owing to their robust character as shown in Figs. \ref{fig:5}(b5), \ref{fig:5}(c5) and \ref{fig:5}(b6), \ref{fig:5}(c6). Vortices located near the periphery of the condensate are displaced further towards the boundary of the condensate. The angular momentum per particle attains a constant value $\expval{L_z}/N\hbar=2$ [see Fig. \ref{fig:4}(b)]. Therefore, in this dynamical process, starting from a superfluid state without any vortex, one can generate vortices in the superfluid state below its critical rotation frequency ($\Omega<\Omega_c^{\rm SF}$).\par 
 Interestingly, we find that vortex nucleation occurs even at a lower trap rotation frequency of $\Omega=0.2\omega$. The angular momentum per particle then increases to $\expval{L_z}/N\hbar=1$, as illustrated in Fig. \ref{fig:4}(b). During this dynamical process also, two vortices are nucleated. However, the difference in angular momentum for $\Omega=0.2\omega$ arises because the vortices are formed at a larger distance from the center of the trap compared to the case  
{\unskip\parfillskip 0pt\par}
\onecolumngrid

\begin{figure}[H]
	\centering
	\includegraphics[width=0.96\textwidth]{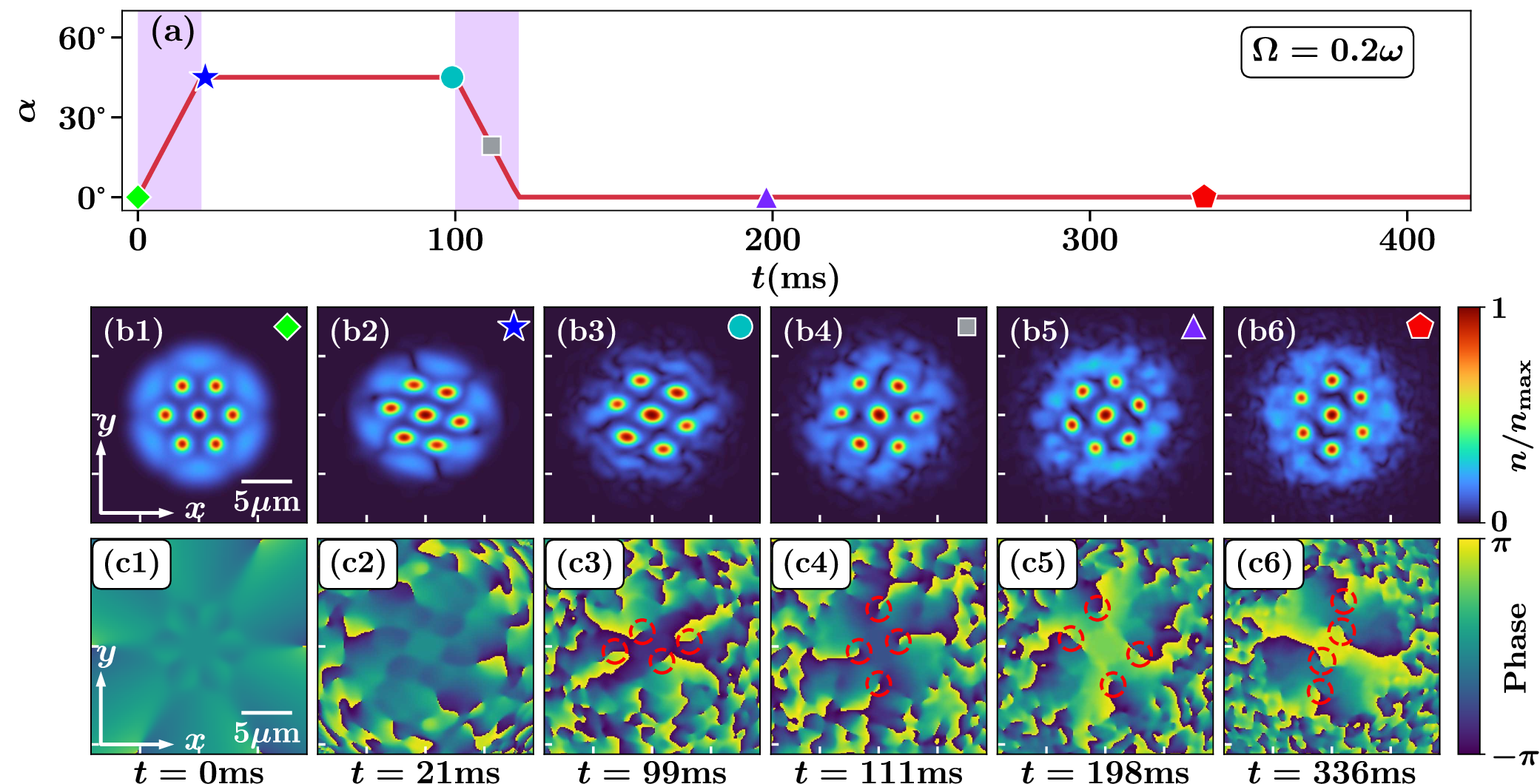}
	\caption{Nucleation of vortices in a supersolid state by changing the polarizing angle from $\alpha=0\degree$ to an intermediate value of $\alpha=45\degree$, and then back to the initial polarization direction of $\alpha=0\degree$, for a fixed trap rotation frequency $\Omega=0.2\omega<\eval{\Omega_c^{\rm SS}}_{\alpha=0\degree}$. Panel (a) shows the instantaneous polarization direction $\alpha(t)$. Panels ($\rm b1-b6$) and ($\rm c1-c6$) display snapshots of the density and phase profiles in the $x$-$y$ plane, respectively, marked by the corresponding marker as marked in panel (a). The red circular patches in the phase profile indicate the positions of the vortices. Results are presented for the case of $N=6\times10^4$ $^{164}$Dy atoms with $a=90a_B$ trapped in an axially symmetric harmonic trap with $(\omega,\omega_z)=2\pi\times(45,133)$Hz.}\label{fig:6}
\end{figure}
\twocolumngrid
\noindent of $\Omega=0.3\omega$ (not shown here).\par 
 Note that following the same protocol, vortices can be induced for a tilt angle $\alpha\neq0\degree$ up to a critical tilt angle $\alpha=60\degree$ (see Appendix \ref{secvi}). For $\alpha=0\degree$, the commutator $[H,L_z]=0$, where $H$ denotes the Hamiltonian of the system. Consequently, the angular momentum remains conserved throughout the transition from superfluid to supersolid states. Hence, if the magnetic field polarizes along the $z$-axis $(\alpha=0\degree)$, a superfluid state without any vortices transitioning to a supersolid state cannot initiate vortex nucleation to maintain the angular momentum $\expval{L_z}=0$. Furthermore, with the increase in tilt angle $\alpha$, the time needed for vortex nucleation decreases (see Appendix \ref{secvi}). Beyond $\alpha>60\degree$, the intermediate state exhibits droplets state, and we observe that droplets demonstrate a solid body response to the external rotation, rendering them incapable of vortex nucleation.

 \subsection{Tuning polarizing angle}\label{ssecivb}
 In this case, we begin with a supersolid state [see Figs. \ref{fig:6}(b1), \ref{fig:6}(c1)] characterized by $a=90a_B$ and $N=6 \times 10^4$ atoms, confined within the identical axially symmetric harmonic trapping configuration as the previous instance. Initially, the atoms are polarized by an external magnetic field along the $z$-direction $(\alpha=0\degree$). We consider that the trap is rotating at a frequency $\Omega=0.2\omega$, which falls below the critical threshold 
 {\unskip\parfillskip 0pt\par}
 \begin{figure}[H]
	\centering
	\includegraphics[width=0.46\textwidth]{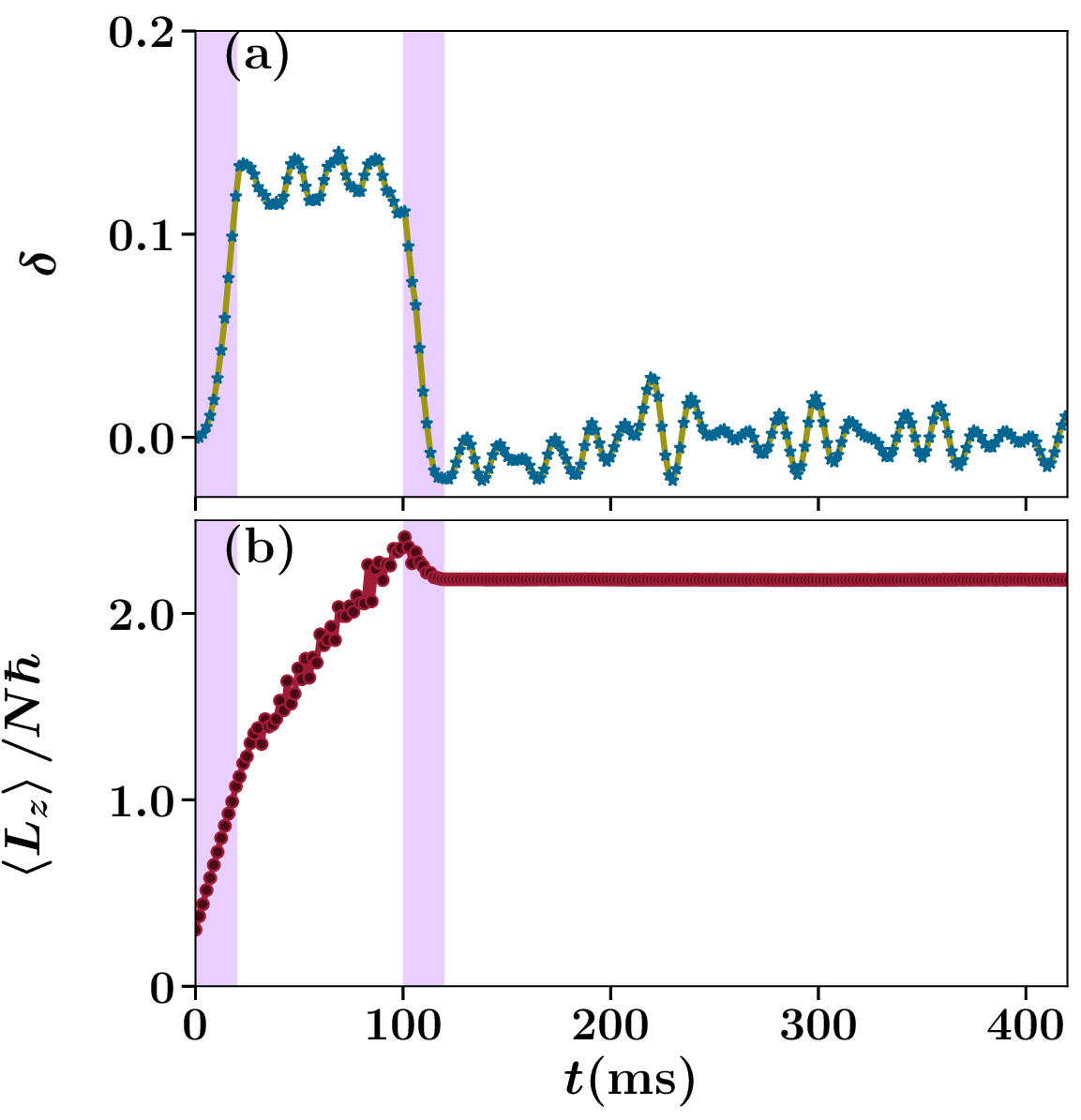}
	\caption{Time evolution of (a) the deformation parameter $\delta$ and (b) the angular momentum per particle $\expval{L_z}/N\hbar$ following the dynamics as shown in \ref{fig:6}(a). The violet-shaded region indicates the period during which the polarization direction changes. Other parameters are same as of Fig. \ref{fig:6}.}\label{fig:7}
\end{figure}

\onecolumngrid

\begin{figure}[H]
	\centering
	\includegraphics[width=\textwidth]{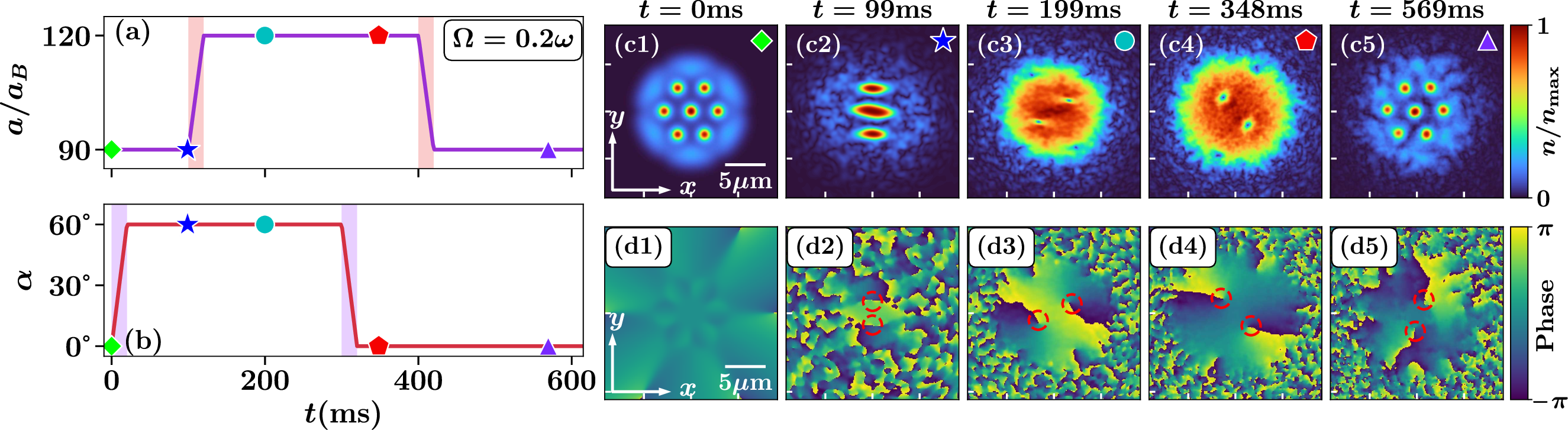}
	\caption{Nucleation of vortices by successive changes in scattering length and polarizing direction as depicted in panels (a, b) for a fixed trap rotation frequency $\Omega=0.2\omega$. The violet-shaded region denotes the period during which the polarization direction changes, while the red-shaded region marks the period of change in the $s$-wave scattering length. Panels ($\rm c1-c5$) and ($\rm d1-d5$) display snapshots of the density and phase profiles in the $x$-$y$ plane, respectively, marked by the corresponding marker as marked in panel (a, b). The red circular patches in the phase profile indicate the positions of the vortices. Results are presented for the case of $N=6\times10^4$ $^{164}$Dy atoms trapped in an axially symmetric pancake-shaped harmonic trap with trapping frequencies $(\omega,\omega_z)=2\pi\times(45,133)$Hz.}\label{fig:8}
\end{figure}
\twocolumngrid
 \noindent $\Omega_c^{\rm SS}=0.225\omega$, and thus insufficient to nucleate vortices in the supersolid state. However, this rotation frequency is high enough to nucleate vortices for a tilt angle $\alpha > 15\degree$ [see Fig. \ref{fig:2}(a)]. Therefore, to nucleate vortices, we linearly increase the tilt angle from $\alpha=0\degree$ to an intermediate tilt angle $\alpha_{\rm int}=45\degree$ over a ramp time $20~\rm ms$ [see Fig. \ref{fig:6}(a)]. As we increase the tilt angle, the DDI between the atoms changes, resulting in an increase in deformation $\delta$ of the condensate leading to dynamical instability. Once $\delta$ reaches its maximum extent, it undergoes a simple periodic oscillation [see Fig. \ref{fig:7}(a)]. Additionally, due to the dynamical instability, surface ripples are generated in the form of density fluctuations and phase singularities [see Figs. \ref{fig:6}(b2) and \ref{fig:6}(c2)]. Because of the intermediate low-density regions between droplets, these surface ripples are promptly pulled towards the center of the condensate by the rotating drive. Meanwhile, the increased polarizing angle results in a decrease in the superfluid fraction of the condensate, resulting in a linear increase in the system's angular momentum $\expval{L_z}$ [see Fig. \ref{fig:7}(b)]. The surface ripples then transform into deformed vortex cores, which settle in the interstitial regions between the droplets as demonstrated in Figs. \ref{fig:6}(b3) and \ref{fig:6}(c3). \par 
 After vortex nucleation at around $t=100~\rm{ms}$, we decrease the polarizing angle from $\alpha_{\rm int}=45\degree$ to $\alpha=0\degree$, restoring the system back to its initial condition [see Fig. \ref{fig:6}(a)]. As we start decreasing the tilt angle, the deformation of the condensate also drops and after reaching the tilt angle $\alpha=0\degree$ it oscillates around $\delta=0$ as shown in Fig. \ref{fig:7}(a). In this case too, we observe that vortices survive as a consequence of their topologically robust nature, and snake between the droplets from one interstitial low-density region to another [see Figs. \ref{fig:6}(b4-b6) and \ref{fig:6}(c4-c6)]. Consequently, the angular momentum of the condensate attains a constant value $\expval{L_z}/N\hbar=2.2$ [see Fig. \ref{fig:7}(b)]. Thus, starting from a supersolid state without any vortices, by employing this protocol one can achieve vortex nucleation in a supersolid state below the critical rotation frequency $\Omega<\Omega_c^{\rm SS}$.\par
 Note that the number of vortices generated using this protocol can vary depending on the intermediate polarizing angle, $\alpha_{\rm int}$ and the trap rotation frequency $\Omega$. With the increase in $\alpha_{\rm int}$, the superfluid fraction decreases, enabling vortex nucleation to be triggered by a lower rotational frequency. We find that by employing this protocol and adjusting the polarizing angle to an intermediate value of $\alpha_{\rm int}=60\degree$, vortex nucleation can be induced even at a rotation frequency of $\Omega=0.15\omega$. Evidently, it is interesting to note that this dynamic protocol for vortex nucleation involving an intermediate state with a larger polarizing angle $45\degree<\alpha<60\degree$, results in a supersolid state with fewer droplets. Moreover, similar to the previous protocol, we find that in this process also, one can vary the value of the intermediate tilt angle maximum up to $\alpha_{\rm int}=60\degree$.

 \begin{figure}[htb!]
	\centering
	\includegraphics[width=0.46\textwidth]{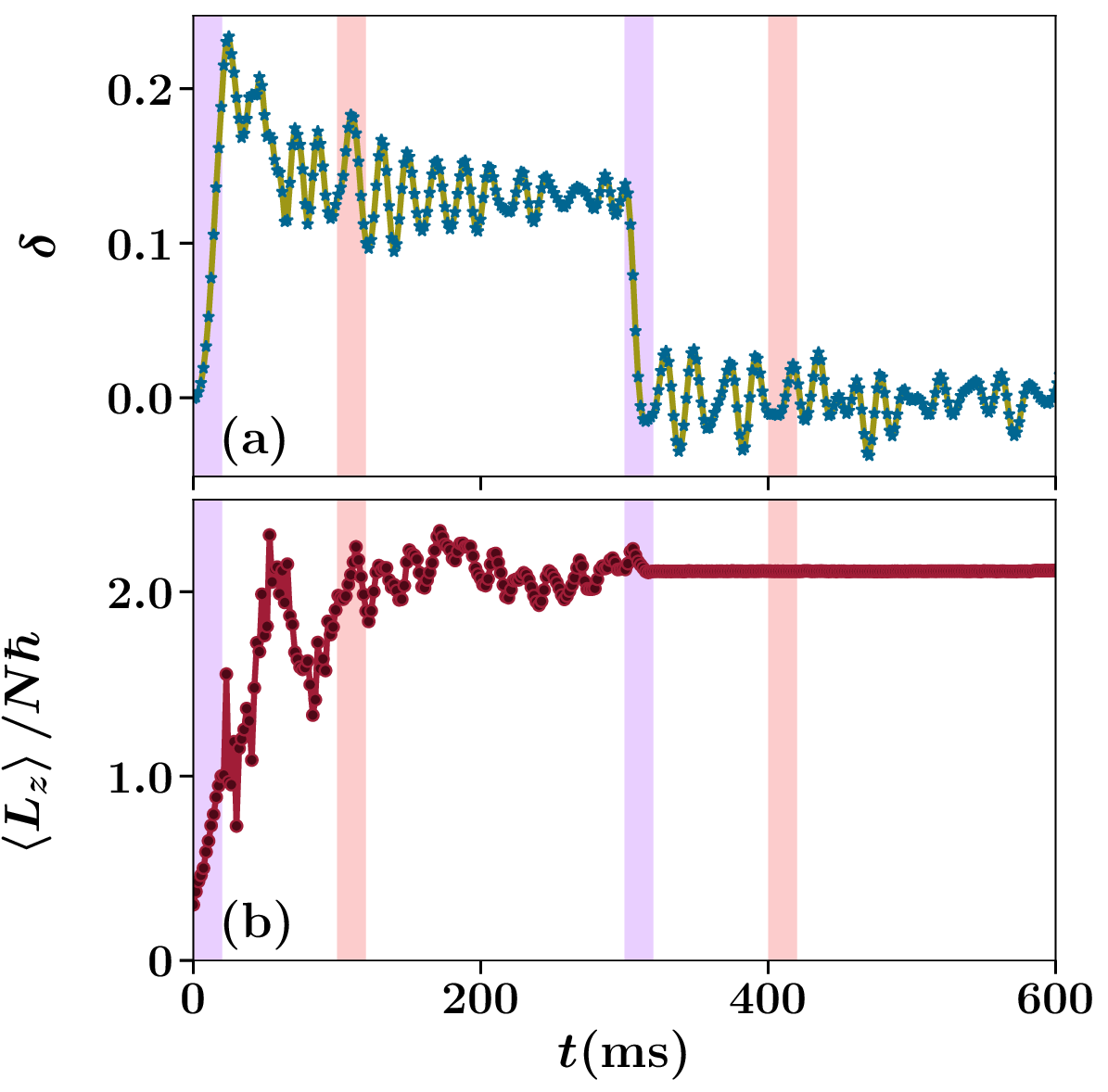}
	\caption{Time evolution of (a) the deformation parameter $\delta$ and (b) the angular momentum per particle $\expval{L_z}/N\hbar$ in response to successive changes in polarizing direction and scattering length, as depicted in Figs. \ref{fig:8} (a) and \ref{fig:8} (b). The violet-shaded region denotes the period during which the polarization direction changes, while the red-shaded region marks the period of change in the s-wave scattering length. Other parameters are the same as in Fig. \ref{fig:8}.}\label{fig:9}
\end{figure}

 \subsection{Varying both polarization direction and scattering length}\label{ssecivc}
In this case, we initially consider a supersolid state with $a=90a_B$ confined in the axially symmetric harmonic trap rotating with a rotation frequency $\Omega=0.2\omega<\Omega_c^{\rm SS}$. Therefore, the supersolid state is unable to nucleate vortices with this trap rotation frequency [see Figs. \ref{fig:8} (c1) and \ref{fig:8}(d1)]. The superfluid fraction of the condensate decreases with the tilt angle $\alpha$, leading to the formation of vortices at lower rotational frequency [see Figs. \ref{fig:2} (a) and \ref{fig:2} (b)]. Thus, we linearly vary the tilt angle from $\alpha=0\degree$ to $\alpha=60\degree$ over a ramp time $t=20~\rm ms$ while keeping the other parameters constant [see Figs. \ref{fig:8}(a) and \ref{fig:8}(b)]. With the increase in the tilt angle, the number of droplets changes as a consequence of the change in the DDI between the particles. At $\alpha=60\degree$, to minimize the DDI energy, the atoms orient themselves in a head-to-tail arrangement along the polarization direction, resulting in an increase in the deformation of the condensate to its maximum extent $\delta=0.24$ [see Fig. \ref{fig:9}(a)] and the condensate exhibits three droplets interlinked by background superfluid as shown in Fig. \ref{fig:8}(c2). Moreover, changing the direction of polarization creates dynamical instability within the system in the form of density fluctuations and phase singularities. Following this dynamical instability, the deformation of the trap decreases and undergoes a quadrupolar shape oscillation around $\delta=0.13$ [Fig. \ref{fig:9}(a)]. The phase singularities at the lower-density region do not contribute to the angular momentum of the condensate. However, as these vortices are formed at the intermediate position between the three droplets, the angular momentum of the condensate increases [see Figs. \ref{fig:8}(c2) and \ref{fig:8}(d2), and Fig. \ref{fig:9}(b)]. Detection of these vortices is very difficult in this phase since the vortices are formed in the low-density region and are cloaked by the high-density droplets.\par 
 Thereafter, forming the vortices in a supersolid state with $\alpha=60\degree$ at $t=100~\rm ms$, we linearly increase the scattering length from $a=90a_B$ to $a=120a_B$ in $20~\rm ms$, transforming the system from a supersolid to a superfluid state [see Figs. \ref{fig:8}(a) and \ref{fig:8}(b)]. Due to the topological robustness, vortices at comparatively higher-density regions survive after the phase transition, and those vortices formed at substantially low-density regions are pushed to the boundary of the condensate. Although the trap rotates with a frequency $\Omega=0.2\omega<\eval{\Omega_c^{\rm SF}}_{\alpha=60\degree}=0.38\omega$, we find that the superfluid with $a=120a_B$ and $\alpha=60\degree$ possess two vortices with elliptical core as shown in Fig. \ref{fig:8}(c3) and \ref{fig:8}(d3). This vortex-vortex pair co-rotates around the central point of the trap in the $x$-$y$ plane in an anisotropic path due to the anisotropic DDI energy between the vortex pair. The distance between the vortices is minimized when the vortex pair aligns along the $y=0$ line and maximized when they align along the $x=0$ line due to the attractive and repulsive DDI, respectively.\par 
At $t=300~\rm ms$, we slowly change the polarization direction over a duration of $20~\rm ms$ and align the external magnetic field along the $z$-direction [see Figs. \ref{fig:8}(a) and \ref{fig:8}(b)]. Consequently, the deformation of the condensate reduces to zero, and the condensate becomes axisymmetric as demonstrated in Fig. \ref{fig:9}(a). Furthermore, even though the rotation frequency $\Omega=0.2\omega$ is less than the critical threshold value $\eval{\Omega_c^{\rm SF}}_{\alpha=0\degree}=0.34\omega$, the condensate still retains the vortices, and the vortex core becomes circularly symmetric [see Figs. \ref{fig:8}(c4) and \ref{fig:8}(d4)]. Thereafter, at $t=400~\rm ms$, we linearly reduce the scattering length from $a=120a_B$ to $a=90a_B$, keeping the polarization fixed along $z$-direction [Figs. \ref{fig:8}(a) and \ref{fig:8}(b)]. We find that following this interaction quench due to the increase in the anisotropic DDI energy, the density of the condensate gets modulated. First, two rings of high density emerge around the vortices, connected at the center. Then, these rings break into seven droplets and are arranged in a triangular lattice pattern, which is connected by the background superfluid. Most significantly, the vortices in the superfluid state also survive this interaction quench and demonstrate the topologically robust character of the vortices [see Figs. \ref{fig:8}(c5) and \ref{fig:8}(d5)]. The angular momentum remains constant at $\expval{L_z}/N\hbar\approx 2$ after making the polarization direction along $z$-axis [see Fig. \ref{fig:9}(b)].\par 

Hence, employing this dynamic protocol, we successively transition the supersolid state through three phases, where we demonstrate the emergence of vortices within the condensate despite having a lower rotational frequency than its critical frequency.

\section{Conclusion}\label{secv}
In conclusion, we exploit the very anisotropic nature of DDI to propose dynamic protocols to generate vortices in both superfluid and supersolid phases of a dipolar BEC, even when rotating below the critical rotation frequency. We demonstrate that once a vortex forms, it remains robustly preserved during transitions of the condensate between superfluid and supersolid states, as well as during changes in the polarization direction of the dipoles. The reduced superfluidity in the supersolid phase lowers the critical rotation frequency compared to the superfluid state. Moreover, the transition from superfluid to a supersolid state in a quasi-two-dimensional trap, induced by decreasing the scattering length, represents a first-order phase transition that instigates dynamical instability into the condensate. Exploiting these characteristics, we demonstrate the formation of vortices below the critical rotation frequency $\Omega<\Omega_c^{\rm SF}$ in a superfluid state under the influence of a tilted magnetic field ranging from $0\degree$ to $60\degree$. This is achieved by transitioning the condensate from superfluid to supersolid phase and then back to its initial superfluid phase. \par

In the presence of a tilted magnetic field, magnetostriction causes the density profile of a dipolar BEC to deform into an ellipsoid \cite{stuhler_2005_observation}, thereby breaking the angular symmetry of an otherwise axisymmetrically trapped system. Furthermore, the superfluid fraction of the condensate varies with the tilt angle, especially in the context of a supersolid state, where the superfluid fraction undergoes significant changes as the polarizing direction is altered. These features cause the critical rotation frequency, required to nucleate vortices, to vary based on the angle between the axial direction of the trap and the polarization direction of the dipoles, ranging from $0$ to $90\degree$. Interestingly, the critical rotation frequency increases with the tilt angle in the superfluid state, while decreasing in the supersolid state. Using this distinctive property in the supersolid phase, we induce a quadrupolar shape oscillation by increasing the tilt angle to an intermediate value and then returning to the initial polarization direction. This dynamical process enables vortex nucleation with a deformed core in the supersolid state, even when rotating below the critical rotation frequency.\par 

Finally, by combining the above two protocols, we present a third approach in which successive changes in the polarization direction and scattering length facilitate the nucleation of vortices in both the superfluid and supersolid phases, below the critical rotation frequency threshold required for vortex nucleation in the supersolid state.  \par

We believe our proposals could serve as a significant benchmark for future experiments on the nucleation of vortices in dipolar BECs rotating at a significantly smaller rotation frequency even below the critical rotation frequency and pave the way for numerous promising research directions for future endeavors. In this work, we investigate a rotating dipolar BEC under the influence of a magnetic field polarized at an arbitrary angle. Instead of using a rotating trap, the same protocol can be applied by magnetostirring the dipolar BEC below the critical rotation frequency to generate vortices \cite{klaus_2022_observationa, casotti_2024_observation}. Another intriguing avenue would be to explore the effects of a higher rotation frequency of the trap or a polarizing magnetic field that exceeds the transverse trapping frequency but still lower than the Larmor frequency \cite{baillie_2020_rotational, tang_2018_tuning}. This work also opens the door for examining the anisotropic nature of the vortex core \cite{mulkerin_2013_anisotropic, li_2024_strongly}, the dynamics of vortex-vortex and vortex-antivortex pairs \cite{das_2024_unveiling, prasad_2024_vortexpair}, the interplay between vortices and droplets \cite{ancilotto_2021_vortex, poli_2023_glitches, yogurt_2023_vortex}, the formation of exotic vortex lattice patterns \cite{klaus_2022_observationa, cai_2018_vortex, prasad_2019_vortex}, and the anisotropic turbulence \cite{sabari_2024_vortex} in the superfluid and supersolid phases of a dipolar BEC under the influence of an arbitrary angle polarized magnetic field.

\begin{acknowledgments}
	We thank Subrata Das for his technical assistance. We acknowledge the National Supercomputing Mission (NSM) for providing computing resources of `PARAM Shakti' at IIT Kharagpur, which is implemented by C-DAC and supported by the Ministry of Electronics and Information Technology (MeitY) and Department of Science and Technology (DST), Government of India. S. H acknowledges the MHRD Govt. of India for the research fellowship. H. S. G and A. S gratefully acknowledge the support from the Prime Minister's Research Fellowship (PMRF), India. S. M gratefully acknowledges the financial support from the Science and Engineering Research Board (SERB) MATRICS project under grant number MTR/2023/000457.
\end{acknowledgments}

\appendix

\begin{figure}[b]
	\centering
	\includegraphics[width=0.46\textwidth]{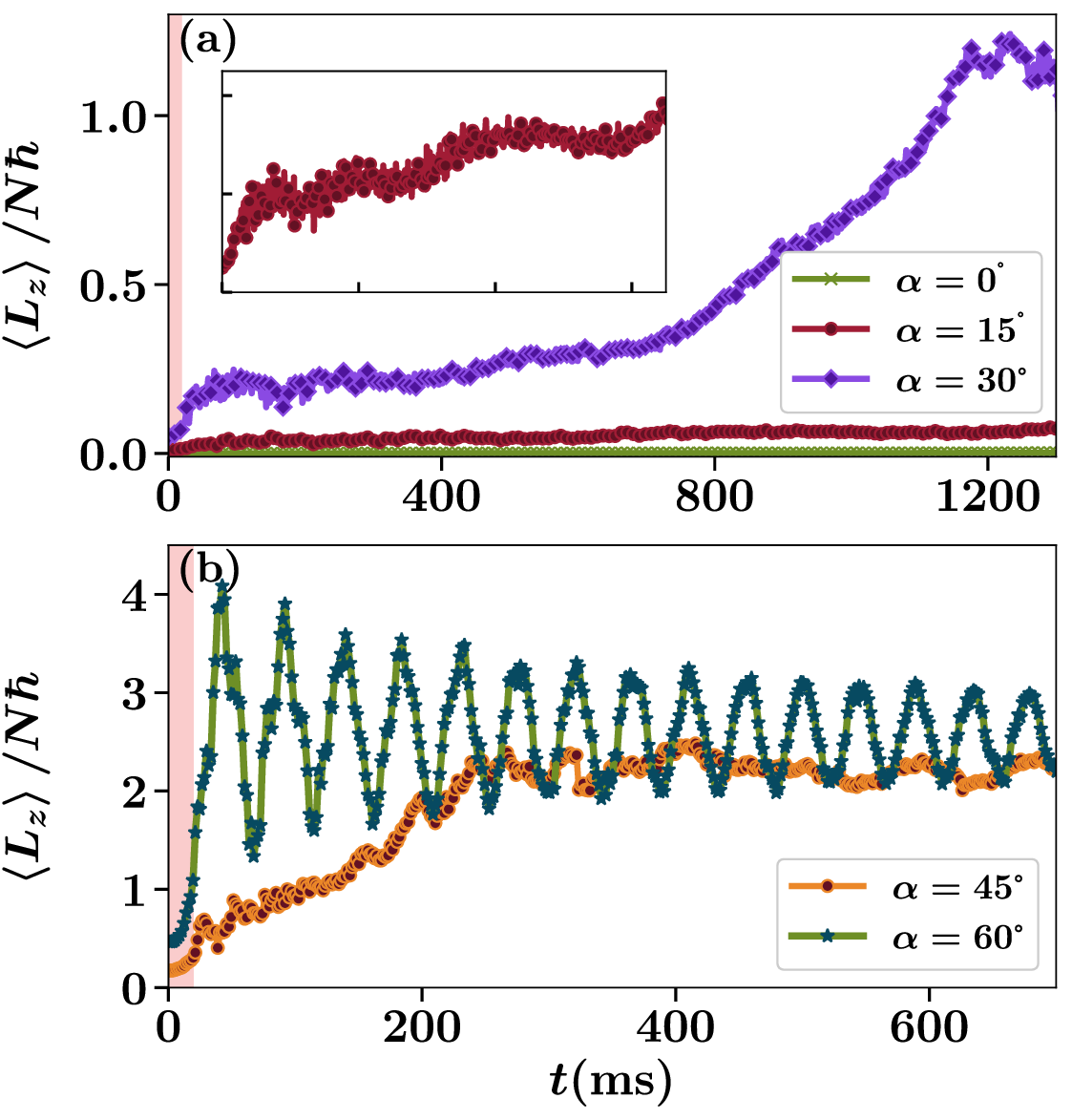}
	\caption{Time evolution of the angular momentum per particle $\expval{L_z}/N\hbar$ following a change in the scattering length from $a=120a_B$ to $a=90a_B$. The red-shaded region indicates the period during which the $s$-wave scattering length changes, for different polarizing angles: (a) $\alpha=0\degree,~15\degree,~30\degree$ and (b) $\alpha=45\degree, ~60\degree$, respectively. The inset of (a) represents the growth of $\expval{L_z}$ for $\alpha=15\degree$. Results are presented for the case of $N=6\times10^4$ $^{164}$Dy atoms rotating with $\Omega=0.3\omega$ and trapped in an axially symmetric pancake-shaped harmonic trap with trapping frequencies $(\omega,\omega_z)=2\pi\times(45,133)$Hz.}\label{fig:10}
\end{figure}
\section{Nucleation of vortices by decreasing scattering length}\label{secvi}
\begin{figure}[tb!]
	\centering
	\includegraphics[width=\textwidth]{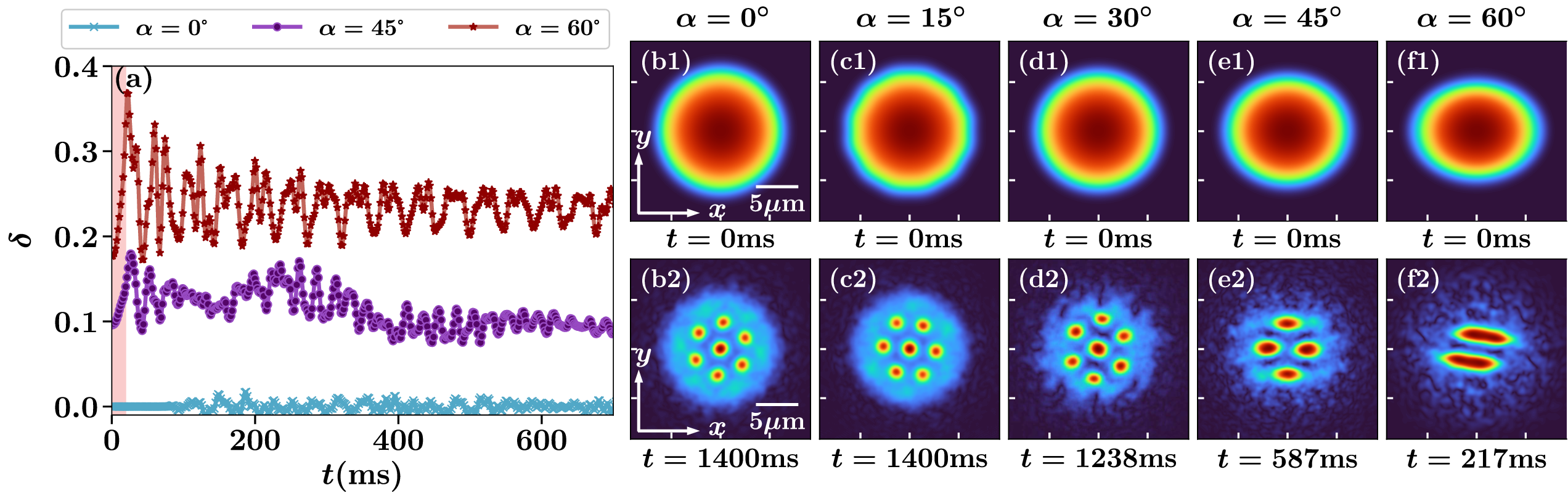}
	\caption{(a) Time evolution of the deformation parameter $\delta$ following the change in the $s$-wave scattering length from $a=120a_B$ to $a=90a_B$. The red-shaded region indicates the period during which the $s$-wave scattering length changes. Panels ($\rm b1-f2$) display snapshots of the density profiles in the $x$-$y$ plane. Each column represents the result of the dynamics for different fixed polarizing angles: ($\rm b1, b2$) $\alpha=0\degree,~ (\rm c1, c2)~\alpha=15\degree,~ (\rm d1, d2)~\alpha=30\degree ~ (\rm e1, e2)~\alpha=45\degree$ and $ ~ (\rm f1, f2)~\alpha=60\degree$, respectively. Results are presented for the case of $N=6\times10^4$ $^{164}$Dy atoms rotating with $\Omega=0.3\omega$ and trapped in an axially symmetric pancake-shaped harmonic trap with trapping frequencies $(\omega,\omega_z)=2\pi\times(45,133)$Hz.}\label{fig:11}
\end{figure}

As we have discussed in the main text (Sec. \ref{sseciva}), a supersolid state requires a lower critical rotational frequency threshold for vortex nucleation than a superfluid state. Here, we demonstrate vortex nucleation starting from a superfluid state with a fixed rotation frequency $\Omega=0.3\omega<\Omega_c^{\rm SF}$ and for different polarizing angles by transitioning from superfluid to supersolid state. We quench the scattering length from $a=120a_B$ to $a=90a_B$ following a linear ramp with ramp time $t=20~\rm ms$. After which, we let the system evolve for a long time. In Figs. \ref{fig:10} (a) and \ref{fig:10} (b), we show the time evolution of the angular momentum. Following the superfluid to supersolid phase transition at a tilt angle $\alpha=0\degree$, the angular momentum of the condensate remains fixed at $\expval{L_z}/N\hbar=0$. This is consistent with the fact that for $\alpha=0$, the commutator $[H,L_z]=0$. For a tilt angle $\alpha=15\degree$, the angular momentum shows an increasing trend [see the inset of Fig. \ref{fig:10}(a)]. However, the value of the angular momentum remains very small and is unable to nucleate vortices within our simulation time $t=1400~\rm ms$. With the increase in the tilt angle to $\alpha=30\degree$ the angular momentum slowly increases, following the superfluid to supersolid transition and after $t=1000~\rm ms$ the condensate induces vortex nucleation. As we further increase the tilt angle to $\alpha=45\degree$ and $\alpha=60\degree$, the value of the angular momentum per particle increases rapidly [see Fig. \ref{fig:10} (b)]. Thus the vortex nucleation time reduces with increasing the tilt angle. For $\alpha=60\degree$, the angular momentum of the condensate fluctuates a lot due to lower superfluid fraction, resulting in spontaneous vortex pinning and unpinning. In Fig. \ref{fig:11} (a), we showcase the variation of $\delta$ with time. We observe that for larger tilt angles, the deformation of the condensate increases following the superfluid to supersolid phase transition. As the deformation reaches its maximum extent, the condensate undergoes a quadrupolar oscillation. Following the quadrupolar shape oscillation and as a consequence of the lower superfluid fraction in the supersolid state, the condensate exhibits vortex nucleation. However, for a smaller tilt angle $\alpha \leq 15\degree$, the condensate remains almost axially symmetric. Therefore, the condensate does not undergo any quadrupolar oscillation and due to the very small gain of angular momentum at a smaller tilt angle, the condensate is unable to nucleate vortices. In Figs. \ref{fig:11}(b1-f2), we show snapshots of the density profile of the condensate following an interaction quench from a superfluid state to a supersolid state in the $x$-$y$ plane.
{\unskip\parfillskip 0pt\par}
\onecolumngrid

\twocolumngrid

\bibliographystyle{apsrev4-2}
\bibliography{reference} 

\begin{thebibliography}{123}%
\makeatletter
\providecommand \@ifxundefined [1]{%
 \@ifx{#1\undefined}
}%
\providecommand \@ifnum [1]{%
 \ifnum #1\expandafter \@firstoftwo
 \else \expandafter \@secondoftwo
 \fi
}%
\providecommand \@ifx [1]{%
 \ifx #1\expandafter \@firstoftwo
 \else \expandafter \@secondoftwo
 \fi
}%
\providecommand \natexlab [1]{#1}%
\providecommand \enquote  [1]{``#1''}%
\providecommand \bibnamefont  [1]{#1}%
\providecommand \bibfnamefont [1]{#1}%
\providecommand \citenamefont [1]{#1}%
\providecommand \href@noop [0]{\@secondoftwo}%
\providecommand \href [0]{\begingroup \@sanitize@url \@href}%
\providecommand \@href[1]{\@@startlink{#1}\@@href}%
\providecommand \@@href[1]{\endgroup#1\@@endlink}%
\providecommand \@sanitize@url [0]{\catcode `\\12\catcode `\$12\catcode
  `\&12\catcode `\#12\catcode `\^12\catcode `\_12\catcode `\%12\relax}%
\providecommand \@@startlink[1]{}%
\providecommand \@@endlink[0]{}%
\providecommand \url  [0]{\begingroup\@sanitize@url \@url }%
\providecommand \@url [1]{\endgroup\@href {#1}{\urlprefix }}%
\providecommand \urlprefix  [0]{URL }%
\providecommand \Eprint [0]{\href }%
\providecommand \doibase [0]{https://doi.org/}%
\providecommand \selectlanguage [0]{\@gobble}%
\providecommand \bibinfo  [0]{\@secondoftwo}%
\providecommand \bibfield  [0]{\@secondoftwo}%
\providecommand \translation [1]{[#1]}%
\providecommand \BibitemOpen [0]{}%
\providecommand \bibitemStop [0]{}%
\providecommand \bibitemNoStop [0]{.\EOS\space}%
\providecommand \EOS [0]{\spacefactor3000\relax}%
\providecommand \BibitemShut  [1]{\csname bibitem#1\endcsname}%
\let\auto@bib@innerbib\@empty
\bibitem [{\citenamefont
  {London}(1938{\natexlab{a}})}]{london_1938_boseeinstein}%
  \BibitemOpen
  \bibfield  {author} {\bibinfo {author} {\bibfnamefont {F.}~\bibnamefont
  {London}},\ }\href {https://doi.org/10.1103/PhysRev.54.947} {\bibfield
  {journal} {\bibinfo  {journal} {Phys. Rev.}\ }\textbf {\bibinfo {volume}
  {54}},\ \bibinfo {pages} {947} (\bibinfo {year}
  {1938}{\natexlab{a}})}\BibitemShut {NoStop}%
\bibitem [{\citenamefont
  {London}(1938{\natexlab{b}})}]{london_1938_lphenomenon}%
  \BibitemOpen
  \bibfield  {author} {\bibinfo {author} {\bibfnamefont {F.}~\bibnamefont
  {London}},\ }\href {https://doi.org/10.1038/141643a0} {\bibfield  {journal}
  {\bibinfo  {journal} {Nature}\ }\textbf {\bibinfo {volume} {141}},\ \bibinfo
  {pages} {643} (\bibinfo {year} {1938}{\natexlab{b}})}\BibitemShut {NoStop}%
\bibitem [{\citenamefont {Landau}(1941)}]{landau_1941_theory}%
  \BibitemOpen
  \bibfield  {author} {\bibinfo {author} {\bibfnamefont {L.}~\bibnamefont
  {Landau}},\ }\href {https://doi.org/10.1103/PhysRev.60.356} {\bibfield
  {journal} {\bibinfo  {journal} {Phys. Rev.}\ }\textbf {\bibinfo {volume}
  {60}},\ \bibinfo {pages} {356} (\bibinfo {year} {1941})}\BibitemShut
  {NoStop}%
\bibitem [{\citenamefont {Anderson}\ \emph {et~al.}(1995)\citenamefont
  {Anderson}, \citenamefont {Ensher}, \citenamefont {Matthews}, \citenamefont
  {Wieman},\ and\ \citenamefont {Cornell}}]{anderson_1995_observation}%
  \BibitemOpen
  \bibfield  {author} {\bibinfo {author} {\bibfnamefont {M.~H.}\ \bibnamefont
  {Anderson}}, \bibinfo {author} {\bibfnamefont {J.~R.}\ \bibnamefont
  {Ensher}}, \bibinfo {author} {\bibfnamefont {M.~R.}\ \bibnamefont
  {Matthews}}, \bibinfo {author} {\bibfnamefont {C.~E.}\ \bibnamefont
  {Wieman}},\ and\ \bibinfo {author} {\bibfnamefont {E.~A.}\ \bibnamefont
  {Cornell}},\ }\href {https://doi.org/10.1126/science.269.5221.198} {\bibfield
   {journal} {\bibinfo  {journal} {Science}\ }\textbf {\bibinfo {volume}
  {269}},\ \bibinfo {pages} {198} (\bibinfo {year} {1995})}\BibitemShut
  {NoStop}%
\bibitem [{\citenamefont {Yarmchuk}\ \emph {et~al.}(1979)\citenamefont
  {Yarmchuk}, \citenamefont {Gordon},\ and\ \citenamefont
  {Packard}}]{yarmchuk_1979_observation}%
  \BibitemOpen
  \bibfield  {author} {\bibinfo {author} {\bibfnamefont {E.~J.}\ \bibnamefont
  {Yarmchuk}}, \bibinfo {author} {\bibfnamefont {M.~J.~V.}\ \bibnamefont
  {Gordon}},\ and\ \bibinfo {author} {\bibfnamefont {R.~E.}\ \bibnamefont
  {Packard}},\ }\href {https://doi.org/10.1103/PhysRevLett.43.214} {\bibfield
  {journal} {\bibinfo  {journal} {Phys. Rev. Lett.}\ }\textbf {\bibinfo
  {volume} {43}},\ \bibinfo {pages} {214} (\bibinfo {year} {1979})}\BibitemShut
  {NoStop}%
\bibitem [{\citenamefont {Abo-Shaeer}\ \emph {et~al.}(2001)\citenamefont
  {Abo-Shaeer}, \citenamefont {Raman}, \citenamefont {Vogels},\ and\
  \citenamefont {Ketterle}}]{abo-shaeer_2001_observation}%
  \BibitemOpen
  \bibfield  {author} {\bibinfo {author} {\bibfnamefont {J.~R.}\ \bibnamefont
  {Abo-Shaeer}}, \bibinfo {author} {\bibfnamefont {C.}~\bibnamefont {Raman}},
  \bibinfo {author} {\bibfnamefont {J.~M.}\ \bibnamefont {Vogels}},\ and\
  \bibinfo {author} {\bibfnamefont {W.}~\bibnamefont {Ketterle}},\ }\href
  {https://doi.org/10.1126/science.1060182} {\bibfield  {journal} {\bibinfo
  {journal} {Science}\ }\textbf {\bibinfo {volume} {292}},\ \bibinfo {pages}
  {476} (\bibinfo {year} {2001})}\BibitemShut {NoStop}%
\bibitem [{\citenamefont {Madison}\ \emph {et~al.}(2000)\citenamefont
  {Madison}, \citenamefont {Chevy}, \citenamefont {Wohlleben},\ and\
  \citenamefont {Dalibard}}]{madison_2000_vortex}%
  \BibitemOpen
  \bibfield  {author} {\bibinfo {author} {\bibfnamefont {K.~W.}\ \bibnamefont
  {Madison}}, \bibinfo {author} {\bibfnamefont {F.}~\bibnamefont {Chevy}},
  \bibinfo {author} {\bibfnamefont {W.}~\bibnamefont {Wohlleben}},\ and\
  \bibinfo {author} {\bibfnamefont {J.}~\bibnamefont {Dalibard}},\ }\href
  {https://doi.org/10.1103/PhysRevLett.84.806} {\bibfield  {journal} {\bibinfo
  {journal} {Phys. Rev. Lett.}\ }\textbf {\bibinfo {volume} {84}},\ \bibinfo
  {pages} {806} (\bibinfo {year} {2000})}\BibitemShut {NoStop}%
\bibitem [{\citenamefont {Onsager}(1949)}]{onsager_1949_statistical}%
  \BibitemOpen
  \bibfield  {author} {\bibinfo {author} {\bibfnamefont {L.}~\bibnamefont
  {Onsager}},\ }\href {https://doi.org/10.1007/BF02780991} {\bibfield
  {journal} {\bibinfo  {journal} {Il Nuovo Cimento (1943-1954)}\ }\textbf
  {\bibinfo {volume} {6}},\ \bibinfo {pages} {279} (\bibinfo {year}
  {1949})}\BibitemShut {NoStop}%
\bibitem [{\citenamefont {Feynman}(1955)}]{feynman_1955_chapter}%
  \BibitemOpen
  \bibfield  {author} {\bibinfo {author} {\bibfnamefont {R.~P.}\ \bibnamefont
  {Feynman}},\ }in\ \href {https://doi.org/10.1016/S0079-6417(08)60077-3}
  {\emph {\bibinfo {booktitle} {Progress in {{Low Temperature Physics}}}}},\
  Vol.~\bibinfo {volume} {1},\ \bibinfo {editor} {edited by\ \bibinfo {editor}
  {\bibfnamefont {C.~J.}\ \bibnamefont {Gorter}}}\ (\bibinfo  {publisher}
  {Elsevier},\ \bibinfo {year} {1955})\ pp.\ \bibinfo {pages}
  {17--53}\BibitemShut {NoStop}%
\bibitem [{\citenamefont {Haljan}\ \emph {et~al.}(2001)\citenamefont {Haljan},
  \citenamefont {Coddington}, \citenamefont {Engels},\ and\ \citenamefont
  {Cornell}}]{haljan_2001_driving}%
  \BibitemOpen
  \bibfield  {author} {\bibinfo {author} {\bibfnamefont {P.~C.}\ \bibnamefont
  {Haljan}}, \bibinfo {author} {\bibfnamefont {I.}~\bibnamefont {Coddington}},
  \bibinfo {author} {\bibfnamefont {P.}~\bibnamefont {Engels}},\ and\ \bibinfo
  {author} {\bibfnamefont {E.~A.}\ \bibnamefont {Cornell}},\ }\href
  {https://doi.org/10.1103/PhysRevLett.87.210403} {\bibfield  {journal}
  {\bibinfo  {journal} {Phys. Rev. Lett.}\ }\textbf {\bibinfo {volume} {87}},\
  \bibinfo {pages} {210403} (\bibinfo {year} {2001})}\BibitemShut {NoStop}%
\bibitem [{\citenamefont {Raman}\ \emph {et~al.}(2001)\citenamefont {Raman},
  \citenamefont {Abo-Shaeer}, \citenamefont {Vogels}, \citenamefont {Xu},\ and\
  \citenamefont {Ketterle}}]{raman_2001_vortex}%
  \BibitemOpen
  \bibfield  {author} {\bibinfo {author} {\bibfnamefont {C.}~\bibnamefont
  {Raman}}, \bibinfo {author} {\bibfnamefont {J.~R.}\ \bibnamefont
  {Abo-Shaeer}}, \bibinfo {author} {\bibfnamefont {J.~M.}\ \bibnamefont
  {Vogels}}, \bibinfo {author} {\bibfnamefont {K.}~\bibnamefont {Xu}},\ and\
  \bibinfo {author} {\bibfnamefont {W.}~\bibnamefont {Ketterle}},\ }\href
  {https://doi.org/10.1103/PhysRevLett.87.210402} {\bibfield  {journal}
  {\bibinfo  {journal} {Phys. Rev. Lett.}\ }\textbf {\bibinfo {volume} {87}},\
  \bibinfo {pages} {210402} (\bibinfo {year} {2001})}\BibitemShut {NoStop}%
\bibitem [{\citenamefont {Neely}\ \emph {et~al.}(2010)\citenamefont {Neely},
  \citenamefont {Samson}, \citenamefont {Bradley}, \citenamefont {Davis},\ and\
  \citenamefont {Anderson}}]{neely_2010_observation}%
  \BibitemOpen
  \bibfield  {author} {\bibinfo {author} {\bibfnamefont {T.~W.}\ \bibnamefont
  {Neely}}, \bibinfo {author} {\bibfnamefont {E.~C.}\ \bibnamefont {Samson}},
  \bibinfo {author} {\bibfnamefont {A.~S.}\ \bibnamefont {Bradley}}, \bibinfo
  {author} {\bibfnamefont {M.~J.}\ \bibnamefont {Davis}},\ and\ \bibinfo
  {author} {\bibfnamefont {B.~P.}\ \bibnamefont {Anderson}},\ }\href
  {https://doi.org/10.1103/PhysRevLett.104.160401} {\bibfield  {journal}
  {\bibinfo  {journal} {Phys. Rev. Lett.}\ }\textbf {\bibinfo {volume} {104}},\
  \bibinfo {pages} {160401} (\bibinfo {year} {2010})}\BibitemShut {NoStop}%
\bibitem [{\citenamefont {Kwon}\ \emph {et~al.}(2021)\citenamefont {Kwon},
  \citenamefont {Del~Pace}, \citenamefont {Xhani}, \citenamefont {Galantucci},
  \citenamefont {Muzi~Falconi}, \citenamefont {Inguscio}, \citenamefont
  {Scazza},\ and\ \citenamefont {Roati}}]{kwon_2021_sound}%
  \BibitemOpen
  \bibfield  {author} {\bibinfo {author} {\bibfnamefont {W.~J.}\ \bibnamefont
  {Kwon}}, \bibinfo {author} {\bibfnamefont {G.}~\bibnamefont {Del~Pace}},
  \bibinfo {author} {\bibfnamefont {K.}~\bibnamefont {Xhani}}, \bibinfo
  {author} {\bibfnamefont {L.}~\bibnamefont {Galantucci}}, \bibinfo {author}
  {\bibfnamefont {A.}~\bibnamefont {Muzi~Falconi}}, \bibinfo {author}
  {\bibfnamefont {M.}~\bibnamefont {Inguscio}}, \bibinfo {author}
  {\bibfnamefont {F.}~\bibnamefont {Scazza}},\ and\ \bibinfo {author}
  {\bibfnamefont {G.}~\bibnamefont {Roati}},\ }\href
  {https://doi.org/10.1038/s41586-021-04047-4} {\bibfield  {journal} {\bibinfo
  {journal} {Nature}\ }\textbf {\bibinfo {volume} {600}},\ \bibinfo {pages}
  {64} (\bibinfo {year} {2021})}\BibitemShut {NoStop}%
\bibitem [{\citenamefont {Kibble}(1976)}]{kibble_1976_topology}%
  \BibitemOpen
  \bibfield  {author} {\bibinfo {author} {\bibfnamefont {T.~W.}\ \bibnamefont
  {Kibble}},\ }\href@noop {} {\bibfield  {journal} {\bibinfo  {journal}
  {Journal of Physics A: Mathematical and General}\ }\textbf {\bibinfo {volume}
  {9}},\ \bibinfo {pages} {1387} (\bibinfo {year} {1976})}\BibitemShut
  {NoStop}%
\bibitem [{\citenamefont {Zurek}(1985)}]{zurek_1985_cosmological}%
  \BibitemOpen
  \bibfield  {author} {\bibinfo {author} {\bibfnamefont {W.~H.}\ \bibnamefont
  {Zurek}},\ }\href {https://doi.org/10.1038/317505a0} {\bibfield  {journal}
  {\bibinfo  {journal} {Nature}\ }\textbf {\bibinfo {volume} {317}},\ \bibinfo
  {pages} {505} (\bibinfo {year} {1985})}\BibitemShut {NoStop}%
\bibitem [{\citenamefont {Weiler}\ \emph {et~al.}(2008)\citenamefont {Weiler},
  \citenamefont {Neely}, \citenamefont {Scherer}, \citenamefont {Bradley},
  \citenamefont {Davis},\ and\ \citenamefont
  {Anderson}}]{weiler_2008_spontaneous}%
  \BibitemOpen
  \bibfield  {author} {\bibinfo {author} {\bibfnamefont {C.~N.}\ \bibnamefont
  {Weiler}}, \bibinfo {author} {\bibfnamefont {T.~W.}\ \bibnamefont {Neely}},
  \bibinfo {author} {\bibfnamefont {D.~R.}\ \bibnamefont {Scherer}}, \bibinfo
  {author} {\bibfnamefont {A.~S.}\ \bibnamefont {Bradley}}, \bibinfo {author}
  {\bibfnamefont {M.~J.}\ \bibnamefont {Davis}},\ and\ \bibinfo {author}
  {\bibfnamefont {B.~P.}\ \bibnamefont {Anderson}},\ }\href
  {https://doi.org/10.1038/nature07334} {\bibfield  {journal} {\bibinfo
  {journal} {Nature}\ }\textbf {\bibinfo {volume} {455}},\ \bibinfo {pages}
  {948} (\bibinfo {year} {2008})}\BibitemShut {NoStop}%
\bibitem [{\citenamefont {Donadello}\ \emph {et~al.}(2014)\citenamefont
  {Donadello}, \citenamefont {Serafini}, \citenamefont {Tylutki}, \citenamefont
  {Pitaevskii}, \citenamefont {Dalfovo}, \citenamefont {Lamporesi},\ and\
  \citenamefont {Ferrari}}]{donadello_2014_observation}%
  \BibitemOpen
  \bibfield  {author} {\bibinfo {author} {\bibfnamefont {S.}~\bibnamefont
  {Donadello}}, \bibinfo {author} {\bibfnamefont {S.}~\bibnamefont {Serafini}},
  \bibinfo {author} {\bibfnamefont {M.}~\bibnamefont {Tylutki}}, \bibinfo
  {author} {\bibfnamefont {L.~P.}\ \bibnamefont {Pitaevskii}}, \bibinfo
  {author} {\bibfnamefont {F.}~\bibnamefont {Dalfovo}}, \bibinfo {author}
  {\bibfnamefont {G.}~\bibnamefont {Lamporesi}},\ and\ \bibinfo {author}
  {\bibfnamefont {G.}~\bibnamefont {Ferrari}},\ }\href
  {https://doi.org/10.1103/PhysRevLett.113.065302} {\bibfield  {journal}
  {\bibinfo  {journal} {Phys. Rev. Lett.}\ }\textbf {\bibinfo {volume} {113}},\
  \bibinfo {pages} {065302} (\bibinfo {year} {2014})}\BibitemShut {NoStop}%
\bibitem [{\citenamefont {Matthews}\ \emph {et~al.}(1999)\citenamefont
  {Matthews}, \citenamefont {Anderson}, \citenamefont {Haljan}, \citenamefont
  {Hall}, \citenamefont {Wieman},\ and\ \citenamefont
  {Cornell}}]{matthews_1999_vortices}%
  \BibitemOpen
  \bibfield  {author} {\bibinfo {author} {\bibfnamefont {M.~R.}\ \bibnamefont
  {Matthews}}, \bibinfo {author} {\bibfnamefont {B.~P.}\ \bibnamefont
  {Anderson}}, \bibinfo {author} {\bibfnamefont {P.~C.}\ \bibnamefont
  {Haljan}}, \bibinfo {author} {\bibfnamefont {D.~S.}\ \bibnamefont {Hall}},
  \bibinfo {author} {\bibfnamefont {C.~E.}\ \bibnamefont {Wieman}},\ and\
  \bibinfo {author} {\bibfnamefont {E.~A.}\ \bibnamefont {Cornell}},\ }\href
  {https://doi.org/10.1103/PhysRevLett.83.2498} {\bibfield  {journal} {\bibinfo
   {journal} {Phys. Rev. Lett.}\ }\textbf {\bibinfo {volume} {83}},\ \bibinfo
  {pages} {2498} (\bibinfo {year} {1999})}\BibitemShut {NoStop}%
\bibitem [{\citenamefont {Leanhardt}\ \emph {et~al.}(2002)\citenamefont
  {Leanhardt}, \citenamefont {G\"orlitz}, \citenamefont {Chikkatur},
  \citenamefont {Kielpinski}, \citenamefont {Shin}, \citenamefont {Pritchard},\
  and\ \citenamefont {Ketterle}}]{leanhardt_2002_imprinting}%
  \BibitemOpen
  \bibfield  {author} {\bibinfo {author} {\bibfnamefont {A.~E.}\ \bibnamefont
  {Leanhardt}}, \bibinfo {author} {\bibfnamefont {A.}~\bibnamefont
  {G\"orlitz}}, \bibinfo {author} {\bibfnamefont {A.~P.}\ \bibnamefont
  {Chikkatur}}, \bibinfo {author} {\bibfnamefont {D.}~\bibnamefont
  {Kielpinski}}, \bibinfo {author} {\bibfnamefont {Y.}~\bibnamefont {Shin}},
  \bibinfo {author} {\bibfnamefont {D.~E.}\ \bibnamefont {Pritchard}},\ and\
  \bibinfo {author} {\bibfnamefont {W.}~\bibnamefont {Ketterle}},\ }\href
  {https://doi.org/10.1103/PhysRevLett.89.190403} {\bibfield  {journal}
  {\bibinfo  {journal} {Phys. Rev. Lett.}\ }\textbf {\bibinfo {volume} {89}},\
  \bibinfo {pages} {190403} (\bibinfo {year} {2002})}\BibitemShut {NoStop}%
\bibitem [{\citenamefont {Kumar}\ \emph {et~al.}(2018)\citenamefont {Kumar},
  \citenamefont {Dubessy}, \citenamefont {Badr}, \citenamefont {De~Rossi},
  \citenamefont {de~Go\"er~de Herve}, \citenamefont {Longchambon},\ and\
  \citenamefont {Perrin}}]{kumar_2018_producing}%
  \BibitemOpen
  \bibfield  {author} {\bibinfo {author} {\bibfnamefont {A.}~\bibnamefont
  {Kumar}}, \bibinfo {author} {\bibfnamefont {R.}~\bibnamefont {Dubessy}},
  \bibinfo {author} {\bibfnamefont {T.}~\bibnamefont {Badr}}, \bibinfo {author}
  {\bibfnamefont {C.}~\bibnamefont {De~Rossi}}, \bibinfo {author}
  {\bibfnamefont {M.}~\bibnamefont {de~Go\"er~de Herve}}, \bibinfo {author}
  {\bibfnamefont {L.}~\bibnamefont {Longchambon}},\ and\ \bibinfo {author}
  {\bibfnamefont {H.}~\bibnamefont {Perrin}},\ }\href
  {https://doi.org/10.1103/PhysRevA.97.043615} {\bibfield  {journal} {\bibinfo
  {journal} {Phys. Rev. A}\ }\textbf {\bibinfo {volume} {97}},\ \bibinfo
  {pages} {043615} (\bibinfo {year} {2018})}\BibitemShut {NoStop}%
\bibitem [{\citenamefont {Andersen}\ \emph {et~al.}(2006)\citenamefont
  {Andersen}, \citenamefont {Ryu}, \citenamefont {Clad\'e}, \citenamefont
  {Natarajan}, \citenamefont {Vaziri}, \citenamefont {Helmerson},\ and\
  \citenamefont {Phillips}}]{andersen_2006_quantized}%
  \BibitemOpen
  \bibfield  {author} {\bibinfo {author} {\bibfnamefont {M.~F.}\ \bibnamefont
  {Andersen}}, \bibinfo {author} {\bibfnamefont {C.}~\bibnamefont {Ryu}},
  \bibinfo {author} {\bibfnamefont {P.}~\bibnamefont {Clad\'e}}, \bibinfo
  {author} {\bibfnamefont {V.}~\bibnamefont {Natarajan}}, \bibinfo {author}
  {\bibfnamefont {A.}~\bibnamefont {Vaziri}}, \bibinfo {author} {\bibfnamefont
  {K.}~\bibnamefont {Helmerson}},\ and\ \bibinfo {author} {\bibfnamefont
  {W.~D.}\ \bibnamefont {Phillips}},\ }\href
  {https://doi.org/10.1103/PhysRevLett.97.170406} {\bibfield  {journal}
  {\bibinfo  {journal} {Phys. Rev. Lett.}\ }\textbf {\bibinfo {volume} {97}},\
  \bibinfo {pages} {170406} (\bibinfo {year} {2006})}\BibitemShut {NoStop}%
\bibitem [{\citenamefont {Aikawa}\ \emph {et~al.}(2012)\citenamefont {Aikawa},
  \citenamefont {Frisch}, \citenamefont {Mark}, \citenamefont {Baier},
  \citenamefont {Rietzler}, \citenamefont {Grimm},\ and\ \citenamefont
  {Ferlaino}}]{aikawa_2012_boseeinstein}%
  \BibitemOpen
  \bibfield  {author} {\bibinfo {author} {\bibfnamefont {K.}~\bibnamefont
  {Aikawa}}, \bibinfo {author} {\bibfnamefont {A.}~\bibnamefont {Frisch}},
  \bibinfo {author} {\bibfnamefont {M.}~\bibnamefont {Mark}}, \bibinfo {author}
  {\bibfnamefont {S.}~\bibnamefont {Baier}}, \bibinfo {author} {\bibfnamefont
  {A.}~\bibnamefont {Rietzler}}, \bibinfo {author} {\bibfnamefont
  {R.}~\bibnamefont {Grimm}},\ and\ \bibinfo {author} {\bibfnamefont
  {F.}~\bibnamefont {Ferlaino}},\ }\href
  {https://doi.org/10.1103/PhysRevLett.108.210401} {\bibfield  {journal}
  {\bibinfo  {journal} {Phys. Rev. Lett.}\ }\textbf {\bibinfo {volume} {108}},\
  \bibinfo {pages} {210401} (\bibinfo {year} {2012})}\BibitemShut {NoStop}%
\bibitem [{\citenamefont {Tang}\ \emph
  {et~al.}(2015{\natexlab{a}})\citenamefont {Tang}, \citenamefont {Burdick},
  \citenamefont {Baumann},\ and\ \citenamefont {Lev}}]{tang_2015_bose}%
  \BibitemOpen
  \bibfield  {author} {\bibinfo {author} {\bibfnamefont {Y.}~\bibnamefont
  {Tang}}, \bibinfo {author} {\bibfnamefont {N.~Q.}\ \bibnamefont {Burdick}},
  \bibinfo {author} {\bibfnamefont {K.}~\bibnamefont {Baumann}},\ and\ \bibinfo
  {author} {\bibfnamefont {B.~L.}\ \bibnamefont {Lev}},\ }\href
  {https://doi.org/10.1088/1367-2630/17/4/045006} {\bibfield  {journal}
  {\bibinfo  {journal} {New Journal of Physics}\ }\textbf {\bibinfo {volume}
  {17}},\ \bibinfo {pages} {045006} (\bibinfo {year}
  {2015}{\natexlab{a}})}\BibitemShut {NoStop}%
\bibitem [{\citenamefont {Tang}\ \emph
  {et~al.}(2015{\natexlab{b}})\citenamefont {Tang}, \citenamefont {Sykes},
  \citenamefont {Burdick}, \citenamefont {Bohn},\ and\ \citenamefont
  {Lev}}]{tang_2015_wave}%
  \BibitemOpen
  \bibfield  {author} {\bibinfo {author} {\bibfnamefont {Y.}~\bibnamefont
  {Tang}}, \bibinfo {author} {\bibfnamefont {A.}~\bibnamefont {Sykes}},
  \bibinfo {author} {\bibfnamefont {N.~Q.}\ \bibnamefont {Burdick}}, \bibinfo
  {author} {\bibfnamefont {J.~L.}\ \bibnamefont {Bohn}},\ and\ \bibinfo
  {author} {\bibfnamefont {B.~L.}\ \bibnamefont {Lev}},\ }\href
  {https://doi.org/10.1103/PhysRevA.92.022703} {\bibfield  {journal} {\bibinfo
  {journal} {Physical Review A}\ }\textbf {\bibinfo {volume} {92}},\ \bibinfo
  {pages} {022703} (\bibinfo {year} {2015}{\natexlab{b}})}\BibitemShut
  {NoStop}%
\bibitem [{\citenamefont {Lucioni}\ \emph {et~al.}(2018)\citenamefont
  {Lucioni}, \citenamefont {Tanzi}, \citenamefont {Fregosi}, \citenamefont
  {Catani}, \citenamefont {Gozzini}, \citenamefont {Inguscio}, \citenamefont
  {Fioretti}, \citenamefont {Gabbanini},\ and\ \citenamefont
  {Modugno}}]{lucioni_2018_dysprosium}%
  \BibitemOpen
  \bibfield  {author} {\bibinfo {author} {\bibfnamefont {E.}~\bibnamefont
  {Lucioni}}, \bibinfo {author} {\bibfnamefont {L.}~\bibnamefont {Tanzi}},
  \bibinfo {author} {\bibfnamefont {A.}~\bibnamefont {Fregosi}}, \bibinfo
  {author} {\bibfnamefont {J.}~\bibnamefont {Catani}}, \bibinfo {author}
  {\bibfnamefont {S.}~\bibnamefont {Gozzini}}, \bibinfo {author} {\bibfnamefont
  {M.}~\bibnamefont {Inguscio}}, \bibinfo {author} {\bibfnamefont
  {A.}~\bibnamefont {Fioretti}}, \bibinfo {author} {\bibfnamefont
  {C.}~\bibnamefont {Gabbanini}},\ and\ \bibinfo {author} {\bibfnamefont
  {G.}~\bibnamefont {Modugno}},\ }\href
  {https://doi.org/10.1103/PhysRevA.97.060701} {\bibfield  {journal} {\bibinfo
  {journal} {Physical Review A}\ }\textbf {\bibinfo {volume} {97}},\ \bibinfo
  {pages} {060701} (\bibinfo {year} {2018})}\BibitemShut {NoStop}%
\bibitem [{\citenamefont {Chomaz}\ \emph {et~al.}(2022)\citenamefont {Chomaz},
  \citenamefont {{Ferrier-Barbut}}, \citenamefont {Ferlaino}, \citenamefont
  {{Laburthe-Tolra}}, \citenamefont {Lev},\ and\ \citenamefont
  {Pfau}}]{chomaz_2022_dipolar}%
  \BibitemOpen
  \bibfield  {author} {\bibinfo {author} {\bibfnamefont {L.}~\bibnamefont
  {Chomaz}}, \bibinfo {author} {\bibfnamefont {I.}~\bibnamefont
  {{Ferrier-Barbut}}}, \bibinfo {author} {\bibfnamefont {F.}~\bibnamefont
  {Ferlaino}}, \bibinfo {author} {\bibfnamefont {B.}~\bibnamefont
  {{Laburthe-Tolra}}}, \bibinfo {author} {\bibfnamefont {B.~L.}\ \bibnamefont
  {Lev}},\ and\ \bibinfo {author} {\bibfnamefont {T.}~\bibnamefont {Pfau}},\
  }\href {https://doi.org/10.1088/1361-6633/aca814} {\bibfield  {journal}
  {\bibinfo  {journal} {Reports on Progress in Physics}\ }\textbf {\bibinfo
  {volume} {86}},\ \bibinfo {pages} {026401} (\bibinfo {year}
  {2022})}\BibitemShut {NoStop}%
\bibitem [{\citenamefont {Prasad}\ \emph {et~al.}(2019)\citenamefont {Prasad},
  \citenamefont {Bland}, \citenamefont {Mulkerin}, \citenamefont {Parker},\
  and\ \citenamefont {Martin}}]{prasad_2019_vortex}%
  \BibitemOpen
  \bibfield  {author} {\bibinfo {author} {\bibfnamefont {S.~B.}\ \bibnamefont
  {Prasad}}, \bibinfo {author} {\bibfnamefont {T.}~\bibnamefont {Bland}},
  \bibinfo {author} {\bibfnamefont {B.~C.}\ \bibnamefont {Mulkerin}}, \bibinfo
  {author} {\bibfnamefont {N.~G.}\ \bibnamefont {Parker}},\ and\ \bibinfo
  {author} {\bibfnamefont {A.~M.}\ \bibnamefont {Martin}},\ }\href
  {https://doi.org/10.1103/PhysRevA.100.023625} {\bibfield  {journal} {\bibinfo
   {journal} {Phys. Rev. A}\ }\textbf {\bibinfo {volume} {100}},\ \bibinfo
  {pages} {023625} (\bibinfo {year} {2019})}\BibitemShut {NoStop}%
\bibitem [{\citenamefont {Prasad}\ \emph {et~al.}(2021)\citenamefont {Prasad},
  \citenamefont {Mulkerin},\ and\ \citenamefont
  {Martin}}]{prasad_2021_arbitraryangle}%
  \BibitemOpen
  \bibfield  {author} {\bibinfo {author} {\bibfnamefont {S.~B.}\ \bibnamefont
  {Prasad}}, \bibinfo {author} {\bibfnamefont {B.~C.}\ \bibnamefont
  {Mulkerin}},\ and\ \bibinfo {author} {\bibfnamefont {A.~M.}\ \bibnamefont
  {Martin}},\ }\href {https://doi.org/10.1103/PhysRevA.103.033322} {\bibfield
  {journal} {\bibinfo  {journal} {Phys. Rev. A}\ }\textbf {\bibinfo {volume}
  {103}},\ \bibinfo {pages} {033322} (\bibinfo {year} {2021})}\BibitemShut
  {NoStop}%
\bibitem [{\citenamefont {Klaus}\ \emph {et~al.}(2022)\citenamefont {Klaus},
  \citenamefont {Bland}, \citenamefont {Poli}, \citenamefont {Politi},
  \citenamefont {Lamporesi}, \citenamefont {Casotti}, \citenamefont {Bisset},
  \citenamefont {Mark},\ and\ \citenamefont
  {Ferlaino}}]{klaus_2022_observationa}%
  \BibitemOpen
  \bibfield  {author} {\bibinfo {author} {\bibfnamefont {L.}~\bibnamefont
  {Klaus}}, \bibinfo {author} {\bibfnamefont {T.}~\bibnamefont {Bland}},
  \bibinfo {author} {\bibfnamefont {E.}~\bibnamefont {Poli}}, \bibinfo {author}
  {\bibfnamefont {C.}~\bibnamefont {Politi}}, \bibinfo {author} {\bibfnamefont
  {G.}~\bibnamefont {Lamporesi}}, \bibinfo {author} {\bibfnamefont
  {E.}~\bibnamefont {Casotti}}, \bibinfo {author} {\bibfnamefont {R.~N.}\
  \bibnamefont {Bisset}}, \bibinfo {author} {\bibfnamefont {M.~J.}\
  \bibnamefont {Mark}},\ and\ \bibinfo {author} {\bibfnamefont
  {F.}~\bibnamefont {Ferlaino}},\ }\href
  {https://doi.org/10.1038/s41567-022-01793-8} {\bibfield  {journal} {\bibinfo
  {journal} {Nature Physics}\ }\textbf {\bibinfo {volume} {18}},\ \bibinfo
  {pages} {1453} (\bibinfo {year} {2022})}\BibitemShut {NoStop}%
\bibitem [{\citenamefont {Mulkerin}\ \emph {et~al.}(2013)\citenamefont
  {Mulkerin}, \citenamefont {van Bijnen}, \citenamefont {O'Dell}, \citenamefont
  {Martin},\ and\ \citenamefont {Parker}}]{mulkerin_2013_anisotropic}%
  \BibitemOpen
  \bibfield  {author} {\bibinfo {author} {\bibfnamefont {B.~C.}\ \bibnamefont
  {Mulkerin}}, \bibinfo {author} {\bibfnamefont {R.~M.~W.}\ \bibnamefont {van
  Bijnen}}, \bibinfo {author} {\bibfnamefont {D.~H.~J.}\ \bibnamefont
  {O'Dell}}, \bibinfo {author} {\bibfnamefont {A.~M.}\ \bibnamefont {Martin}},\
  and\ \bibinfo {author} {\bibfnamefont {N.~G.}\ \bibnamefont {Parker}},\
  }\href {https://doi.org/10.1103/PhysRevLett.111.170402} {\bibfield  {journal}
  {\bibinfo  {journal} {Phys. Rev. Lett.}\ }\textbf {\bibinfo {volume} {111}},\
  \bibinfo {pages} {170402} (\bibinfo {year} {2013})}\BibitemShut {NoStop}%
\bibitem [{\citenamefont {Gautam}(2014)}]{gautam_2014_dynamics}%
  \BibitemOpen
  \bibfield  {author} {\bibinfo {author} {\bibfnamefont {S.}~\bibnamefont
  {Gautam}},\ }\href@noop {} {\bibfield  {journal} {\bibinfo  {journal}
  {Journal of Physics B: Atomic, Molecular and Optical Physics}\ }\textbf
  {\bibinfo {volume} {47}},\ \bibinfo {pages} {165301} (\bibinfo {year}
  {2014})}\BibitemShut {NoStop}%
\bibitem [{\citenamefont {Yi}\ and\ \citenamefont {Pu}(2006)}]{yi_2006_vortex}%
  \BibitemOpen
  \bibfield  {author} {\bibinfo {author} {\bibfnamefont {S.}~\bibnamefont
  {Yi}}\ and\ \bibinfo {author} {\bibfnamefont {H.}~\bibnamefont {Pu}},\ }\href
  {https://doi.org/10.1103/PhysRevA.73.061602} {\bibfield  {journal} {\bibinfo
  {journal} {Phys. Rev. A}\ }\textbf {\bibinfo {volume} {73}},\ \bibinfo
  {pages} {061602} (\bibinfo {year} {2006})}\BibitemShut {NoStop}%
\bibitem [{\citenamefont {Cai}\ \emph {et~al.}(2018)\citenamefont {Cai},
  \citenamefont {Yuan}, \citenamefont {Rosenkranz}, \citenamefont {Pu},\ and\
  \citenamefont {Bao}}]{cai_2018_vortex}%
  \BibitemOpen
  \bibfield  {author} {\bibinfo {author} {\bibfnamefont {Y.}~\bibnamefont
  {Cai}}, \bibinfo {author} {\bibfnamefont {Y.}~\bibnamefont {Yuan}}, \bibinfo
  {author} {\bibfnamefont {M.}~\bibnamefont {Rosenkranz}}, \bibinfo {author}
  {\bibfnamefont {H.}~\bibnamefont {Pu}},\ and\ \bibinfo {author}
  {\bibfnamefont {W.}~\bibnamefont {Bao}},\ }\href
  {https://doi.org/10.1103/PhysRevA.98.023610} {\bibfield  {journal} {\bibinfo
  {journal} {Phys. Rev. A}\ }\textbf {\bibinfo {volume} {98}},\ \bibinfo
  {pages} {023610} (\bibinfo {year} {2018})}\BibitemShut {NoStop}%
\bibitem [{\citenamefont {Ticknor}\ \emph {et~al.}(2011)\citenamefont
  {Ticknor}, \citenamefont {Wilson},\ and\ \citenamefont
  {Bohn}}]{ticknor_2011_anisotropic}%
  \BibitemOpen
  \bibfield  {author} {\bibinfo {author} {\bibfnamefont {C.}~\bibnamefont
  {Ticknor}}, \bibinfo {author} {\bibfnamefont {R.~M.}\ \bibnamefont
  {Wilson}},\ and\ \bibinfo {author} {\bibfnamefont {J.~L.}\ \bibnamefont
  {Bohn}},\ }\href {https://doi.org/10.1103/PhysRevLett.106.065301} {\bibfield
  {journal} {\bibinfo  {journal} {Phys. Rev. Lett.}\ }\textbf {\bibinfo
  {volume} {106}},\ \bibinfo {pages} {065301} (\bibinfo {year}
  {2011})}\BibitemShut {NoStop}%
\bibitem [{\citenamefont {Bismut}\ \emph {et~al.}(2012)\citenamefont {Bismut},
  \citenamefont {Laburthe-Tolra}, \citenamefont {Mar\'echal}, \citenamefont
  {Pedri}, \citenamefont {Gorceix},\ and\ \citenamefont
  {Vernac}}]{bismut_2012_anisotropic}%
  \BibitemOpen
  \bibfield  {author} {\bibinfo {author} {\bibfnamefont {G.}~\bibnamefont
  {Bismut}}, \bibinfo {author} {\bibfnamefont {B.}~\bibnamefont
  {Laburthe-Tolra}}, \bibinfo {author} {\bibfnamefont {E.}~\bibnamefont
  {Mar\'echal}}, \bibinfo {author} {\bibfnamefont {P.}~\bibnamefont {Pedri}},
  \bibinfo {author} {\bibfnamefont {O.}~\bibnamefont {Gorceix}},\ and\ \bibinfo
  {author} {\bibfnamefont {L.}~\bibnamefont {Vernac}},\ }\href
  {https://doi.org/10.1103/PhysRevLett.109.155302} {\bibfield  {journal}
  {\bibinfo  {journal} {Phys. Rev. Lett.}\ }\textbf {\bibinfo {volume} {109}},\
  \bibinfo {pages} {155302} (\bibinfo {year} {2012})}\BibitemShut {NoStop}%
\bibitem [{\citenamefont {Wenzel}\ \emph {et~al.}(2018)\citenamefont {Wenzel},
  \citenamefont {B\"ottcher}, \citenamefont {Schmidt}, \citenamefont
  {Eisenmann}, \citenamefont {Langen}, \citenamefont {Pfau},\ and\
  \citenamefont {Ferrier-Barbut}}]{wenzel_2018_anisotropic}%
  \BibitemOpen
  \bibfield  {author} {\bibinfo {author} {\bibfnamefont {M.}~\bibnamefont
  {Wenzel}}, \bibinfo {author} {\bibfnamefont {F.}~\bibnamefont {B\"ottcher}},
  \bibinfo {author} {\bibfnamefont {J.-N.}\ \bibnamefont {Schmidt}}, \bibinfo
  {author} {\bibfnamefont {M.}~\bibnamefont {Eisenmann}}, \bibinfo {author}
  {\bibfnamefont {T.}~\bibnamefont {Langen}}, \bibinfo {author} {\bibfnamefont
  {T.}~\bibnamefont {Pfau}},\ and\ \bibinfo {author} {\bibfnamefont
  {I.}~\bibnamefont {Ferrier-Barbut}},\ }\href
  {https://doi.org/10.1103/PhysRevLett.121.030401} {\bibfield  {journal}
  {\bibinfo  {journal} {Phys. Rev. Lett.}\ }\textbf {\bibinfo {volume} {121}},\
  \bibinfo {pages} {030401} (\bibinfo {year} {2018})}\BibitemShut {NoStop}%
\bibitem [{\citenamefont {Kadau}\ \emph {et~al.}(2016)\citenamefont {Kadau},
  \citenamefont {Schmitt}, \citenamefont {Wenzel}, \citenamefont {Wink},
  \citenamefont {Maier}, \citenamefont {Ferrier-Barbut},\ and\ \citenamefont
  {Pfau}}]{kadau_2016_observing}%
  \BibitemOpen
  \bibfield  {author} {\bibinfo {author} {\bibfnamefont {H.}~\bibnamefont
  {Kadau}}, \bibinfo {author} {\bibfnamefont {M.}~\bibnamefont {Schmitt}},
  \bibinfo {author} {\bibfnamefont {M.}~\bibnamefont {Wenzel}}, \bibinfo
  {author} {\bibfnamefont {C.}~\bibnamefont {Wink}}, \bibinfo {author}
  {\bibfnamefont {T.}~\bibnamefont {Maier}}, \bibinfo {author} {\bibfnamefont
  {I.}~\bibnamefont {Ferrier-Barbut}},\ and\ \bibinfo {author} {\bibfnamefont
  {T.}~\bibnamefont {Pfau}},\ }\href {https://doi.org/10.1038/nature16485}
  {\bibfield  {journal} {\bibinfo  {journal} {Nature}\ }\textbf {\bibinfo
  {volume} {530}},\ \bibinfo {pages} {194} (\bibinfo {year}
  {2016})}\BibitemShut {NoStop}%
\bibitem [{\citenamefont {Bisset}\ \emph {et~al.}(2013)\citenamefont {Bisset},
  \citenamefont {Baillie},\ and\ \citenamefont {Blakie}}]{bisset_2013_roton}%
  \BibitemOpen
  \bibfield  {author} {\bibinfo {author} {\bibfnamefont {R.~N.}\ \bibnamefont
  {Bisset}}, \bibinfo {author} {\bibfnamefont {D.}~\bibnamefont {Baillie}},\
  and\ \bibinfo {author} {\bibfnamefont {P.~B.}\ \bibnamefont {Blakie}},\
  }\href {https://doi.org/10.1103/PhysRevA.88.043606} {\bibfield  {journal}
  {\bibinfo  {journal} {Phys. Rev. A}\ }\textbf {\bibinfo {volume} {88}},\
  \bibinfo {pages} {043606} (\bibinfo {year} {2013})}\BibitemShut {NoStop}%
\bibitem [{\citenamefont {Schmidt}\ \emph {et~al.}(2021)\citenamefont
  {Schmidt}, \citenamefont {Hertkorn}, \citenamefont {Guo}, \citenamefont
  {B\"ottcher}, \citenamefont {Schmidt}, \citenamefont {Ng}, \citenamefont
  {Graham}, \citenamefont {Langen}, \citenamefont {Zwierlein},\ and\
  \citenamefont {Pfau}}]{schmidt_2021_roton}%
  \BibitemOpen
  \bibfield  {author} {\bibinfo {author} {\bibfnamefont {J.-N.}\ \bibnamefont
  {Schmidt}}, \bibinfo {author} {\bibfnamefont {J.}~\bibnamefont {Hertkorn}},
  \bibinfo {author} {\bibfnamefont {M.}~\bibnamefont {Guo}}, \bibinfo {author}
  {\bibfnamefont {F.}~\bibnamefont {B\"ottcher}}, \bibinfo {author}
  {\bibfnamefont {M.}~\bibnamefont {Schmidt}}, \bibinfo {author} {\bibfnamefont
  {K.~S.~H.}\ \bibnamefont {Ng}}, \bibinfo {author} {\bibfnamefont {S.~D.}\
  \bibnamefont {Graham}}, \bibinfo {author} {\bibfnamefont {T.}~\bibnamefont
  {Langen}}, \bibinfo {author} {\bibfnamefont {M.}~\bibnamefont {Zwierlein}},\
  and\ \bibinfo {author} {\bibfnamefont {T.}~\bibnamefont {Pfau}},\ }\href
  {https://doi.org/10.1103/PhysRevLett.126.193002} {\bibfield  {journal}
  {\bibinfo  {journal} {Phys. Rev. Lett.}\ }\textbf {\bibinfo {volume} {126}},\
  \bibinfo {pages} {193002} (\bibinfo {year} {2021})}\BibitemShut {NoStop}%
\bibitem [{\citenamefont {Blakie}\ \emph {et~al.}(2012)\citenamefont {Blakie},
  \citenamefont {Baillie},\ and\ \citenamefont {Bisset}}]{blakie_2012_roton}%
  \BibitemOpen
  \bibfield  {author} {\bibinfo {author} {\bibfnamefont {P.~B.}\ \bibnamefont
  {Blakie}}, \bibinfo {author} {\bibfnamefont {D.}~\bibnamefont {Baillie}},\
  and\ \bibinfo {author} {\bibfnamefont {R.~N.}\ \bibnamefont {Bisset}},\
  }\href {https://doi.org/10.1103/PhysRevA.86.021604} {\bibfield  {journal}
  {\bibinfo  {journal} {Phys. Rev. A}\ }\textbf {\bibinfo {volume} {86}},\
  \bibinfo {pages} {021604} (\bibinfo {year} {2012})}\BibitemShut {NoStop}%
\bibitem [{\citenamefont {Petter}\ \emph {et~al.}(2019)\citenamefont {Petter},
  \citenamefont {Natale}, \citenamefont {van Bijnen}, \citenamefont
  {Patscheider}, \citenamefont {Mark}, \citenamefont {Chomaz},\ and\
  \citenamefont {Ferlaino}}]{petter_2019_probing}%
  \BibitemOpen
  \bibfield  {author} {\bibinfo {author} {\bibfnamefont {D.}~\bibnamefont
  {Petter}}, \bibinfo {author} {\bibfnamefont {G.}~\bibnamefont {Natale}},
  \bibinfo {author} {\bibfnamefont {R.~M.~W.}\ \bibnamefont {van Bijnen}},
  \bibinfo {author} {\bibfnamefont {A.}~\bibnamefont {Patscheider}}, \bibinfo
  {author} {\bibfnamefont {M.~J.}\ \bibnamefont {Mark}}, \bibinfo {author}
  {\bibfnamefont {L.}~\bibnamefont {Chomaz}},\ and\ \bibinfo {author}
  {\bibfnamefont {F.}~\bibnamefont {Ferlaino}},\ }\href
  {https://doi.org/10.1103/PhysRevLett.122.183401} {\bibfield  {journal}
  {\bibinfo  {journal} {Phys. Rev. Lett.}\ }\textbf {\bibinfo {volume} {122}},\
  \bibinfo {pages} {183401} (\bibinfo {year} {2019})}\BibitemShut {NoStop}%
\bibitem [{\citenamefont {Santos}\ \emph {et~al.}(2003)\citenamefont {Santos},
  \citenamefont {Shlyapnikov},\ and\ \citenamefont
  {Lewenstein}}]{santos_2003_rotonmaxon}%
  \BibitemOpen
  \bibfield  {author} {\bibinfo {author} {\bibfnamefont {L.}~\bibnamefont
  {Santos}}, \bibinfo {author} {\bibfnamefont {G.~V.}\ \bibnamefont
  {Shlyapnikov}},\ and\ \bibinfo {author} {\bibfnamefont {M.}~\bibnamefont
  {Lewenstein}},\ }\href {https://doi.org/10.1103/PhysRevLett.90.250403}
  {\bibfield  {journal} {\bibinfo  {journal} {Phys. Rev. Lett.}\ }\textbf
  {\bibinfo {volume} {90}},\ \bibinfo {pages} {250403} (\bibinfo {year}
  {2003})}\BibitemShut {NoStop}%
\bibitem [{\citenamefont {Natale}\ \emph {et~al.}(2019)\citenamefont {Natale},
  \citenamefont {van Bijnen}, \citenamefont {Patscheider}, \citenamefont
  {Petter}, \citenamefont {Mark}, \citenamefont {Chomaz},\ and\ \citenamefont
  {Ferlaino}}]{natale_2019_excitation}%
  \BibitemOpen
  \bibfield  {author} {\bibinfo {author} {\bibfnamefont {G.}~\bibnamefont
  {Natale}}, \bibinfo {author} {\bibfnamefont {R.~M.~W.}\ \bibnamefont {van
  Bijnen}}, \bibinfo {author} {\bibfnamefont {A.}~\bibnamefont {Patscheider}},
  \bibinfo {author} {\bibfnamefont {D.}~\bibnamefont {Petter}}, \bibinfo
  {author} {\bibfnamefont {M.~J.}\ \bibnamefont {Mark}}, \bibinfo {author}
  {\bibfnamefont {L.}~\bibnamefont {Chomaz}},\ and\ \bibinfo {author}
  {\bibfnamefont {F.}~\bibnamefont {Ferlaino}},\ }\href
  {https://doi.org/10.1103/PhysRevLett.123.050402} {\bibfield  {journal}
  {\bibinfo  {journal} {Phys. Rev. Lett.}\ }\textbf {\bibinfo {volume} {123}},\
  \bibinfo {pages} {050402} (\bibinfo {year} {2019})}\BibitemShut {NoStop}%
\bibitem [{\citenamefont {Kirkby}\ \emph {et~al.}(2023)\citenamefont {Kirkby},
  \citenamefont {Bland}, \citenamefont {Ferlaino},\ and\ \citenamefont
  {Bisset}}]{kirkby_2023_spin}%
  \BibitemOpen
  \bibfield  {author} {\bibinfo {author} {\bibfnamefont {W.}~\bibnamefont
  {Kirkby}}, \bibinfo {author} {\bibfnamefont {T.}~\bibnamefont {Bland}},
  \bibinfo {author} {\bibfnamefont {F.}~\bibnamefont {Ferlaino}},\ and\
  \bibinfo {author} {\bibfnamefont {R.~N.}\ \bibnamefont {Bisset}},\ }\href
  {https://doi.org/10.21468/SciPostPhysCore.6.4.084} {\bibfield  {journal}
  {\bibinfo  {journal} {SciPost Phys. Core}\ }\textbf {\bibinfo {volume} {6}},\
  \bibinfo {pages} {084} (\bibinfo {year} {2023})}\BibitemShut {NoStop}%
\bibitem [{\citenamefont {Chomaz}\ \emph {et~al.}(2018)\citenamefont {Chomaz},
  \citenamefont {van Bijnen}, \citenamefont {Petter}, \citenamefont {Faraoni},
  \citenamefont {Baier}, \citenamefont {Becher}, \citenamefont {Mark},
  \citenamefont {Waechtler}, \citenamefont {Santos},\ and\ \citenamefont
  {Ferlaino}}]{chomaz_2018_observation}%
  \BibitemOpen
  \bibfield  {author} {\bibinfo {author} {\bibfnamefont {L.}~\bibnamefont
  {Chomaz}}, \bibinfo {author} {\bibfnamefont {R.~M.}\ \bibnamefont {van
  Bijnen}}, \bibinfo {author} {\bibfnamefont {D.}~\bibnamefont {Petter}},
  \bibinfo {author} {\bibfnamefont {G.}~\bibnamefont {Faraoni}}, \bibinfo
  {author} {\bibfnamefont {S.}~\bibnamefont {Baier}}, \bibinfo {author}
  {\bibfnamefont {J.~H.}\ \bibnamefont {Becher}}, \bibinfo {author}
  {\bibfnamefont {M.~J.}\ \bibnamefont {Mark}}, \bibinfo {author}
  {\bibfnamefont {F.}~\bibnamefont {Waechtler}}, \bibinfo {author}
  {\bibfnamefont {L.}~\bibnamefont {Santos}},\ and\ \bibinfo {author}
  {\bibfnamefont {F.}~\bibnamefont {Ferlaino}},\ }\href
  {https://doi.org/https://doi.org/10.1038/s41567-018-0054-7} {\bibfield
  {journal} {\bibinfo  {journal} {Nature physics}\ }\textbf {\bibinfo {volume}
  {14}},\ \bibinfo {pages} {442} (\bibinfo {year} {2018})}\BibitemShut
  {NoStop}%
\bibitem [{\citenamefont {Lee}\ \emph {et~al.}(2022)\citenamefont {Lee},
  \citenamefont {Baillie},\ and\ \citenamefont {Blakie}}]{lee_2022_stability}%
  \BibitemOpen
  \bibfield  {author} {\bibinfo {author} {\bibfnamefont {A.-C.}\ \bibnamefont
  {Lee}}, \bibinfo {author} {\bibfnamefont {D.}~\bibnamefont {Baillie}},\ and\
  \bibinfo {author} {\bibfnamefont {P.~B.}\ \bibnamefont {Blakie}},\ }\href
  {https://doi.org/10.1103/PhysRevResearch.4.033153} {\bibfield  {journal}
  {\bibinfo  {journal} {Physical Review Research}\ }\textbf {\bibinfo {volume}
  {4}},\ \bibinfo {pages} {033153} (\bibinfo {year} {2022})}\BibitemShut
  {NoStop}%
\bibitem [{\citenamefont {Baillie}\ \emph {et~al.}(2016)\citenamefont
  {Baillie}, \citenamefont {Wilson}, \citenamefont {Bisset},\ and\
  \citenamefont {Blakie}}]{baillie_2016_selfbound}%
  \BibitemOpen
  \bibfield  {author} {\bibinfo {author} {\bibfnamefont {D.}~\bibnamefont
  {Baillie}}, \bibinfo {author} {\bibfnamefont {R.~M.}\ \bibnamefont {Wilson}},
  \bibinfo {author} {\bibfnamefont {R.~N.}\ \bibnamefont {Bisset}},\ and\
  \bibinfo {author} {\bibfnamefont {P.~B.}\ \bibnamefont {Blakie}},\ }\href
  {https://doi.org/10.1103/PhysRevA.94.021602} {\bibfield  {journal} {\bibinfo
  {journal} {Phys. Rev. A}\ }\textbf {\bibinfo {volume} {94}},\ \bibinfo
  {pages} {021602} (\bibinfo {year} {2016})}\BibitemShut {NoStop}%
\bibitem [{\citenamefont {Ferrier-Barbut}\ \emph {et~al.}(2016)\citenamefont
  {Ferrier-Barbut}, \citenamefont {Kadau}, \citenamefont {Schmitt},
  \citenamefont {Wenzel},\ and\ \citenamefont
  {Pfau}}]{ferrier-barbut_2016_observation}%
  \BibitemOpen
  \bibfield  {author} {\bibinfo {author} {\bibfnamefont {I.}~\bibnamefont
  {Ferrier-Barbut}}, \bibinfo {author} {\bibfnamefont {H.}~\bibnamefont
  {Kadau}}, \bibinfo {author} {\bibfnamefont {M.}~\bibnamefont {Schmitt}},
  \bibinfo {author} {\bibfnamefont {M.}~\bibnamefont {Wenzel}},\ and\ \bibinfo
  {author} {\bibfnamefont {T.}~\bibnamefont {Pfau}},\ }\href
  {https://doi.org/10.1103/PhysRevLett.116.215301} {\bibfield  {journal}
  {\bibinfo  {journal} {Phys. Rev. Lett.}\ }\textbf {\bibinfo {volume} {116}},\
  \bibinfo {pages} {215301} (\bibinfo {year} {2016})}\BibitemShut {NoStop}%
\bibitem [{\citenamefont {Baillie}\ \emph {et~al.}(2017)\citenamefont
  {Baillie}, \citenamefont {Wilson},\ and\ \citenamefont
  {Blakie}}]{baillie_2017_collective}%
  \BibitemOpen
  \bibfield  {author} {\bibinfo {author} {\bibfnamefont {D.}~\bibnamefont
  {Baillie}}, \bibinfo {author} {\bibfnamefont {R.~M.}\ \bibnamefont
  {Wilson}},\ and\ \bibinfo {author} {\bibfnamefont {P.~B.}\ \bibnamefont
  {Blakie}},\ }\href {https://doi.org/10.1103/PhysRevLett.119.255302}
  {\bibfield  {journal} {\bibinfo  {journal} {Phys. Rev. Lett.}\ }\textbf
  {\bibinfo {volume} {119}},\ \bibinfo {pages} {255302} (\bibinfo {year}
  {2017})}\BibitemShut {NoStop}%
\bibitem [{\citenamefont {Schmitt}\ \emph {et~al.}(2016)\citenamefont
  {Schmitt}, \citenamefont {Wenzel}, \citenamefont {B{\"o}ttcher},
  \citenamefont {Ferrier-Barbut},\ and\ \citenamefont
  {Pfau}}]{schmitt_2016_selfbound}%
  \BibitemOpen
  \bibfield  {author} {\bibinfo {author} {\bibfnamefont {M.}~\bibnamefont
  {Schmitt}}, \bibinfo {author} {\bibfnamefont {M.}~\bibnamefont {Wenzel}},
  \bibinfo {author} {\bibfnamefont {F.}~\bibnamefont {B{\"o}ttcher}}, \bibinfo
  {author} {\bibfnamefont {I.}~\bibnamefont {Ferrier-Barbut}},\ and\ \bibinfo
  {author} {\bibfnamefont {T.}~\bibnamefont {Pfau}},\ }\href
  {https://doi.org/10.1038/nature20126} {\bibfield  {journal} {\bibinfo
  {journal} {Nature}\ }\textbf {\bibinfo {volume} {539}},\ \bibinfo {pages}
  {259} (\bibinfo {year} {2016})}\BibitemShut {NoStop}%
\bibitem [{\citenamefont {Chomaz}\ \emph {et~al.}(2016)\citenamefont {Chomaz},
  \citenamefont {Baier}, \citenamefont {Petter}, \citenamefont {Mark},
  \citenamefont {W\"achtler}, \citenamefont {Santos},\ and\ \citenamefont
  {Ferlaino}}]{chomaz_2016_quantumfluctuationdriven}%
  \BibitemOpen
  \bibfield  {author} {\bibinfo {author} {\bibfnamefont {L.}~\bibnamefont
  {Chomaz}}, \bibinfo {author} {\bibfnamefont {S.}~\bibnamefont {Baier}},
  \bibinfo {author} {\bibfnamefont {D.}~\bibnamefont {Petter}}, \bibinfo
  {author} {\bibfnamefont {M.~J.}\ \bibnamefont {Mark}}, \bibinfo {author}
  {\bibfnamefont {F.}~\bibnamefont {W\"achtler}}, \bibinfo {author}
  {\bibfnamefont {L.}~\bibnamefont {Santos}},\ and\ \bibinfo {author}
  {\bibfnamefont {F.}~\bibnamefont {Ferlaino}},\ }\href
  {https://doi.org/10.1103/PhysRevX.6.041039} {\bibfield  {journal} {\bibinfo
  {journal} {Phys. Rev. X}\ }\textbf {\bibinfo {volume} {6}},\ \bibinfo {pages}
  {041039} (\bibinfo {year} {2016})}\BibitemShut {NoStop}%
\bibitem [{\citenamefont {W\"achtler}\ and\ \citenamefont
  {Santos}(2016{\natexlab{a}})}]{wachtler_2016_groundstate}%
  \BibitemOpen
  \bibfield  {author} {\bibinfo {author} {\bibfnamefont {F.}~\bibnamefont
  {W\"achtler}}\ and\ \bibinfo {author} {\bibfnamefont {L.}~\bibnamefont
  {Santos}},\ }\href {https://doi.org/10.1103/PhysRevA.94.043618} {\bibfield
  {journal} {\bibinfo  {journal} {Phys. Rev. A}\ }\textbf {\bibinfo {volume}
  {94}},\ \bibinfo {pages} {043618} (\bibinfo {year}
  {2016}{\natexlab{a}})}\BibitemShut {NoStop}%
\bibitem [{\citenamefont {W\"achtler}\ and\ \citenamefont
  {Santos}(2016{\natexlab{b}})}]{wachtler_2016_quantum}%
  \BibitemOpen
  \bibfield  {author} {\bibinfo {author} {\bibfnamefont {F.}~\bibnamefont
  {W\"achtler}}\ and\ \bibinfo {author} {\bibfnamefont {L.}~\bibnamefont
  {Santos}},\ }\href {https://doi.org/10.1103/PhysRevA.93.061603} {\bibfield
  {journal} {\bibinfo  {journal} {Phys. Rev. A}\ }\textbf {\bibinfo {volume}
  {93}},\ \bibinfo {pages} {061603} (\bibinfo {year}
  {2016}{\natexlab{b}})}\BibitemShut {NoStop}%
\bibitem [{\citenamefont {Mishra}\ \emph {et~al.}(2020)\citenamefont {Mishra},
  \citenamefont {Santos},\ and\ \citenamefont {Nath}}]{mishra_2020_selfbound}%
  \BibitemOpen
  \bibfield  {author} {\bibinfo {author} {\bibfnamefont {C.}~\bibnamefont
  {Mishra}}, \bibinfo {author} {\bibfnamefont {L.}~\bibnamefont {Santos}},\
  and\ \bibinfo {author} {\bibfnamefont {R.}~\bibnamefont {Nath}},\ }\href
  {https://doi.org/10.1103/PhysRevLett.124.073402} {\bibfield  {journal}
  {\bibinfo  {journal} {Phys. Rev. Lett.}\ }\textbf {\bibinfo {volume} {124}},\
  \bibinfo {pages} {073402} (\bibinfo {year} {2020})}\BibitemShut {NoStop}%
\bibitem [{\citenamefont {Bisset}\ \emph {et~al.}(2021)\citenamefont {Bisset},
  \citenamefont {Ardila},\ and\ \citenamefont {Santos}}]{bisset_2021_quantum}%
  \BibitemOpen
  \bibfield  {author} {\bibinfo {author} {\bibfnamefont {R.~N.}\ \bibnamefont
  {Bisset}}, \bibinfo {author} {\bibfnamefont {L.~A. P.~n.}\ \bibnamefont
  {Ardila}},\ and\ \bibinfo {author} {\bibfnamefont {L.}~\bibnamefont
  {Santos}},\ }\href {https://doi.org/10.1103/PhysRevLett.126.025301}
  {\bibfield  {journal} {\bibinfo  {journal} {Phys. Rev. Lett.}\ }\textbf
  {\bibinfo {volume} {126}},\ \bibinfo {pages} {025301} (\bibinfo {year}
  {2021})}\BibitemShut {NoStop}%
\bibitem [{\citenamefont {Hertkorn}\ \emph {et~al.}(2021)\citenamefont
  {Hertkorn}, \citenamefont {Schmidt}, \citenamefont {Guo}, \citenamefont
  {B\"ottcher}, \citenamefont {Ng}, \citenamefont {Graham}, \citenamefont
  {Uerlings}, \citenamefont {B\"uchler}, \citenamefont {Langen}, \citenamefont
  {Zwierlein},\ and\ \citenamefont {Pfau}}]{hertkorn_2021_supersolidity}%
  \BibitemOpen
  \bibfield  {author} {\bibinfo {author} {\bibfnamefont {J.}~\bibnamefont
  {Hertkorn}}, \bibinfo {author} {\bibfnamefont {J.-N.}\ \bibnamefont
  {Schmidt}}, \bibinfo {author} {\bibfnamefont {M.}~\bibnamefont {Guo}},
  \bibinfo {author} {\bibfnamefont {F.}~\bibnamefont {B\"ottcher}}, \bibinfo
  {author} {\bibfnamefont {K.~S.~H.}\ \bibnamefont {Ng}}, \bibinfo {author}
  {\bibfnamefont {S.~D.}\ \bibnamefont {Graham}}, \bibinfo {author}
  {\bibfnamefont {P.}~\bibnamefont {Uerlings}}, \bibinfo {author}
  {\bibfnamefont {H.~P.}\ \bibnamefont {B\"uchler}}, \bibinfo {author}
  {\bibfnamefont {T.}~\bibnamefont {Langen}}, \bibinfo {author} {\bibfnamefont
  {M.}~\bibnamefont {Zwierlein}},\ and\ \bibinfo {author} {\bibfnamefont
  {T.}~\bibnamefont {Pfau}},\ }\href
  {https://doi.org/10.1103/PhysRevLett.127.155301} {\bibfield  {journal}
  {\bibinfo  {journal} {Phys. Rev. Lett.}\ }\textbf {\bibinfo {volume} {127}},\
  \bibinfo {pages} {155301} (\bibinfo {year} {2021})}\BibitemShut {NoStop}%
\bibitem [{\citenamefont {Tanzi}\ \emph {et~al.}(2019)\citenamefont {Tanzi},
  \citenamefont {Lucioni}, \citenamefont {Fam\`a}, \citenamefont {Catani},
  \citenamefont {Fioretti}, \citenamefont {Gabbanini}, \citenamefont {Bisset},
  \citenamefont {Santos},\ and\ \citenamefont
  {Modugno}}]{tanzi_2019_observation}%
  \BibitemOpen
  \bibfield  {author} {\bibinfo {author} {\bibfnamefont {L.}~\bibnamefont
  {Tanzi}}, \bibinfo {author} {\bibfnamefont {E.}~\bibnamefont {Lucioni}},
  \bibinfo {author} {\bibfnamefont {F.}~\bibnamefont {Fam\`a}}, \bibinfo
  {author} {\bibfnamefont {J.}~\bibnamefont {Catani}}, \bibinfo {author}
  {\bibfnamefont {A.}~\bibnamefont {Fioretti}}, \bibinfo {author}
  {\bibfnamefont {C.}~\bibnamefont {Gabbanini}}, \bibinfo {author}
  {\bibfnamefont {R.~N.}\ \bibnamefont {Bisset}}, \bibinfo {author}
  {\bibfnamefont {L.}~\bibnamefont {Santos}},\ and\ \bibinfo {author}
  {\bibfnamefont {G.}~\bibnamefont {Modugno}},\ }\href
  {https://doi.org/10.1103/PhysRevLett.122.130405} {\bibfield  {journal}
  {\bibinfo  {journal} {Phys. Rev. Lett.}\ }\textbf {\bibinfo {volume} {122}},\
  \bibinfo {pages} {130405} (\bibinfo {year} {2019})}\BibitemShut {NoStop}%
\bibitem [{\citenamefont {Smith}\ \emph {et~al.}(2023)\citenamefont {Smith},
  \citenamefont {Baillie},\ and\ \citenamefont
  {Blakie}}]{smith_2023_supersolidity}%
  \BibitemOpen
  \bibfield  {author} {\bibinfo {author} {\bibfnamefont {J.~C.}\ \bibnamefont
  {Smith}}, \bibinfo {author} {\bibfnamefont {D.}~\bibnamefont {Baillie}},\
  and\ \bibinfo {author} {\bibfnamefont {P.~B.}\ \bibnamefont {Blakie}},\
  }\href {https://doi.org/10.1103/PhysRevA.107.033301} {\bibfield  {journal}
  {\bibinfo  {journal} {Phys. Rev. A}\ }\textbf {\bibinfo {volume} {107}},\
  \bibinfo {pages} {033301} (\bibinfo {year} {2023})}\BibitemShut {NoStop}%
\bibitem [{\citenamefont {Blakie}\ \emph {et~al.}(2020)\citenamefont {Blakie},
  \citenamefont {Baillie}, \citenamefont {Chomaz},\ and\ \citenamefont
  {Ferlaino}}]{blakie_2020_supersolidity}%
  \BibitemOpen
  \bibfield  {author} {\bibinfo {author} {\bibfnamefont {P.~B.}\ \bibnamefont
  {Blakie}}, \bibinfo {author} {\bibfnamefont {D.}~\bibnamefont {Baillie}},
  \bibinfo {author} {\bibfnamefont {L.}~\bibnamefont {Chomaz}},\ and\ \bibinfo
  {author} {\bibfnamefont {F.}~\bibnamefont {Ferlaino}},\ }\href
  {https://doi.org/10.1103/PhysRevResearch.2.043318} {\bibfield  {journal}
  {\bibinfo  {journal} {Phys. Rev. Res.}\ }\textbf {\bibinfo {volume} {2}},\
  \bibinfo {pages} {043318} (\bibinfo {year} {2020})}\BibitemShut {NoStop}%
\bibitem [{\citenamefont {Poli}\ \emph {et~al.}(2021)\citenamefont {Poli},
  \citenamefont {Bland}, \citenamefont {Politi}, \citenamefont {Klaus},
  \citenamefont {Norcia}, \citenamefont {Ferlaino}, \citenamefont {Bisset},\
  and\ \citenamefont {Santos}}]{poli_2021_maintaining}%
  \BibitemOpen
  \bibfield  {author} {\bibinfo {author} {\bibfnamefont {E.}~\bibnamefont
  {Poli}}, \bibinfo {author} {\bibfnamefont {T.}~\bibnamefont {Bland}},
  \bibinfo {author} {\bibfnamefont {C.}~\bibnamefont {Politi}}, \bibinfo
  {author} {\bibfnamefont {L.}~\bibnamefont {Klaus}}, \bibinfo {author}
  {\bibfnamefont {M.~A.}\ \bibnamefont {Norcia}}, \bibinfo {author}
  {\bibfnamefont {F.}~\bibnamefont {Ferlaino}}, \bibinfo {author}
  {\bibfnamefont {R.~N.}\ \bibnamefont {Bisset}},\ and\ \bibinfo {author}
  {\bibfnamefont {L.}~\bibnamefont {Santos}},\ }\href
  {https://doi.org/10.1103/PhysRevA.104.063307} {\bibfield  {journal} {\bibinfo
   {journal} {Phys. Rev. A}\ }\textbf {\bibinfo {volume} {104}},\ \bibinfo
  {pages} {063307} (\bibinfo {year} {2021})}\BibitemShut {NoStop}%
\bibitem [{\citenamefont {Roccuzzo}\ and\ \citenamefont
  {Ancilotto}(2019)}]{roccuzzo_2019_supersolid}%
  \BibitemOpen
  \bibfield  {author} {\bibinfo {author} {\bibfnamefont {S.~M.}\ \bibnamefont
  {Roccuzzo}}\ and\ \bibinfo {author} {\bibfnamefont {F.}~\bibnamefont
  {Ancilotto}},\ }\href {https://doi.org/10.1103/PhysRevA.99.041601} {\bibfield
   {journal} {\bibinfo  {journal} {Phys. Rev. A}\ }\textbf {\bibinfo {volume}
  {99}},\ \bibinfo {pages} {041601} (\bibinfo {year} {2019})}\BibitemShut
  {NoStop}%
\bibitem [{\citenamefont {Norcia}\ \emph {et~al.}(2021)\citenamefont {Norcia},
  \citenamefont {Politi}, \citenamefont {Klaus}, \citenamefont {Poli},
  \citenamefont {Sohmen}, \citenamefont {Mark}, \citenamefont {Bisset},
  \citenamefont {Santos},\ and\ \citenamefont
  {Ferlaino}}]{norcia_2021_twodimensional}%
  \BibitemOpen
  \bibfield  {author} {\bibinfo {author} {\bibfnamefont {M.~A.}\ \bibnamefont
  {Norcia}}, \bibinfo {author} {\bibfnamefont {C.}~\bibnamefont {Politi}},
  \bibinfo {author} {\bibfnamefont {L.}~\bibnamefont {Klaus}}, \bibinfo
  {author} {\bibfnamefont {E.}~\bibnamefont {Poli}}, \bibinfo {author}
  {\bibfnamefont {M.}~\bibnamefont {Sohmen}}, \bibinfo {author} {\bibfnamefont
  {M.~J.}\ \bibnamefont {Mark}}, \bibinfo {author} {\bibfnamefont {R.~N.}\
  \bibnamefont {Bisset}}, \bibinfo {author} {\bibfnamefont {L.}~\bibnamefont
  {Santos}},\ and\ \bibinfo {author} {\bibfnamefont {F.}~\bibnamefont
  {Ferlaino}},\ }\href {https://doi.org/10.1038/s41586-021-03725-7} {\bibfield
  {journal} {\bibinfo  {journal} {Nature}\ }\textbf {\bibinfo {volume} {596}},\
  \bibinfo {pages} {357} (\bibinfo {year} {2021})}\BibitemShut {NoStop}%
\bibitem [{\citenamefont {Bland}\ \emph
  {et~al.}(2022{\natexlab{a}})\citenamefont {Bland}, \citenamefont {Poli},
  \citenamefont {Politi}, \citenamefont {Klaus}, \citenamefont {Norcia},
  \citenamefont {Ferlaino}, \citenamefont {Santos},\ and\ \citenamefont
  {Bisset}}]{bland_2022_twodimensional}%
  \BibitemOpen
  \bibfield  {author} {\bibinfo {author} {\bibfnamefont {T.}~\bibnamefont
  {Bland}}, \bibinfo {author} {\bibfnamefont {E.}~\bibnamefont {Poli}},
  \bibinfo {author} {\bibfnamefont {C.}~\bibnamefont {Politi}}, \bibinfo
  {author} {\bibfnamefont {L.}~\bibnamefont {Klaus}}, \bibinfo {author}
  {\bibfnamefont {M.~A.}\ \bibnamefont {Norcia}}, \bibinfo {author}
  {\bibfnamefont {F.}~\bibnamefont {Ferlaino}}, \bibinfo {author}
  {\bibfnamefont {L.}~\bibnamefont {Santos}},\ and\ \bibinfo {author}
  {\bibfnamefont {R.~N.}\ \bibnamefont {Bisset}},\ }\href
  {https://doi.org/10.1103/PhysRevLett.128.195302} {\bibfield  {journal}
  {\bibinfo  {journal} {Phys. Rev. Lett.}\ }\textbf {\bibinfo {volume} {128}},\
  \bibinfo {pages} {195302} (\bibinfo {year} {2022}{\natexlab{a}})}\BibitemShut
  {NoStop}%
\bibitem [{\citenamefont {Halder}\ \emph {et~al.}(2022)\citenamefont {Halder},
  \citenamefont {Mukherjee}, \citenamefont {Mistakidis}, \citenamefont {Das},
  \citenamefont {Kevrekidis}, \citenamefont {Panigrahi}, \citenamefont
  {Majumder},\ and\ \citenamefont {Sadeghpour}}]{halder_2022_control}%
  \BibitemOpen
  \bibfield  {author} {\bibinfo {author} {\bibfnamefont {S.}~\bibnamefont
  {Halder}}, \bibinfo {author} {\bibfnamefont {K.}~\bibnamefont {Mukherjee}},
  \bibinfo {author} {\bibfnamefont {S.~I.}\ \bibnamefont {Mistakidis}},
  \bibinfo {author} {\bibfnamefont {S.}~\bibnamefont {Das}}, \bibinfo {author}
  {\bibfnamefont {P.~G.}\ \bibnamefont {Kevrekidis}}, \bibinfo {author}
  {\bibfnamefont {P.~K.}\ \bibnamefont {Panigrahi}}, \bibinfo {author}
  {\bibfnamefont {S.}~\bibnamefont {Majumder}},\ and\ \bibinfo {author}
  {\bibfnamefont {H.~R.}\ \bibnamefont {Sadeghpour}},\ }\href
  {https://doi.org/10.1103/PhysRevResearch.4.043124} {\bibfield  {journal}
  {\bibinfo  {journal} {Phys. Rev. Res.}\ }\textbf {\bibinfo {volume} {4}},\
  \bibinfo {pages} {043124} (\bibinfo {year} {2022})}\BibitemShut {NoStop}%
\bibitem [{\citenamefont {Schmidt}\ \emph {et~al.}(2022)\citenamefont
  {Schmidt}, \citenamefont {Lassabli\`ere}, \citenamefont {Qu\'em\'ener},\ and\
  \citenamefont {Langen}}]{schmidt_2022_selfbound}%
  \BibitemOpen
  \bibfield  {author} {\bibinfo {author} {\bibfnamefont {M.}~\bibnamefont
  {Schmidt}}, \bibinfo {author} {\bibfnamefont {L.}~\bibnamefont
  {Lassabli\`ere}}, \bibinfo {author} {\bibfnamefont {G.}~\bibnamefont
  {Qu\'em\'ener}},\ and\ \bibinfo {author} {\bibfnamefont {T.}~\bibnamefont
  {Langen}},\ }\href {https://doi.org/10.1103/PhysRevResearch.4.013235}
  {\bibfield  {journal} {\bibinfo  {journal} {Phys. Rev. Res.}\ }\textbf
  {\bibinfo {volume} {4}},\ \bibinfo {pages} {013235} (\bibinfo {year}
  {2022})}\BibitemShut {NoStop}%
\bibitem [{\citenamefont {Langen}\ \emph {et~al.}(2024)\citenamefont {Langen},
  \citenamefont {Valtolina}, \citenamefont {Wang},\ and\ \citenamefont
  {Ye}}]{langen_2024_quantum}%
  \BibitemOpen
  \bibfield  {author} {\bibinfo {author} {\bibfnamefont {T.}~\bibnamefont
  {Langen}}, \bibinfo {author} {\bibfnamefont {G.}~\bibnamefont {Valtolina}},
  \bibinfo {author} {\bibfnamefont {D.}~\bibnamefont {Wang}},\ and\ \bibinfo
  {author} {\bibfnamefont {J.}~\bibnamefont {Ye}},\ }\href
  {https://doi.org/10.1038/s41567-024-02423-1} {\bibfield  {journal} {\bibinfo
  {journal} {Nature Physics}\ }\textbf {\bibinfo {volume} {20}},\ \bibinfo
  {pages} {702} (\bibinfo {year} {2024})}\BibitemShut {NoStop}%
\bibitem [{\citenamefont {Bigagli}\ \emph {et~al.}(2024)\citenamefont
  {Bigagli}, \citenamefont {Yuan}, \citenamefont {Zhang}, \citenamefont
  {Bulatovic}, \citenamefont {Karman}, \citenamefont {Stevenson},\ and\
  \citenamefont {Will}}]{bigagli_2024_observation}%
  \BibitemOpen
  \bibfield  {author} {\bibinfo {author} {\bibfnamefont {N.}~\bibnamefont
  {Bigagli}}, \bibinfo {author} {\bibfnamefont {W.}~\bibnamefont {Yuan}},
  \bibinfo {author} {\bibfnamefont {S.}~\bibnamefont {Zhang}}, \bibinfo
  {author} {\bibfnamefont {B.}~\bibnamefont {Bulatovic}}, \bibinfo {author}
  {\bibfnamefont {T.}~\bibnamefont {Karman}}, \bibinfo {author} {\bibfnamefont
  {I.}~\bibnamefont {Stevenson}},\ and\ \bibinfo {author} {\bibfnamefont
  {S.}~\bibnamefont {Will}},\ }\href
  {https://doi.org/10.1038/s41586-024-07492-z} {\bibfield  {journal} {\bibinfo
  {journal} {Nature}\ ,\ \bibinfo {pages} {1}} (\bibinfo {year}
  {2024})}\BibitemShut {NoStop}%
\bibitem [{\citenamefont {Ghosh}\ \emph {et~al.}(2024)\citenamefont {Ghosh},
  \citenamefont {Halder}, \citenamefont {Das},\ and\ \citenamefont
  {Majumder}}]{ghosh_2024_induced}%
  \BibitemOpen
  \bibfield  {author} {\bibinfo {author} {\bibfnamefont {H.~S.}\ \bibnamefont
  {Ghosh}}, \bibinfo {author} {\bibfnamefont {S.}~\bibnamefont {Halder}},
  \bibinfo {author} {\bibfnamefont {S.}~\bibnamefont {Das}},\ and\ \bibinfo
  {author} {\bibfnamefont {S.}~\bibnamefont {Majumder}},\ }\href@noop {}
  {\bibinfo {title} {Induced supersolidity and hypersonic flow of a dipolar
  bose-einstein condensate in a rotating bubble trap}} (\bibinfo {year}
  {2024}),\ \Eprint {https://arxiv.org/abs/2402.13422} {arXiv:2402.13422
  [cond-mat.quant-gas]} \BibitemShut {NoStop}%
\bibitem [{\citenamefont {S\'anchez-Baena}\ \emph {et~al.}(2024)\citenamefont
  {S\'anchez-Baena}, \citenamefont {Bomb\'{\i}n},\ and\ \citenamefont
  {Boronat}}]{sanchez-baena_2024_ring}%
  \BibitemOpen
  \bibfield  {author} {\bibinfo {author} {\bibfnamefont {J.}~\bibnamefont
  {S\'anchez-Baena}}, \bibinfo {author} {\bibfnamefont {R.}~\bibnamefont
  {Bomb\'{\i}n}},\ and\ \bibinfo {author} {\bibfnamefont {J.}~\bibnamefont
  {Boronat}},\ }\href {https://doi.org/10.1103/PhysRevResearch.6.033116}
  {\bibfield  {journal} {\bibinfo  {journal} {Phys. Rev. Res.}\ }\textbf
  {\bibinfo {volume} {6}},\ \bibinfo {pages} {033116} (\bibinfo {year}
  {2024})}\BibitemShut {NoStop}%
\bibitem [{\citenamefont {Ciardi}\ \emph {et~al.}(2024)\citenamefont {Ciardi},
  \citenamefont {Cinti}, \citenamefont {Pellicane},\ and\ \citenamefont
  {Prestipino}}]{ciardi_2024_supersolid}%
  \BibitemOpen
  \bibfield  {author} {\bibinfo {author} {\bibfnamefont {M.}~\bibnamefont
  {Ciardi}}, \bibinfo {author} {\bibfnamefont {F.}~\bibnamefont {Cinti}},
  \bibinfo {author} {\bibfnamefont {G.}~\bibnamefont {Pellicane}},\ and\
  \bibinfo {author} {\bibfnamefont {S.}~\bibnamefont {Prestipino}},\ }\href
  {https://doi.org/10.1103/PhysRevLett.132.026001} {\bibfield  {journal}
  {\bibinfo  {journal} {Phys. Rev. Lett.}\ }\textbf {\bibinfo {volume} {132}},\
  \bibinfo {pages} {026001} (\bibinfo {year} {2024})}\BibitemShut {NoStop}%
\bibitem [{\citenamefont {Prestipino}\ and\ \citenamefont
  {Giaquinta}(2019)}]{prestipino_2019_ground}%
  \BibitemOpen
  \bibfield  {author} {\bibinfo {author} {\bibfnamefont {S.}~\bibnamefont
  {Prestipino}}\ and\ \bibinfo {author} {\bibfnamefont {P.~V.}\ \bibnamefont
  {Giaquinta}},\ }\href {https://doi.org/10.1103/PhysRevA.99.063619} {\bibfield
   {journal} {\bibinfo  {journal} {Phys. Rev. A}\ }\textbf {\bibinfo {volume}
  {99}},\ \bibinfo {pages} {063619} (\bibinfo {year} {2019})}\BibitemShut
  {NoStop}%
\bibitem [{\citenamefont {Halder}\ \emph {et~al.}(2023)\citenamefont {Halder},
  \citenamefont {Das},\ and\ \citenamefont
  {Majumder}}]{halder_2023_twodimensional}%
  \BibitemOpen
  \bibfield  {author} {\bibinfo {author} {\bibfnamefont {S.}~\bibnamefont
  {Halder}}, \bibinfo {author} {\bibfnamefont {S.}~\bibnamefont {Das}},\ and\
  \bibinfo {author} {\bibfnamefont {S.}~\bibnamefont {Majumder}},\ }\href
  {https://doi.org/10.1103/PhysRevA.107.063303} {\bibfield  {journal} {\bibinfo
   {journal} {Phys. Rev. A}\ }\textbf {\bibinfo {volume} {107}},\ \bibinfo
  {pages} {063303} (\bibinfo {year} {2023})}\BibitemShut {NoStop}%
\bibitem [{\citenamefont {Halder}\ \emph {et~al.}(2024)\citenamefont {Halder},
  \citenamefont {Das},\ and\ \citenamefont {Majumder}}]{halder_2024_induced}%
  \BibitemOpen
  \bibfield  {author} {\bibinfo {author} {\bibfnamefont {S.}~\bibnamefont
  {Halder}}, \bibinfo {author} {\bibfnamefont {S.}~\bibnamefont {Das}},\ and\
  \bibinfo {author} {\bibfnamefont {S.}~\bibnamefont {Majumder}},\ }\href
  {https://doi.org/10.1103/PhysRevA.109.063321} {\bibfield  {journal} {\bibinfo
   {journal} {Phys. Rev. A}\ }\textbf {\bibinfo {volume} {109}},\ \bibinfo
  {pages} {063321} (\bibinfo {year} {2024})}\BibitemShut {NoStop}%
\bibitem [{\citenamefont {Scheiermann}\ \emph {et~al.}(2023)\citenamefont
  {Scheiermann}, \citenamefont {Ardila}, \citenamefont {Bland}, \citenamefont
  {Bisset},\ and\ \citenamefont {Santos}}]{scheiermann_2023_catalyzation}%
  \BibitemOpen
  \bibfield  {author} {\bibinfo {author} {\bibfnamefont {D.}~\bibnamefont
  {Scheiermann}}, \bibinfo {author} {\bibfnamefont {L.~A. P.~n.}\ \bibnamefont
  {Ardila}}, \bibinfo {author} {\bibfnamefont {T.}~\bibnamefont {Bland}},
  \bibinfo {author} {\bibfnamefont {R.~N.}\ \bibnamefont {Bisset}},\ and\
  \bibinfo {author} {\bibfnamefont {L.}~\bibnamefont {Santos}},\ }\href
  {https://doi.org/10.1103/PhysRevA.107.L021302} {\bibfield  {journal}
  {\bibinfo  {journal} {Phys. Rev. A}\ }\textbf {\bibinfo {volume} {107}},\
  \bibinfo {pages} {L021302} (\bibinfo {year} {2023})}\BibitemShut {NoStop}%
\bibitem [{\citenamefont {Bland}\ \emph
  {et~al.}(2022{\natexlab{b}})\citenamefont {Bland}, \citenamefont {Poli},
  \citenamefont {Ardila}, \citenamefont {Santos}, \citenamefont {Ferlaino},\
  and\ \citenamefont {Bisset}}]{bland_2022_alternatingdomain}%
  \BibitemOpen
  \bibfield  {author} {\bibinfo {author} {\bibfnamefont {T.}~\bibnamefont
  {Bland}}, \bibinfo {author} {\bibfnamefont {E.}~\bibnamefont {Poli}},
  \bibinfo {author} {\bibfnamefont {L.~A. P.~n.}\ \bibnamefont {Ardila}},
  \bibinfo {author} {\bibfnamefont {L.}~\bibnamefont {Santos}}, \bibinfo
  {author} {\bibfnamefont {F.}~\bibnamefont {Ferlaino}},\ and\ \bibinfo
  {author} {\bibfnamefont {R.~N.}\ \bibnamefont {Bisset}},\ }\href
  {https://doi.org/10.1103/PhysRevA.106.053322} {\bibfield  {journal} {\bibinfo
   {journal} {Phys. Rev. A}\ }\textbf {\bibinfo {volume} {106}},\ \bibinfo
  {pages} {053322} (\bibinfo {year} {2022}{\natexlab{b}})}\BibitemShut
  {NoStop}%
\bibitem [{\citenamefont {Li}\ \emph {et~al.}(2022)\citenamefont {Li},
  \citenamefont {Le},\ and\ \citenamefont {Saito}}]{li_2022_longlifetime}%
  \BibitemOpen
  \bibfield  {author} {\bibinfo {author} {\bibfnamefont {S.}~\bibnamefont
  {Li}}, \bibinfo {author} {\bibfnamefont {U.~N.}\ \bibnamefont {Le}},\ and\
  \bibinfo {author} {\bibfnamefont {H.}~\bibnamefont {Saito}},\ }\href
  {https://doi.org/10.1103/PhysRevA.105.L061302} {\bibfield  {journal}
  {\bibinfo  {journal} {Phys. Rev. A}\ }\textbf {\bibinfo {volume} {105}},\
  \bibinfo {pages} {L061302} (\bibinfo {year} {2022})}\BibitemShut {NoStop}%
\bibitem [{\citenamefont {Arazo}\ \emph {et~al.}(2023)\citenamefont {Arazo},
  \citenamefont {Gallem\'{\i}}, \citenamefont {Guilleumas}, \citenamefont
  {Mayol},\ and\ \citenamefont {Santos}}]{arazo_2023_selfbound}%
  \BibitemOpen
  \bibfield  {author} {\bibinfo {author} {\bibfnamefont {M.}~\bibnamefont
  {Arazo}}, \bibinfo {author} {\bibfnamefont {A.}~\bibnamefont {Gallem\'{\i}}},
  \bibinfo {author} {\bibfnamefont {M.}~\bibnamefont {Guilleumas}}, \bibinfo
  {author} {\bibfnamefont {R.}~\bibnamefont {Mayol}},\ and\ \bibinfo {author}
  {\bibfnamefont {L.}~\bibnamefont {Santos}},\ }\href
  {https://doi.org/10.1103/PhysRevResearch.5.043038} {\bibfield  {journal}
  {\bibinfo  {journal} {Phys. Rev. Res.}\ }\textbf {\bibinfo {volume} {5}},\
  \bibinfo {pages} {043038} (\bibinfo {year} {2023})}\BibitemShut {NoStop}%
\bibitem [{\citenamefont {Cinti}\ \emph {et~al.}(2010)\citenamefont {Cinti},
  \citenamefont {Jain}, \citenamefont {Boninsegni}, \citenamefont {Micheli},
  \citenamefont {Zoller},\ and\ \citenamefont
  {Pupillo}}]{cinti_2010_supersolid}%
  \BibitemOpen
  \bibfield  {author} {\bibinfo {author} {\bibfnamefont {F.}~\bibnamefont
  {Cinti}}, \bibinfo {author} {\bibfnamefont {P.}~\bibnamefont {Jain}},
  \bibinfo {author} {\bibfnamefont {M.}~\bibnamefont {Boninsegni}}, \bibinfo
  {author} {\bibfnamefont {A.}~\bibnamefont {Micheli}}, \bibinfo {author}
  {\bibfnamefont {P.}~\bibnamefont {Zoller}},\ and\ \bibinfo {author}
  {\bibfnamefont {G.}~\bibnamefont {Pupillo}},\ }\href
  {https://doi.org/10.1103/PhysRevLett.105.135301} {\bibfield  {journal}
  {\bibinfo  {journal} {Physical Review Letters}\ }\textbf {\bibinfo {volume}
  {105}},\ \bibinfo {pages} {135301} (\bibinfo {year} {2010})}\BibitemShut
  {NoStop}%
\bibitem [{\citenamefont {Henkel}\ \emph {et~al.}(2010)\citenamefont {Henkel},
  \citenamefont {Nath},\ and\ \citenamefont
  {Pohl}}]{henkel_2010_threedimensional}%
  \BibitemOpen
  \bibfield  {author} {\bibinfo {author} {\bibfnamefont {N.}~\bibnamefont
  {Henkel}}, \bibinfo {author} {\bibfnamefont {R.}~\bibnamefont {Nath}},\ and\
  \bibinfo {author} {\bibfnamefont {T.}~\bibnamefont {Pohl}},\ }\href
  {https://doi.org/10.1103/PhysRevLett.104.195302} {\bibfield  {journal}
  {\bibinfo  {journal} {Physical Review Letters}\ }\textbf {\bibinfo {volume}
  {104}},\ \bibinfo {pages} {195302} (\bibinfo {year} {2010})}\BibitemShut
  {NoStop}%
\bibitem [{\citenamefont {Henkel}\ \emph {et~al.}(2012)\citenamefont {Henkel},
  \citenamefont {Cinti}, \citenamefont {Jain}, \citenamefont {Pupillo},\ and\
  \citenamefont {Pohl}}]{henkel_2012_supersolid}%
  \BibitemOpen
  \bibfield  {author} {\bibinfo {author} {\bibfnamefont {N.}~\bibnamefont
  {Henkel}}, \bibinfo {author} {\bibfnamefont {F.}~\bibnamefont {Cinti}},
  \bibinfo {author} {\bibfnamefont {P.}~\bibnamefont {Jain}}, \bibinfo {author}
  {\bibfnamefont {G.}~\bibnamefont {Pupillo}},\ and\ \bibinfo {author}
  {\bibfnamefont {T.}~\bibnamefont {Pohl}},\ }\href
  {https://doi.org/10.1103/PhysRevLett.108.265301} {\bibfield  {journal}
  {\bibinfo  {journal} {Physical Review Letters}\ }\textbf {\bibinfo {volume}
  {108}},\ \bibinfo {pages} {265301} (\bibinfo {year} {2012})}\BibitemShut
  {NoStop}%
\bibitem [{\citenamefont {Wang}\ \emph {et~al.}(2010)\citenamefont {Wang},
  \citenamefont {Gao}, \citenamefont {Jian},\ and\ \citenamefont
  {Zhai}}]{wang_2010_spinorbit}%
  \BibitemOpen
  \bibfield  {author} {\bibinfo {author} {\bibfnamefont {C.}~\bibnamefont
  {Wang}}, \bibinfo {author} {\bibfnamefont {C.}~\bibnamefont {Gao}}, \bibinfo
  {author} {\bibfnamefont {C.-M.}\ \bibnamefont {Jian}},\ and\ \bibinfo
  {author} {\bibfnamefont {H.}~\bibnamefont {Zhai}},\ }\href
  {https://doi.org/10.1103/PhysRevLett.105.160403} {\bibfield  {journal}
  {\bibinfo  {journal} {Physical Review Letters}\ }\textbf {\bibinfo {volume}
  {105}},\ \bibinfo {pages} {160403} (\bibinfo {year} {2010})}\BibitemShut
  {NoStop}%
\bibitem [{\citenamefont {Li}\ \emph {et~al.}(2013)\citenamefont {Li},
  \citenamefont {Martone}, \citenamefont {Pitaevskii},\ and\ \citenamefont
  {Stringari}}]{li_2013_superstripes}%
  \BibitemOpen
  \bibfield  {author} {\bibinfo {author} {\bibfnamefont {Y.}~\bibnamefont
  {Li}}, \bibinfo {author} {\bibfnamefont {G.~I.}\ \bibnamefont {Martone}},
  \bibinfo {author} {\bibfnamefont {L.~P.}\ \bibnamefont {Pitaevskii}},\ and\
  \bibinfo {author} {\bibfnamefont {S.}~\bibnamefont {Stringari}},\ }\href
  {https://doi.org/10.1103/PhysRevLett.110.235302} {\bibfield  {journal}
  {\bibinfo  {journal} {Physical Review Letters}\ }\textbf {\bibinfo {volume}
  {110}},\ \bibinfo {pages} {235302} (\bibinfo {year} {2013})}\BibitemShut
  {NoStop}%
\bibitem [{\citenamefont {Li}\ \emph {et~al.}(2016)\citenamefont {Li},
  \citenamefont {Huang}, \citenamefont {Shteynas}, \citenamefont {Burchesky},
  \citenamefont {Top}, \citenamefont {Su}, \citenamefont {Lee}, \citenamefont
  {Jamison},\ and\ \citenamefont {Ketterle}}]{li_2016_spinorbit}%
  \BibitemOpen
  \bibfield  {author} {\bibinfo {author} {\bibfnamefont {J.}~\bibnamefont
  {Li}}, \bibinfo {author} {\bibfnamefont {W.}~\bibnamefont {Huang}}, \bibinfo
  {author} {\bibfnamefont {B.}~\bibnamefont {Shteynas}}, \bibinfo {author}
  {\bibfnamefont {S.}~\bibnamefont {Burchesky}}, \bibinfo {author}
  {\bibfnamefont {F.~{\c C}.}\ \bibnamefont {Top}}, \bibinfo {author}
  {\bibfnamefont {E.}~\bibnamefont {Su}}, \bibinfo {author} {\bibfnamefont
  {J.}~\bibnamefont {Lee}}, \bibinfo {author} {\bibfnamefont {A.~O.}\
  \bibnamefont {Jamison}},\ and\ \bibinfo {author} {\bibfnamefont
  {W.}~\bibnamefont {Ketterle}},\ }\href
  {https://doi.org/10.1103/PhysRevLett.117.185301} {\bibfield  {journal}
  {\bibinfo  {journal} {Physical Review Letters}\ }\textbf {\bibinfo {volume}
  {117}},\ \bibinfo {pages} {185301} (\bibinfo {year} {2016})}\BibitemShut
  {NoStop}%
\bibitem [{\citenamefont {Li}\ \emph {et~al.}(2017)\citenamefont {Li},
  \citenamefont {Lee}, \citenamefont {Huang}, \citenamefont {Burchesky},
  \citenamefont {Shteynas}, \citenamefont {Top}, \citenamefont {Jamison},\ and\
  \citenamefont {Ketterle}}]{li_2017_stripe}%
  \BibitemOpen
  \bibfield  {author} {\bibinfo {author} {\bibfnamefont {J.-R.}\ \bibnamefont
  {Li}}, \bibinfo {author} {\bibfnamefont {J.}~\bibnamefont {Lee}}, \bibinfo
  {author} {\bibfnamefont {W.}~\bibnamefont {Huang}}, \bibinfo {author}
  {\bibfnamefont {S.}~\bibnamefont {Burchesky}}, \bibinfo {author}
  {\bibfnamefont {B.}~\bibnamefont {Shteynas}}, \bibinfo {author}
  {\bibfnamefont {F.~{\c C}.}\ \bibnamefont {Top}}, \bibinfo {author}
  {\bibfnamefont {A.~O.}\ \bibnamefont {Jamison}},\ and\ \bibinfo {author}
  {\bibfnamefont {W.}~\bibnamefont {Ketterle}},\ }\href
  {https://doi.org/10.1038/nature21431} {\bibfield  {journal} {\bibinfo
  {journal} {Nature}\ }\textbf {\bibinfo {volume} {543}},\ \bibinfo {pages}
  {91} (\bibinfo {year} {2017})}\BibitemShut {NoStop}%
\bibitem [{\citenamefont {Bersano}\ \emph {et~al.}(2019)\citenamefont
  {Bersano}, \citenamefont {Hou}, \citenamefont {Mossman}, \citenamefont
  {Gokhroo}, \citenamefont {Luo}, \citenamefont {Sun}, \citenamefont {Zhang},\
  and\ \citenamefont {Engels}}]{bersano_2019_experimental}%
  \BibitemOpen
  \bibfield  {author} {\bibinfo {author} {\bibfnamefont {T.~M.}\ \bibnamefont
  {Bersano}}, \bibinfo {author} {\bibfnamefont {J.}~\bibnamefont {Hou}},
  \bibinfo {author} {\bibfnamefont {S.}~\bibnamefont {Mossman}}, \bibinfo
  {author} {\bibfnamefont {V.}~\bibnamefont {Gokhroo}}, \bibinfo {author}
  {\bibfnamefont {X.-W.}\ \bibnamefont {Luo}}, \bibinfo {author} {\bibfnamefont
  {K.}~\bibnamefont {Sun}}, \bibinfo {author} {\bibfnamefont {C.}~\bibnamefont
  {Zhang}},\ and\ \bibinfo {author} {\bibfnamefont {P.}~\bibnamefont
  {Engels}},\ }\href {https://doi.org/10.1103/PhysRevA.99.051602} {\bibfield
  {journal} {\bibinfo  {journal} {Physical Review A}\ }\textbf {\bibinfo
  {volume} {99}},\ \bibinfo {pages} {051602} (\bibinfo {year}
  {2019})}\BibitemShut {NoStop}%
\bibitem [{\citenamefont {Putra}\ \emph {et~al.}(2020)\citenamefont {Putra},
  \citenamefont {{Salces-C{\'a}rcoba}}, \citenamefont {Yue}, \citenamefont
  {Sugawa},\ and\ \citenamefont {Spielman}}]{putra_2020_spatial}%
  \BibitemOpen
  \bibfield  {author} {\bibinfo {author} {\bibfnamefont {A.}~\bibnamefont
  {Putra}}, \bibinfo {author} {\bibfnamefont {F.}~\bibnamefont
  {{Salces-C{\'a}rcoba}}}, \bibinfo {author} {\bibfnamefont {Y.}~\bibnamefont
  {Yue}}, \bibinfo {author} {\bibfnamefont {S.}~\bibnamefont {Sugawa}},\ and\
  \bibinfo {author} {\bibfnamefont {I.~B.}\ \bibnamefont {Spielman}},\ }\href
  {https://doi.org/10.1103/PhysRevLett.124.053605} {\bibfield  {journal}
  {\bibinfo  {journal} {Physical Review Letters}\ }\textbf {\bibinfo {volume}
  {124}},\ \bibinfo {pages} {053605} (\bibinfo {year} {2020})}\BibitemShut
  {NoStop}%
\bibitem [{\citenamefont {Sachdeva}\ \emph {et~al.}(2020)\citenamefont
  {Sachdeva}, \citenamefont {Tengstrand},\ and\ \citenamefont
  {Reimann}}]{sachdeva_2020_selfbound}%
  \BibitemOpen
  \bibfield  {author} {\bibinfo {author} {\bibfnamefont {R.}~\bibnamefont
  {Sachdeva}}, \bibinfo {author} {\bibfnamefont {M.~N.}\ \bibnamefont
  {Tengstrand}},\ and\ \bibinfo {author} {\bibfnamefont {S.~M.}\ \bibnamefont
  {Reimann}},\ }\href {https://doi.org/10.1103/PhysRevA.102.043304} {\bibfield
  {journal} {\bibinfo  {journal} {Physical Review A}\ }\textbf {\bibinfo
  {volume} {102}},\ \bibinfo {pages} {043304} (\bibinfo {year}
  {2020})}\BibitemShut {NoStop}%
\bibitem [{\citenamefont {Geier}\ \emph {et~al.}(2021)\citenamefont {Geier},
  \citenamefont {Martone}, \citenamefont {Hauke},\ and\ \citenamefont
  {Stringari}}]{geier_2021_exciting}%
  \BibitemOpen
  \bibfield  {author} {\bibinfo {author} {\bibfnamefont {K.~T.}\ \bibnamefont
  {Geier}}, \bibinfo {author} {\bibfnamefont {G.~I.}\ \bibnamefont {Martone}},
  \bibinfo {author} {\bibfnamefont {P.}~\bibnamefont {Hauke}},\ and\ \bibinfo
  {author} {\bibfnamefont {S.}~\bibnamefont {Stringari}},\ }\href
  {https://doi.org/10.1103/PhysRevLett.127.115301} {\bibfield  {journal}
  {\bibinfo  {journal} {Phys. Rev. Lett.}\ }\textbf {\bibinfo {volume} {127}},\
  \bibinfo {pages} {115301} (\bibinfo {year} {2021})}\BibitemShut {NoStop}%
\bibitem [{\citenamefont {Roccuzzo}\ \emph {et~al.}(2022)\citenamefont
  {Roccuzzo}, \citenamefont {Recati},\ and\ \citenamefont
  {Stringari}}]{roccuzzo_2022_moment}%
  \BibitemOpen
  \bibfield  {author} {\bibinfo {author} {\bibfnamefont {S.~M.}\ \bibnamefont
  {Roccuzzo}}, \bibinfo {author} {\bibfnamefont {A.}~\bibnamefont {Recati}},\
  and\ \bibinfo {author} {\bibfnamefont {S.}~\bibnamefont {Stringari}},\ }\href
  {https://doi.org/10.1103/PhysRevA.105.023316} {\bibfield  {journal} {\bibinfo
   {journal} {Physical Review A}\ }\textbf {\bibinfo {volume} {105}},\ \bibinfo
  {pages} {023316} (\bibinfo {year} {2022})}\BibitemShut {NoStop}%
\bibitem [{\citenamefont {Pomeau}\ and\ \citenamefont
  {Rica}(1994)}]{pomeau_1994_dynamics}%
  \BibitemOpen
  \bibfield  {author} {\bibinfo {author} {\bibfnamefont {Y.}~\bibnamefont
  {Pomeau}}\ and\ \bibinfo {author} {\bibfnamefont {S.}~\bibnamefont {Rica}},\
  }\href {https://doi.org/10.1103/PhysRevLett.72.2426} {\bibfield  {journal}
  {\bibinfo  {journal} {Physical Review Letters}\ }\textbf {\bibinfo {volume}
  {72}},\ \bibinfo {pages} {2426} (\bibinfo {year} {1994})}\BibitemShut
  {NoStop}%
\bibitem [{\citenamefont {\ifmmode~\check{S}\else \v{S}\fi{}indik}\ \emph
  {et~al.}(2022)\citenamefont {\ifmmode~\check{S}\else \v{S}\fi{}indik},
  \citenamefont {Recati}, \citenamefont {Roccuzzo}, \citenamefont {Santos},\
  and\ \citenamefont {Stringari}}]{sindik_2022_creation}%
  \BibitemOpen
  \bibfield  {author} {\bibinfo {author} {\bibfnamefont {M.}~\bibnamefont
  {\ifmmode~\check{S}\else \v{S}\fi{}indik}}, \bibinfo {author} {\bibfnamefont
  {A.}~\bibnamefont {Recati}}, \bibinfo {author} {\bibfnamefont {S.~M.}\
  \bibnamefont {Roccuzzo}}, \bibinfo {author} {\bibfnamefont {L.}~\bibnamefont
  {Santos}},\ and\ \bibinfo {author} {\bibfnamefont {S.}~\bibnamefont
  {Stringari}},\ }\href {https://doi.org/10.1103/PhysRevA.106.L061303}
  {\bibfield  {journal} {\bibinfo  {journal} {Phys. Rev. A}\ }\textbf {\bibinfo
  {volume} {106}},\ \bibinfo {pages} {L061303} (\bibinfo {year}
  {2022})}\BibitemShut {NoStop}%
\bibitem [{\citenamefont {Gallem\'{\i}}\ \emph {et~al.}(2020)\citenamefont
  {Gallem\'{\i}}, \citenamefont {Roccuzzo}, \citenamefont {Stringari},\ and\
  \citenamefont {Recati}}]{gallemi_2020_quantized}%
  \BibitemOpen
  \bibfield  {author} {\bibinfo {author} {\bibfnamefont {A.}~\bibnamefont
  {Gallem\'{\i}}}, \bibinfo {author} {\bibfnamefont {S.~M.}\ \bibnamefont
  {Roccuzzo}}, \bibinfo {author} {\bibfnamefont {S.}~\bibnamefont
  {Stringari}},\ and\ \bibinfo {author} {\bibfnamefont {A.}~\bibnamefont
  {Recati}},\ }\href {https://doi.org/10.1103/PhysRevA.102.023322} {\bibfield
  {journal} {\bibinfo  {journal} {Phys. Rev. A}\ }\textbf {\bibinfo {volume}
  {102}},\ \bibinfo {pages} {023322} (\bibinfo {year} {2020})}\BibitemShut
  {NoStop}%
\bibitem [{\citenamefont {Roccuzzo}\ \emph {et~al.}(2020)\citenamefont
  {Roccuzzo}, \citenamefont {Gallem{\'i}}, \citenamefont {Recati},\ and\
  \citenamefont {Stringari}}]{roccuzzo_2020_rotating}%
  \BibitemOpen
  \bibfield  {author} {\bibinfo {author} {\bibfnamefont {S.~M.}\ \bibnamefont
  {Roccuzzo}}, \bibinfo {author} {\bibfnamefont {A.}~\bibnamefont
  {Gallem{\'i}}}, \bibinfo {author} {\bibfnamefont {A.}~\bibnamefont
  {Recati}},\ and\ \bibinfo {author} {\bibfnamefont {S.}~\bibnamefont
  {Stringari}},\ }\href {https://doi.org/10.1103/PhysRevLett.124.045702}
  {\bibfield  {journal} {\bibinfo  {journal} {Physical Review Letters}\
  }\textbf {\bibinfo {volume} {124}},\ \bibinfo {pages} {045702} (\bibinfo
  {year} {2020})}\BibitemShut {NoStop}%
\bibitem [{\citenamefont {Ancilotto}\ \emph {et~al.}(2021)\citenamefont
  {Ancilotto}, \citenamefont {Barranco}, \citenamefont {Pi},\ and\
  \citenamefont {Reatto}}]{ancilotto_2021_vortex}%
  \BibitemOpen
  \bibfield  {author} {\bibinfo {author} {\bibfnamefont {F.}~\bibnamefont
  {Ancilotto}}, \bibinfo {author} {\bibfnamefont {M.}~\bibnamefont {Barranco}},
  \bibinfo {author} {\bibfnamefont {M.}~\bibnamefont {Pi}},\ and\ \bibinfo
  {author} {\bibfnamefont {L.}~\bibnamefont {Reatto}},\ }\href
  {https://doi.org/10.1103/PhysRevA.103.033314} {\bibfield  {journal} {\bibinfo
   {journal} {Phys. Rev. A}\ }\textbf {\bibinfo {volume} {103}},\ \bibinfo
  {pages} {033314} (\bibinfo {year} {2021})}\BibitemShut {NoStop}%
\bibitem [{\citenamefont {Poli}\ \emph {et~al.}(2023)\citenamefont {Poli},
  \citenamefont {Bland}, \citenamefont {White}, \citenamefont {Mark},
  \citenamefont {Ferlaino}, \citenamefont {Trabucco},\ and\ \citenamefont
  {Mannarelli}}]{poli_2023_glitches}%
  \BibitemOpen
  \bibfield  {author} {\bibinfo {author} {\bibfnamefont {E.}~\bibnamefont
  {Poli}}, \bibinfo {author} {\bibfnamefont {T.}~\bibnamefont {Bland}},
  \bibinfo {author} {\bibfnamefont {S.~J.~M.}\ \bibnamefont {White}}, \bibinfo
  {author} {\bibfnamefont {M.~J.}\ \bibnamefont {Mark}}, \bibinfo {author}
  {\bibfnamefont {F.}~\bibnamefont {Ferlaino}}, \bibinfo {author}
  {\bibfnamefont {S.}~\bibnamefont {Trabucco}},\ and\ \bibinfo {author}
  {\bibfnamefont {M.}~\bibnamefont {Mannarelli}},\ }\href
  {https://doi.org/10.1103/PhysRevLett.131.223401} {\bibfield  {journal}
  {\bibinfo  {journal} {Phys. Rev. Lett.}\ }\textbf {\bibinfo {volume} {131}},\
  \bibinfo {pages} {223401} (\bibinfo {year} {2023})}\BibitemShut {NoStop}%
\bibitem [{\citenamefont {Casotti}\ \emph {et~al.}(2024)\citenamefont
  {Casotti}, \citenamefont {Poli}, \citenamefont {Klaus}, \citenamefont
  {Litvinov}, \citenamefont {Ulm}, \citenamefont {Politi}, \citenamefont
  {Mark}, \citenamefont {Bland},\ and\ \citenamefont
  {Ferlaino}}]{casotti_2024_observation}%
  \BibitemOpen
  \bibfield  {author} {\bibinfo {author} {\bibfnamefont {E.}~\bibnamefont
  {Casotti}}, \bibinfo {author} {\bibfnamefont {E.}~\bibnamefont {Poli}},
  \bibinfo {author} {\bibfnamefont {L.}~\bibnamefont {Klaus}}, \bibinfo
  {author} {\bibfnamefont {A.}~\bibnamefont {Litvinov}}, \bibinfo {author}
  {\bibfnamefont {C.}~\bibnamefont {Ulm}}, \bibinfo {author} {\bibfnamefont
  {C.}~\bibnamefont {Politi}}, \bibinfo {author} {\bibfnamefont {M.~J.}\
  \bibnamefont {Mark}}, \bibinfo {author} {\bibfnamefont {T.}~\bibnamefont
  {Bland}},\ and\ \bibinfo {author} {\bibfnamefont {F.}~\bibnamefont
  {Ferlaino}},\ }\href@noop {} {\bibinfo {title} {Observation of vortices in a
  dipolar supersolid}} (\bibinfo {year} {2024}),\ \Eprint
  {https://arxiv.org/abs/2403.18510} {arXiv:2403.18510 [cond-mat.quant-gas]}
  \BibitemShut {NoStop}%
\bibitem [{\citenamefont {Dalfovo}\ and\ \citenamefont
  {Stringari}(1996)}]{dalfovo_1996_bosons}%
  \BibitemOpen
  \bibfield  {author} {\bibinfo {author} {\bibfnamefont {F.}~\bibnamefont
  {Dalfovo}}\ and\ \bibinfo {author} {\bibfnamefont {S.}~\bibnamefont
  {Stringari}},\ }\href {https://doi.org/10.1103/PhysRevA.53.2477} {\bibfield
  {journal} {\bibinfo  {journal} {Phys. Rev. A}\ }\textbf {\bibinfo {volume}
  {53}},\ \bibinfo {pages} {2477} (\bibinfo {year} {1996})}\BibitemShut
  {NoStop}%
\bibitem [{\citenamefont {Sinha}(1997)}]{sinha_1997_semiclassical}%
  \BibitemOpen
  \bibfield  {author} {\bibinfo {author} {\bibfnamefont {S.}~\bibnamefont
  {Sinha}},\ }\href {https://doi.org/10.1103/PhysRevA.55.4325} {\bibfield
  {journal} {\bibinfo  {journal} {Phys. Rev. A}\ }\textbf {\bibinfo {volume}
  {55}},\ \bibinfo {pages} {4325} (\bibinfo {year} {1997})}\BibitemShut
  {NoStop}%
\bibitem [{\citenamefont {Lundh}\ \emph {et~al.}(1997)\citenamefont {Lundh},
  \citenamefont {Pethick},\ and\ \citenamefont
  {Smith}}]{lundh_1997_zerotemperature}%
  \BibitemOpen
  \bibfield  {author} {\bibinfo {author} {\bibfnamefont {E.}~\bibnamefont
  {Lundh}}, \bibinfo {author} {\bibfnamefont {C.~J.}\ \bibnamefont {Pethick}},\
  and\ \bibinfo {author} {\bibfnamefont {H.}~\bibnamefont {Smith}},\ }\href
  {https://doi.org/10.1103/PhysRevA.55.2126} {\bibfield  {journal} {\bibinfo
  {journal} {Phys. Rev. A}\ }\textbf {\bibinfo {volume} {55}},\ \bibinfo
  {pages} {2126} (\bibinfo {year} {1997})}\BibitemShut {NoStop}%
\bibitem [{\citenamefont {Sinha}\ and\ \citenamefont
  {Castin}(2001)}]{sinha_2001_dynamic}%
  \BibitemOpen
  \bibfield  {author} {\bibinfo {author} {\bibfnamefont {S.}~\bibnamefont
  {Sinha}}\ and\ \bibinfo {author} {\bibfnamefont {Y.}~\bibnamefont {Castin}},\
  }\href {https://doi.org/10.1103/PhysRevLett.87.190402} {\bibfield  {journal}
  {\bibinfo  {journal} {Phys. Rev. Lett.}\ }\textbf {\bibinfo {volume} {87}},\
  \bibinfo {pages} {190402} (\bibinfo {year} {2001})}\BibitemShut {NoStop}%
\bibitem [{\citenamefont {Kasamatsu}\ \emph {et~al.}(2003)\citenamefont
  {Kasamatsu}, \citenamefont {Tsubota},\ and\ \citenamefont
  {Ueda}}]{kasamatsu_2003_nonlinear}%
  \BibitemOpen
  \bibfield  {author} {\bibinfo {author} {\bibfnamefont {K.}~\bibnamefont
  {Kasamatsu}}, \bibinfo {author} {\bibfnamefont {M.}~\bibnamefont {Tsubota}},\
  and\ \bibinfo {author} {\bibfnamefont {M.}~\bibnamefont {Ueda}},\ }\href
  {https://doi.org/10.1103/PhysRevA.67.033610} {\bibfield  {journal} {\bibinfo
  {journal} {Phys. Rev. A}\ }\textbf {\bibinfo {volume} {67}},\ \bibinfo
  {pages} {033610} (\bibinfo {year} {2003})}\BibitemShut {NoStop}%
\bibitem [{\citenamefont {Madison}\ \emph {et~al.}(2001)\citenamefont
  {Madison}, \citenamefont {Chevy}, \citenamefont {Bretin},\ and\ \citenamefont
  {Dalibard}}]{madison_2001_stationary}%
  \BibitemOpen
  \bibfield  {author} {\bibinfo {author} {\bibfnamefont {K.~W.}\ \bibnamefont
  {Madison}}, \bibinfo {author} {\bibfnamefont {F.}~\bibnamefont {Chevy}},
  \bibinfo {author} {\bibfnamefont {V.}~\bibnamefont {Bretin}},\ and\ \bibinfo
  {author} {\bibfnamefont {J.}~\bibnamefont {Dalibard}},\ }\href
  {https://doi.org/10.1103/PhysRevLett.86.4443} {\bibfield  {journal} {\bibinfo
   {journal} {Phys. Rev. Lett.}\ }\textbf {\bibinfo {volume} {86}},\ \bibinfo
  {pages} {4443} (\bibinfo {year} {2001})}\BibitemShut {NoStop}%
\bibitem [{\citenamefont {Parker}\ \emph {et~al.}(2006)\citenamefont {Parker},
  \citenamefont {van Bijnen},\ and\ \citenamefont
  {Martin}}]{parker_2006_instabilities}%
  \BibitemOpen
  \bibfield  {author} {\bibinfo {author} {\bibfnamefont {N.~G.}\ \bibnamefont
  {Parker}}, \bibinfo {author} {\bibfnamefont {R.~M.~W.}\ \bibnamefont {van
  Bijnen}},\ and\ \bibinfo {author} {\bibfnamefont {A.~M.}\ \bibnamefont
  {Martin}},\ }\href {https://doi.org/10.1103/PhysRevA.73.061603} {\bibfield
  {journal} {\bibinfo  {journal} {Phys. Rev. A}\ }\textbf {\bibinfo {volume}
  {73}},\ \bibinfo {pages} {061603} (\bibinfo {year} {2006})}\BibitemShut
  {NoStop}%
\bibitem [{\citenamefont {O'Dell}\ and\ \citenamefont
  {Eberlein}(2007)}]{odell_2007_vortex}%
  \BibitemOpen
  \bibfield  {author} {\bibinfo {author} {\bibfnamefont {D.~H.~J.}\
  \bibnamefont {O'Dell}}\ and\ \bibinfo {author} {\bibfnamefont
  {C.}~\bibnamefont {Eberlein}},\ }\href
  {https://doi.org/10.1103/PhysRevA.75.013604} {\bibfield  {journal} {\bibinfo
  {journal} {Phys. Rev. A}\ }\textbf {\bibinfo {volume} {75}},\ \bibinfo
  {pages} {013604} (\bibinfo {year} {2007})}\BibitemShut {NoStop}%
\bibitem [{\citenamefont {Martin}\ \emph {et~al.}(2017)\citenamefont {Martin},
  \citenamefont {Marchant}, \citenamefont {O'Dell},\ and\ \citenamefont
  {Parker}}]{martin_2017_vortices}%
  \BibitemOpen
  \bibfield  {author} {\bibinfo {author} {\bibfnamefont {A.~M.}\ \bibnamefont
  {Martin}}, \bibinfo {author} {\bibfnamefont {N.~G.}\ \bibnamefont
  {Marchant}}, \bibinfo {author} {\bibfnamefont {D.~H.~J.}\ \bibnamefont
  {O'Dell}},\ and\ \bibinfo {author} {\bibfnamefont {N.~G.}\ \bibnamefont
  {Parker}},\ }\href {https://doi.org/10.1088/1361-648X/aa53a6} {\bibfield
  {journal} {\bibinfo  {journal} {Journal of Physics: Condensed Matter}\
  }\textbf {\bibinfo {volume} {29}},\ \bibinfo {pages} {103004} (\bibinfo
  {year} {2017})}\BibitemShut {NoStop}%
\bibitem [{\citenamefont {van Bijnen}\ \emph {et~al.}(2007)\citenamefont {van
  Bijnen}, \citenamefont {O'Dell}, \citenamefont {Parker},\ and\ \citenamefont
  {Martin}}]{vanbijnen_2007_dynamical}%
  \BibitemOpen
  \bibfield  {author} {\bibinfo {author} {\bibfnamefont {R.~M.~W.}\
  \bibnamefont {van Bijnen}}, \bibinfo {author} {\bibfnamefont {D.~H.~J.}\
  \bibnamefont {O'Dell}}, \bibinfo {author} {\bibfnamefont {N.~G.}\
  \bibnamefont {Parker}},\ and\ \bibinfo {author} {\bibfnamefont {A.~M.}\
  \bibnamefont {Martin}},\ }\href
  {https://doi.org/10.1103/PhysRevLett.98.150401} {\bibfield  {journal}
  {\bibinfo  {journal} {Phys. Rev. Lett.}\ }\textbf {\bibinfo {volume} {98}},\
  \bibinfo {pages} {150401} (\bibinfo {year} {2007})}\BibitemShut {NoStop}%
\bibitem [{\citenamefont {van Bijnen}\ \emph {et~al.}(2009)\citenamefont {van
  Bijnen}, \citenamefont {Dow}, \citenamefont {O'Dell}, \citenamefont
  {Parker},\ and\ \citenamefont {Martin}}]{vanbijnen_2009_exact}%
  \BibitemOpen
  \bibfield  {author} {\bibinfo {author} {\bibfnamefont {R.~M.~W.}\
  \bibnamefont {van Bijnen}}, \bibinfo {author} {\bibfnamefont {A.~J.}\
  \bibnamefont {Dow}}, \bibinfo {author} {\bibfnamefont {D.~H.~J.}\
  \bibnamefont {O'Dell}}, \bibinfo {author} {\bibfnamefont {N.~G.}\
  \bibnamefont {Parker}},\ and\ \bibinfo {author} {\bibfnamefont {A.~M.}\
  \bibnamefont {Martin}},\ }\href {https://doi.org/10.1103/PhysRevA.80.033617}
  {\bibfield  {journal} {\bibinfo  {journal} {Phys. Rev. A}\ }\textbf {\bibinfo
  {volume} {80}},\ \bibinfo {pages} {033617} (\bibinfo {year}
  {2009})}\BibitemShut {NoStop}%
\bibitem [{\citenamefont {SCH\"{U}TZHOLD}\ \emph {et~al.}(2006)\citenamefont
  {SCH\"{U}TZHOLD}, \citenamefont {UHLMANN}, \citenamefont {XU},\ and\
  \citenamefont {FISCHER}}]{schutzhold_2006_meanfield}%
  \BibitemOpen
  \bibfield  {author} {\bibinfo {author} {\bibfnamefont {R.}~\bibnamefont
  {SCH\"{U}TZHOLD}}, \bibinfo {author} {\bibfnamefont {M.}~\bibnamefont
  {UHLMANN}}, \bibinfo {author} {\bibfnamefont {Y.}~\bibnamefont {XU}},\ and\
  \bibinfo {author} {\bibfnamefont {U.~R.}\ \bibnamefont {FISCHER}},\ }\href
  {https://doi.org/10.1142/S0217979206035631} {\bibfield  {journal} {\bibinfo
  {journal} {International Journal of Modern Physics B}\ }\textbf {\bibinfo
  {volume} {20}},\ \bibinfo {pages} {3555} (\bibinfo {year} {2006})},\ \Eprint
  {https://arxiv.org/abs/https://doi.org/10.1142/S0217979206035631}
  {https://doi.org/10.1142/S0217979206035631} \BibitemShut {NoStop}%
\bibitem [{\citenamefont {Lima}\ and\ \citenamefont
  {Pelster}(2011)}]{lima_2011_quantum}%
  \BibitemOpen
  \bibfield  {author} {\bibinfo {author} {\bibfnamefont {A.~R.~P.}\
  \bibnamefont {Lima}}\ and\ \bibinfo {author} {\bibfnamefont {A.}~\bibnamefont
  {Pelster}},\ }\href {https://doi.org/10.1103/PhysRevA.84.041604} {\bibfield
  {journal} {\bibinfo  {journal} {Phys. Rev. A}\ }\textbf {\bibinfo {volume}
  {84}},\ \bibinfo {pages} {041604} (\bibinfo {year} {2011})}\BibitemShut
  {NoStop}%
\bibitem [{\citenamefont {Lima}\ and\ \citenamefont
  {Pelster}(2012)}]{lima_2012_meanfield}%
  \BibitemOpen
  \bibfield  {author} {\bibinfo {author} {\bibfnamefont {A.~R.~P.}\
  \bibnamefont {Lima}}\ and\ \bibinfo {author} {\bibfnamefont {A.}~\bibnamefont
  {Pelster}},\ }\href {https://doi.org/10.1103/PhysRevA.86.063609} {\bibfield
  {journal} {\bibinfo  {journal} {Phys. Rev. A}\ }\textbf {\bibinfo {volume}
  {86}},\ \bibinfo {pages} {063609} (\bibinfo {year} {2012})}\BibitemShut
  {NoStop}%
\bibitem [{\citenamefont {Petrov}(2015)}]{petrov_2015_quantum}%
  \BibitemOpen
  \bibfield  {author} {\bibinfo {author} {\bibfnamefont {D.~S.}\ \bibnamefont
  {Petrov}},\ }\href {https://doi.org/10.1103/PhysRevLett.115.155302}
  {\bibfield  {journal} {\bibinfo  {journal} {Phys. Rev. Lett.}\ }\textbf
  {\bibinfo {volume} {115}},\ \bibinfo {pages} {155302} (\bibinfo {year}
  {2015})}\BibitemShut {NoStop}%
\bibitem [{\citenamefont {Bisset}\ \emph {et~al.}(2016)\citenamefont {Bisset},
  \citenamefont {Wilson}, \citenamefont {Baillie},\ and\ \citenamefont
  {Blakie}}]{bisset_2016_groundstate}%
  \BibitemOpen
  \bibfield  {author} {\bibinfo {author} {\bibfnamefont {R.~N.}\ \bibnamefont
  {Bisset}}, \bibinfo {author} {\bibfnamefont {R.~M.}\ \bibnamefont {Wilson}},
  \bibinfo {author} {\bibfnamefont {D.}~\bibnamefont {Baillie}},\ and\ \bibinfo
  {author} {\bibfnamefont {P.~B.}\ \bibnamefont {Blakie}},\ }\href
  {https://doi.org/10.1103/PhysRevA.94.033619} {\bibfield  {journal} {\bibinfo
  {journal} {Physical Review A}\ }\textbf {\bibinfo {volume} {94}},\ \bibinfo
  {pages} {033619} (\bibinfo {year} {2016})}\BibitemShut {NoStop}%
\bibitem [{fir()}]{firstnote}%
  \BibitemOpen
  \href@noop {} {}\bibinfo {note} {In an axially symmetric trap geometry the
  superfluid fraction is given by $f_s=1-\lim_{\Omega\to 0}\expval{L_z}/I
  \Omega$ \cite{leggett_1970_can}, where $I=\int
  \dd{r}(x^2+y^2)\abs{\psi(r)}^2$ is the classical rigid body moment of
  inertia.}\BibitemShut {Stop}%
\bibitem [{\citenamefont {Baillie}\ and\ \citenamefont
  {Blakie}(2020)}]{baillie_2020_rotational}%
  \BibitemOpen
  \bibfield  {author} {\bibinfo {author} {\bibfnamefont {D.}~\bibnamefont
  {Baillie}}\ and\ \bibinfo {author} {\bibfnamefont {P.~B.}\ \bibnamefont
  {Blakie}},\ }\href {https://doi.org/10.1103/PhysRevA.101.043606} {\bibfield
  {journal} {\bibinfo  {journal} {Phys. Rev. A}\ }\textbf {\bibinfo {volume}
  {101}},\ \bibinfo {pages} {043606} (\bibinfo {year} {2020})}\BibitemShut
  {NoStop}%
\bibitem [{\citenamefont {Biagioni}\ \emph {et~al.}(2022)\citenamefont
  {Biagioni}, \citenamefont {Antolini}, \citenamefont {Ala\~na}, \citenamefont
  {Modugno}, \citenamefont {Fioretti}, \citenamefont {Gabbanini}, \citenamefont
  {Tanzi},\ and\ \citenamefont {Modugno}}]{biagioni_2022_dimensional}%
  \BibitemOpen
  \bibfield  {author} {\bibinfo {author} {\bibfnamefont {G.}~\bibnamefont
  {Biagioni}}, \bibinfo {author} {\bibfnamefont {N.}~\bibnamefont {Antolini}},
  \bibinfo {author} {\bibfnamefont {A.}~\bibnamefont {Ala\~na}}, \bibinfo
  {author} {\bibfnamefont {M.}~\bibnamefont {Modugno}}, \bibinfo {author}
  {\bibfnamefont {A.}~\bibnamefont {Fioretti}}, \bibinfo {author}
  {\bibfnamefont {C.}~\bibnamefont {Gabbanini}}, \bibinfo {author}
  {\bibfnamefont {L.}~\bibnamefont {Tanzi}},\ and\ \bibinfo {author}
  {\bibfnamefont {G.}~\bibnamefont {Modugno}},\ }\href
  {https://doi.org/10.1103/PhysRevX.12.021019} {\bibfield  {journal} {\bibinfo
  {journal} {Phys. Rev. X}\ }\textbf {\bibinfo {volume} {12}},\ \bibinfo
  {pages} {021019} (\bibinfo {year} {2022})}\BibitemShut {NoStop}%
\bibitem [{\citenamefont {Stuhler}\ \emph {et~al.}(2005)\citenamefont
  {Stuhler}, \citenamefont {Griesmaier}, \citenamefont {Koch}, \citenamefont
  {Fattori}, \citenamefont {Pfau}, \citenamefont {Giovanazzi}, \citenamefont
  {Pedri},\ and\ \citenamefont {Santos}}]{stuhler_2005_observation}%
  \BibitemOpen
  \bibfield  {author} {\bibinfo {author} {\bibfnamefont {J.}~\bibnamefont
  {Stuhler}}, \bibinfo {author} {\bibfnamefont {A.}~\bibnamefont {Griesmaier}},
  \bibinfo {author} {\bibfnamefont {T.}~\bibnamefont {Koch}}, \bibinfo {author}
  {\bibfnamefont {M.}~\bibnamefont {Fattori}}, \bibinfo {author} {\bibfnamefont
  {T.}~\bibnamefont {Pfau}}, \bibinfo {author} {\bibfnamefont {S.}~\bibnamefont
  {Giovanazzi}}, \bibinfo {author} {\bibfnamefont {P.}~\bibnamefont {Pedri}},\
  and\ \bibinfo {author} {\bibfnamefont {L.}~\bibnamefont {Santos}},\ }\href
  {https://doi.org/10.1103/PhysRevLett.95.150406} {\bibfield  {journal}
  {\bibinfo  {journal} {Phys. Rev. Lett.}\ }\textbf {\bibinfo {volume} {95}},\
  \bibinfo {pages} {150406} (\bibinfo {year} {2005})}\BibitemShut {NoStop}%
\bibitem [{\citenamefont {Tang}\ \emph {et~al.}(2018)\citenamefont {Tang},
  \citenamefont {Kao}, \citenamefont {Li},\ and\ \citenamefont
  {Lev}}]{tang_2018_tuning}%
  \BibitemOpen
  \bibfield  {author} {\bibinfo {author} {\bibfnamefont {Y.}~\bibnamefont
  {Tang}}, \bibinfo {author} {\bibfnamefont {W.}~\bibnamefont {Kao}}, \bibinfo
  {author} {\bibfnamefont {K.-Y.}\ \bibnamefont {Li}},\ and\ \bibinfo {author}
  {\bibfnamefont {B.~L.}\ \bibnamefont {Lev}},\ }\href
  {https://doi.org/10.1103/PhysRevLett.120.230401} {\bibfield  {journal}
  {\bibinfo  {journal} {Phys. Rev. Lett.}\ }\textbf {\bibinfo {volume} {120}},\
  \bibinfo {pages} {230401} (\bibinfo {year} {2018})}\BibitemShut {NoStop}%
\bibitem [{\citenamefont {Li}\ \emph {et~al.}(2024)\citenamefont {Li},
  \citenamefont {Zhao}, \citenamefont {Jiang}, \citenamefont {Chen},
  \citenamefont {Liu}, \citenamefont {Malomed},\ and\ \citenamefont
  {Li}}]{li_2024_strongly}%
  \BibitemOpen
  \bibfield  {author} {\bibinfo {author} {\bibfnamefont {G.}~\bibnamefont
  {Li}}, \bibinfo {author} {\bibfnamefont {Z.}~\bibnamefont {Zhao}}, \bibinfo
  {author} {\bibfnamefont {X.}~\bibnamefont {Jiang}}, \bibinfo {author}
  {\bibfnamefont {Z.}~\bibnamefont {Chen}}, \bibinfo {author} {\bibfnamefont
  {B.}~\bibnamefont {Liu}}, \bibinfo {author} {\bibfnamefont {B.~A.}\
  \bibnamefont {Malomed}},\ and\ \bibinfo {author} {\bibfnamefont
  {Y.}~\bibnamefont {Li}},\ }\href@noop {} {\bibinfo {title} {Strongly
  anisotropic vortices in dipolar quantum droplets}} (\bibinfo {year} {2024}),\
  \Eprint {https://arxiv.org/abs/2310.17840} {arXiv:2310.17840 [cond-mat,
  physics:nlin]} \BibitemShut {NoStop}%
\bibitem [{\citenamefont {Das}\ and\ \citenamefont
  {Scarola}(2024)}]{das_2024_unveiling}%
  \BibitemOpen
  \bibfield  {author} {\bibinfo {author} {\bibfnamefont {S.}~\bibnamefont
  {Das}}\ and\ \bibinfo {author} {\bibfnamefont {V.~W.}\ \bibnamefont
  {Scarola}},\ }\href {https://doi.org/10.48550/arXiv.2407.02481} {\bibinfo
  {title} {Unveiling {{Supersolid Order}} via {{Vortex Trajectory
  Correlations}}}} (\bibinfo {year} {2024}),\ \Eprint
  {https://arxiv.org/abs/2407.02481} {arXiv:2407.02481 [cond-mat,
  physics:physics]} \BibitemShut {NoStop}%
\bibitem [{\citenamefont {Prasad}\ \emph {et~al.}(2024)\citenamefont {Prasad},
  \citenamefont {Parker},\ and\ \citenamefont
  {Baggaley}}]{prasad_2024_vortexpair}%
  \BibitemOpen
  \bibfield  {author} {\bibinfo {author} {\bibfnamefont {S.~B.}\ \bibnamefont
  {Prasad}}, \bibinfo {author} {\bibfnamefont {N.~G.}\ \bibnamefont {Parker}},\
  and\ \bibinfo {author} {\bibfnamefont {A.~W.}\ \bibnamefont {Baggaley}},\
  }\href {https://doi.org/10.1103/PhysRevA.109.063323} {\bibfield  {journal}
  {\bibinfo  {journal} {Phys. Rev. A}\ }\textbf {\bibinfo {volume} {109}},\
  \bibinfo {pages} {063323} (\bibinfo {year} {2024})}\BibitemShut {NoStop}%
\bibitem [{\citenamefont {Yo\ifmmode~\breve{g}\else \u{g}\fi{}urt}\ \emph
  {et~al.}(2023)\citenamefont {Yo\ifmmode~\breve{g}\else \u{g}\fi{}urt},
  \citenamefont {Tanyeri}, \citenamefont {Kele\ifmmode~\mbox{\c{s}}\else
  \c{s}\fi{}},\ and\ \citenamefont {Oktel}}]{yogurt_2023_vortex}%
  \BibitemOpen
  \bibfield  {author} {\bibinfo {author} {\bibfnamefont {T.~A.}\ \bibnamefont
  {Yo\ifmmode~\breve{g}\else \u{g}\fi{}urt}}, \bibinfo {author} {\bibfnamefont
  {U.}~\bibnamefont {Tanyeri}}, \bibinfo {author} {\bibfnamefont
  {A.}~\bibnamefont {Kele\ifmmode~\mbox{\c{s}}\else \c{s}\fi{}}},\ and\
  \bibinfo {author} {\bibfnamefont {M.~O.}\ \bibnamefont {Oktel}},\ }\href
  {https://doi.org/10.1103/PhysRevA.108.033315} {\bibfield  {journal} {\bibinfo
   {journal} {Phys. Rev. A}\ }\textbf {\bibinfo {volume} {108}},\ \bibinfo
  {pages} {033315} (\bibinfo {year} {2023})}\BibitemShut {NoStop}%
\bibitem [{\citenamefont {Sabari}\ \emph {et~al.}(2024)\citenamefont {Sabari},
  \citenamefont {Kishor~Kumar},\ and\ \citenamefont
  {Tomio}}]{sabari_2024_vortex}%
  \BibitemOpen
  \bibfield  {author} {\bibinfo {author} {\bibfnamefont {S.}~\bibnamefont
  {Sabari}}, \bibinfo {author} {\bibfnamefont {R.}~\bibnamefont
  {Kishor~Kumar}},\ and\ \bibinfo {author} {\bibfnamefont {L.}~\bibnamefont
  {Tomio}},\ }\href {https://doi.org/10.1103/PhysRevA.109.023313} {\bibfield
  {journal} {\bibinfo  {journal} {Phys. Rev. A}\ }\textbf {\bibinfo {volume}
  {109}},\ \bibinfo {pages} {023313} (\bibinfo {year} {2024})}\BibitemShut
  {NoStop}%
\bibitem [{\citenamefont {Leggett}(1970)}]{leggett_1970_can}%
  \BibitemOpen
  \bibfield  {author} {\bibinfo {author} {\bibfnamefont {A.~J.}\ \bibnamefont
  {Leggett}},\ }\href {https://doi.org/10.1103/PhysRevLett.25.1543} {\bibfield
  {journal} {\bibinfo  {journal} {Physical Review Letters}\ }\textbf {\bibinfo
  {volume} {25}},\ \bibinfo {pages} {1543} (\bibinfo {year}
  {1970})}\BibitemShut {NoStop}%
\end{thebibliography}%

\end{document}